\documentclass[fleqn,usenatbib]{mnras}

\usepackage{newtxtext,newtxmath}

\usepackage[T1]{fontenc}

\DeclareRobustCommand{\VAN}[3]{#2}
\let\VANthebibliography\thebibliography
\def\thebibliography{\DeclareRobustCommand{\VAN}[3]{##3}\VANthebibliography}

\usepackage{longtable}
\usepackage{graphicx}	
\usepackage{amsmath}	
\usepackage{subfig}

\title[Selecting High-z Candidates with Medium Bands]{The impact of medium-width bands on the selection, and subsequent luminosity function measurements, of high-z galaxies}

\author[N. J. Adams et al.]{
N. J. Adams$^{1}$\thanks{E-mail: nathan.adams@manchester.ac.uk},
D. Austin,$^{1}$
T. Harvey,$^{1}$
C. J. Conselice,$^{1}$
J.\@ A.\@ A.\@ Trussler,$^{2}$
Q. Li,$^{1}$
L. Westcott,$^{1}$ \newauthor
L. Ferreira,$^{3}$
V. Rusakov,$^{1}$
C. M. Goolsby,$^{1}$
\\
$^{1}$Jodrell Bank Centre for Astrophysics, University of Manchester, Oxford Road, Manchester, UK\\
$^{2}$ Center for Astrophysics | Harvard \& Smithsonian, 60 Garden St., Cambridge MA 02138 USA \\
$^{3}$School of Physics and Astronomy, University of Victoria, Victoria, BC, Canada
}

\date{Accepted XXX. Received YYY; in original form ZZZ}

\pubyear{2015}

\begin{document}
\label{firstpage}
\pagerange{\pageref{firstpage}--\pageref{lastpage}}
\maketitle

\begin{abstract}

New, ultra-deep medium-width photometric coverage with JWST's NIRCam instrument provides the potential for much improved photo-z reliability at high redshifts. In this study, we conduct a systematic analysis of the JADES Origins Field, which contains 14 broad- and medium-width near-infrared bands, to assess the benefits of medium band photometry on high-z completeness and contamination rates. Using imaging reaching AB mag $29.8-30.35$ depth, we test how high-z selections differ when images are artificially degraded or bands are removed. In parallel, the same experiments are conducted on simulated catalogues from the JAGUAR semi-analytic model to examine if observations can be replicated. We find sample completeness is high ($80\%+$) and contamination low ($<4\%$) when in the $10\sigma+$ regime, even without the use of any medium-width bands. The addition of medium-width bands leads to increases in completeness ($\sim10\%$) but multiple bands are required to improve contamination rates due to the small redshift ranges over which they probe strong emission lines. Incidents of Balmer-Lyman degeneracy increase in the $5-7\sigma$ regime and this can be replicated in both simulated catalogues and degraded real data. We measure the faint-end of the UV LF at $8.5<z<13.5$, finding high number densities that agree with previous JWST observations. Overall, medium bands are effective at increasing completeness and reducing contamination, but investment in achieving comparable depths in the blue ($<1.5\mu$m) as achieved in the red is also found to be key to fully reducing contamination from high-z samples.
\end{abstract}

\begin{keywords}
galaxies: abundances -- galaxies: distances and redshifts -- galaxies: high-redshift -- galaxies: photometry
\end{keywords}



\section{Introduction}

In the first few years since the successful commissioning of the James Webb Space Telescope (JWST), the study of the first billion years of the Universe's history has undergone a revolution. Exquisitely deep, high resolution imaging and spectroscopy has resulted in the photometric selection of $\sim1000$'s of high-z galaxies \citep[$z>6.5$ e.g.][]{Harikane2023,Adams2024,Hainline2024,Finkelstein2024,McLeod2024,Donnan2024,DSilva2025} and spectroscopic confirmation of $\sim100$'s \citep[e.g.][]{Bunker2024,DEugenio2024,Price2024,Mascia2024,Heintz2024}. These observations have revealed higher number densities of UV luminous galaxies than first expected from extrapolated HST observations \citep{Oesch2018,Harikane2022}, pushed the redshift frontier to $z\sim14$ \citep{CurtisLake2022,Wang2023,Carnani2024} and discovered previously unknown classifications of sources \citep[e.g. little red dots][]{labbe2023,matthee2023little,Kokorev2024}.

However, even though JWST provides substantially more information on the Universe than we previously had, and follow up spectroscopy has shown photometric selections have been largely successful, there have been at least two cases where robust looking sources at very high redshifts have turned out to be sample contaminants at lower redshift when examined with NIRSpec \citep{ArrabalHaro2023,Castellano2025}. Degeneracies between the Lyman or Balmer breaks in a galaxy's spectra and combinations of dust attenuation with the chance alignment of strong emission lines can result in cases where low-z galaxies (in this case $2<z<6$) enter photometric samples of high-z galaxies \citep[See also discussion in][for dusty sources]{Zavala2023,Meyer2024,Gandolfi2025}. While spectroscopy is the most powerful tool to obtain precise and accurate redshift measurements, JWST has begun to produce deep fields exceeding AB magnitudes of 30 in depth. In this faint regime, follow-up spectroscopy becomes increasingly time consuming, bordering infeasible in fact due to the anticipated costs to obtain required signal-to-noise ratios \citep{Giardino2022}. The use of medium-width photometric bands, while not producing the level of detail as spectroscopy, may provide a route to increase the confidence of high redshift solutions for ultra-faint sources with smaller overall costs. In Cycle 2, the accepted General Observer programme PID 3215 \citep[PI: D. Eisenstein][]{Eisenstein2024} produced ultra-deep imaging in multiple medium with bands over a region the GOODS-South field now dubbed the JADES Origins Field (JOF), reaching AB magnitudes of 30.3 in broad- and medium-width bands (0.32as diameter apertures), providing a total of 14 NIRCam bands compared to the more standard 7-8 that many other Cycle 1 and 2 programmes utilised.

 In the lead up to the launch of JWST, several studies assessed what systematics may plague the selection of high redshift galaxies and quantified what completeness and purities would be expected from hypothetical survey designs \citep[e.g.][]{Kauffmann2020,Hainline2020}. With both extreme depths and a wide range of photometric bands, the completed JADES Origins Field provides an opportunity to expand upon this early work given what we now know about JWST data and the initial work studying the early Universe using it. In this study, we explore the impacts that medium-band photometry has on the sample completeness and contamination rates when selecting galaxies at $z>7.5$. 
 
 From this we can also assess the potential impact that the use of medium-bands has on measuring the galaxy UV luminosity function, particularly at the faint-end where completeness and contamination rates vary the most. To accomplish this, we conduct two experiments. The first, we conduct a photo-z selection procedure on the JADES Origins Field using different combinations of the photometric bands available to more closely recreate other survey designs. We then repeat this procedure on images which are intentionally degraded with additional noise levels and assess how completeness and contamination behave given our prior knowledge of the original, multi-band, deep imaging. In parallel, we utilise the JAGUAR semi-analytic simulation and its photometric catalogues \citep{Williams2018} to explore if observed trends can be replicated and to trial a few hypothetical scenarios to inform future survey design.  We also examine the UV luminosity function for this field and how the different filter sets gives us different realisations of the LF.

This paper is structured as follows: in Section 2 we describe the data and reduction processes used, including the development of custom templates for `wisp' artefacts between 1.5-2.1 microns; In Section 3, we describe the photo-z and SED fitting processes used in this work and conduct an initial analysis of the JOF field; In Section 4 we repeat the analysis from Section 3 using different combinations of photometric bands in combination with simulated catalogues to assess how completeness and contamination may be affected; In Section 5 we intentionally degrade the quality of the JOF imaging and repeat the analysis again to observe how selections behave in the low SNR regime; In Section 6 we discuss our findings in the context of the faint-end of the UV Luminosity Function before summarising our conclusions in Section 7.

\section{Data}

\begin{table}
\caption{Average 5$\sigma$ depths calculated in 0.32 and 0.2 arcsecond diameter apertures for the combined PID 1210 and 3125 data sets. Quoted values are rounded to nearest 0.05 increment and represent the confidence on the raw measured value, not an aperture corrected value.}
\centering
\begin{tabular}{l|ll}
Filter      & Depth (0.32as) & Depth (0.2as) \\ \hline
F090W & 29.80 & 30.75        \\
F115W & 29.85 & 30.85                 \\ \hline
F150W & 30.10 & 31.10               \\
F162M & 29.90 & 30.90               \\
F182M & 30.30 & 31.25               \\
F200W & 30.10 & 31.05                \\
F210M & 30.10 & 31.05               \\ \hline
F250M          & 29.85 & 30.65     \\
F277W          & 30.35  &  31.20   \\
F300M          & 30.05  &  30.95   \\
F335M         & 30.20   &  31.00  \\
F356W           & 30.35 &  31.15    \\
F410M        & 29.80    &  30.50 \\
F444W        & 30.25    &  30.95 
\end{tabular}
\end{table}

This work focuses on the region dubbed the JADES Origins Field \citep[JOF][]{Eisenstein2024}. This is a single NIRCam pointing sub-region of the GOODS South field which has some of the deepest NIRCam data available from Cycles 1 and 2. In Cycle 1, the field was visited as a parallel to JADES NIRSpec observations of the Hubble Ultra Deep Field \citep[PID 1210, PI: N. Luetzgendorf][]{Eisenstein2023DR2}. In Cycle 2, a large programme (PID 3215, PI: D. Eisenstein) was awarded time to revisit this small patch of sky to add in complementary medium bands to comparable depths to the wide bands already available. This field is not only one of the deepest fields currently available, but also contains some of the widest variety of JWST imaging filters available with a total of 14 available (See also the JEMS \citet{Williams2023}, Technicolour and MegaScience programmes \citet{Suess2024}).

There is additional data available in this field through the shallower tiers of the wider JADES observing programme in the GOODS-S field. However, these data that are currently public do not uniformly cover the footprint of the JOF, and so we omit them on account of providing only marginal depth improvements at the loss of uniformity, which could complicate the experiments conducted in this work.

\begin{figure*}%
    \centering
    \subfloat[\centering Original v2 wisp template downloaded from STScI.]{{\includegraphics[width=5.25cm]{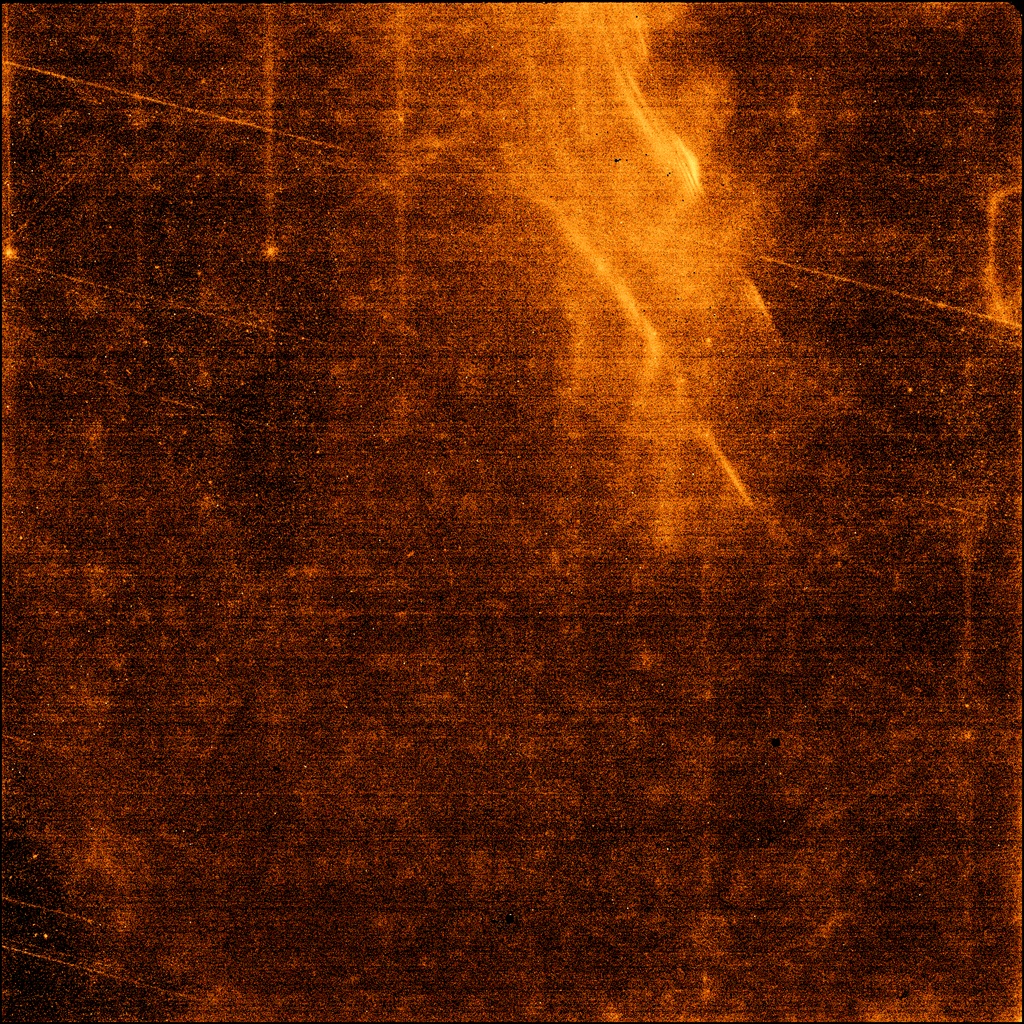} }}%
    \qquad
    \subfloat[\centering New wisp template derived in this work]{{\includegraphics[width=5.25cm]{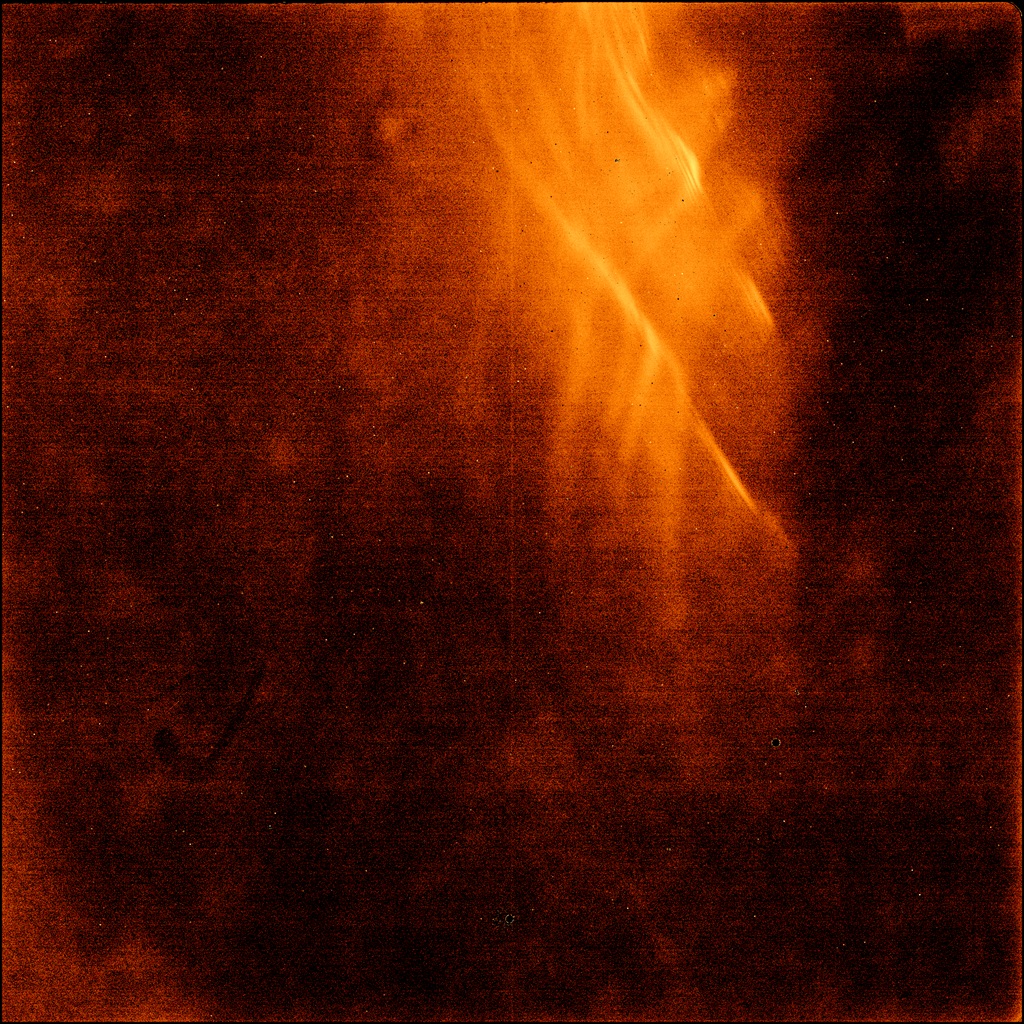} }}%
    \qquad
    \subfloat[\centering Original v3 wisp template downloaded from STScI.]{{\includegraphics[width=5.25cm]{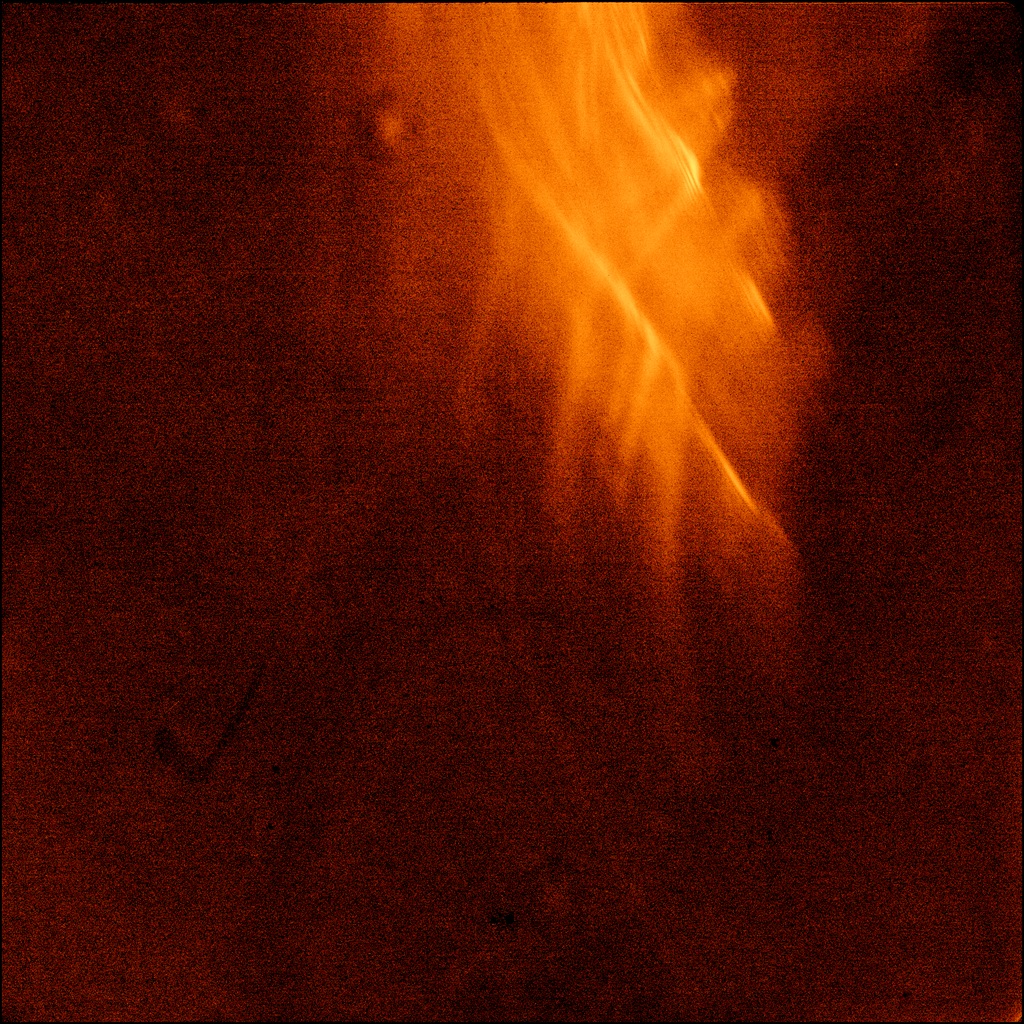} }}%
    \caption{A comparison between NIRCam F150W, b4 module wisp models, with normalisation in the intensity scaling applied to match an aperture placed around the most intense region. Panel a) shows the original wisp template released by STScI in November 2022, panel b) shows the new wisp template developed as part of this work, panel c) shows the latest wisp template model (v3) produced by STScI in Summer 2024. We note that our template closely resembles the newest official template, with the primary difference being that our template has some large scale features in the background which are likely the consequence of the use of slightly older calibrations.}%
    \label{fig:wisp}%
\end{figure*}

We process these images using an adapted version of the official STScI pipeline employed in \citet{Adams2024,Conseliece2024,Harvey2024}. We use version 1.8.2 of the JWST pipeline with CRDS pmap1084. Between stage 2 and stage 3 of the JWST pipeline, we implement our own custom 2D background subtractions on each of the individual calibrated frames before mosaicing. We also implement the subtraction of our own wisp templates (see subsection 2.1). Within Stage 3 of the pipeline, we align our images to the JADES DR2 catalogs \citep{Eisenstein2023DR2} which are in-turn aligned to GAIADR3 \citep{GAIADR3}. After the mosaic is constructed, we conduct a second round of background subtractions, align all of our images to the F277W filter to ensure WCS consistency and finally pixel match the images.

It should be noted that at this stage, the medium band filters F182M and F210M were found to have significant WCS distortion relative to the other bandpasses, particularly in NIRCam module A. This resulted in offsets upwards of 3 pixels towards detector edges, considering the small aperture sizes used in the source extraction of high-z candidates, such offsets can lead to significant errors in flux measurements. This distortion effect was also noted in the work of \citet{Robertson2024} and corrected by replacing the distortion map with the nearest broadband filter F200W. In Feburary of 2024, calibration pmap1197 was released which implements a formal correction to the problem, we subsequently remade the medium-band mosaics from PID 3215 using pmap1210 (post-correction). We find these new images are greatly improved, with the gradients in our estimated residual distortion removed and with a mean WCS offset of less than 0.8 pixels between photometric bands. 

Each source in our catalogue is prescribed a local depth in each photometric band. These are calculated using the normalised median absolute deviation (NMAD) of the nearest 200 empty circular apertures of the same size which are randomly positioned in `empty' regions of the image (defined as being 1 arcsec away from occupied pixels in the segmentation map). We show the mean average depth for the full JOF field in Table 1. These local depths are used in the calculation of the photometric errors and significance of detection for each individual source.

\subsection{New Wisp Templates}

The NIRCam instrument on JWST exhibits an artefact related to stray light which has been dubbed `Wisps'. These artefacts primary impact the central 4 blue imaging chips (A3, A4, B3, B4) between 1.5-2.1 microns. The precise shape and intensity of these artefacts has some minor wavelength and position dependence, though many studies so far have utilised the subtraction of an intensity-scaled template to remove the majority of this rogue light (see \citet{Robotham2023} for an example of a non-template based methodology). To best correct for this artefact (and any other consistent flat fielding features) in our imaging, we produce our own wisp templates using images produced with the same pipeline and calibration version.

Our own templates are created by conducting a median stack of all STAGE 2 products from a wide selection of deep-field observations. The full list of JWST programme ID's used are as follows: 1180 \citep[JADES, PI: D. Eisenstein][]{Eisenstein2023JADES,Bunker2024,Rieke2023,Eisenstein2023DR2,Hainline2024}, 1208 \citep[CANUCS, PI: C. Willott;][]{Willott2024}, 1210 (JADES, PI: N. Luetzgendorf), 1324 \citep[GLASS, PI: T. Treu;][]{Treu2022,Mascia2024}, 1345 \citep[CEERS, PI: S. Finkelstein;][]{Bagley2023,Finkelstein2024}, 1895 \citep[FRESCO, PI: P. Oesch;][]{Oesch2023}, 1963 \citep[JEMS, PI: C. Williams;][]{Williams2023}, 2079 \citep[NGDEEP, PI: S. Finkelstein;][]{Bagley2024}, 2738 \citep[PEARLS, PI: R. Windhorst;][]{Windhorst2023}, 3215 \citep[JOF, PI: D. Eistenstein;][]{Eisenstein2024,Robertson2024}, 4111 \citep[MegaScience, PI: K. Suess;][]{Suess2024}. This dataset provides around 200 individual `\_cal.fits' frames for each NIRCam module over a range of RA and DEC values for the F150W, F200W, F182M and F210M filters. We use Stage 2 outputs in our procedure to ensure our wisps do not duplicate the application of other calibration steps within the pipeline. We switch off the pipeline's own background subtraction step and replace it with a uniform background subtraction to prevent any pre-subtraction of the wisp artefact by a 2D background model. 

\begin{figure*}%
    \centering
    \includegraphics[width=0.75\textwidth]{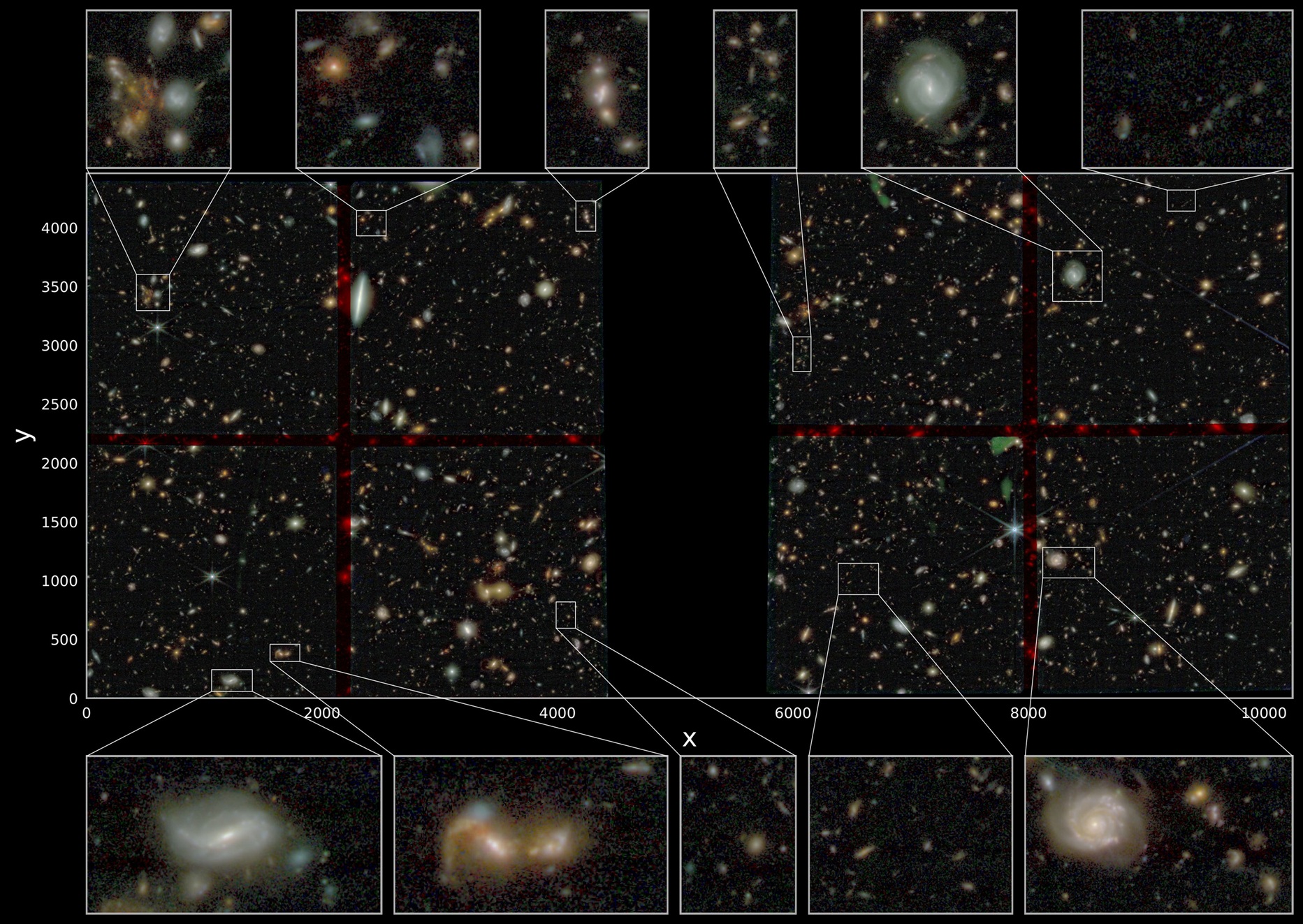} 
    \caption{A tour of an RGB image of the JOF field. The image was generated using the {\tt trilogy} package \citep{Coe2012} with the broadband images of the field. Bright, green regions in the NIRCam b4 module (lower left of the right side module) indicate masking of residual wisp artefacts. The red, cross-shaped pattern in each side of the image (caused by the blue module gaps) are similarly masked in our analysis.}%
    \label{fig:RGB}%
\end{figure*}

Unfortunately the F162M filter is used much less frequently then the F182M and F210M in current observational data of deep fields, resulting in a lower quality template relative to the other bands (average over 50 frames instead of $\sim200$ frames). The usage of the band is also dominated by the JADES Origins Field which utilises a small dithering pattern. To reduce the impact this has on the median stack, we only use six F162M frames from the JOF field in order to prevent significant residuals of real sources (bright galaxies and stars) entering the template. Since the central wavelength is relatively close to the F150W filter, we experiment with reducing the F162M images using our derived medium band template and the F150W template to determine which has the best performance. We find the F150W template produces the lowest average background level for the final F162M mosaic but leaves small regions over or under subtracted where the wavelength dependence has spatially shifted some of the highest intensity wisp features. Since the unique depth of the JADES Origins Field is its primary selling point, we opt for the low background produced by the F150W template in exchange for masking a small amount of improperly handled artefact.

Before subtraction of the wisp from our calibrated images, we first place circular apertures in a grid around the image and the template in order to estimate a scaling factor to be applied to the template before subtraction. We calculate this scaling factor using the brightest 10 per cent of apertures when placed on the wisp template (the most affected regions) for each of the NIRCam modules. Following testing with Epoch 1 of the NGDEEP field (PID: 2079 PI: S. Finkelstein), we find that corrected modules now have the same average depth as those which are unaffected by wisps (A1, A2, B1, B2) within 0.04 magnitudes. Confident this procedure works, we apply our templates to the JADES Origins Field. We find that our templates are also effective here, with the average depth of wisp affected modules within 0.02-0.10 magnitudes of the non-affected modules, with the greatest difference being with the deepest medium band of F182M.

We show an example of our wisp template in F150W module B4 within Figure \ref{fig:wisp}. We show this alongside the V2 (Nov 2022) and V3 (Summer 2024) wisp templates produced by STScI \footnote[1]{\url{https://jwst-docs.stsci.edu/known-issues-with-jwst-data/nircam-known-issues/nircam-scattered-light-artifacts\#NIRCamScatteredLightArtifacts-wispsWisps\&gsc.tab=0}}. We find our template has a significantly lower background compared to the V2 template, has removed features consistent with residuals of real sources in the median stack (e.g. stellar diffraction spikes) and reveals the wisp artefact to be more extended. The shape and intensity of the artefact is very similar to that of STScI's V3 template, though our template exhibits larger scale features in the background which may be a consequence of residual flat fielding due to our usage of the older pmap versions. Since we know our template was constructed using the same pipeline and calibration versions as our science images, we proceed with the use of this template. The final RGB mosaic is presented in Figure \ref{fig:RGB}.

\section{Galaxy Selection Procedures}

A catalogue of the sources in the Jades Origins Field is produced using the {\tt SExtractor} software in dual image mode. Here, a weighted stack of the red broad band images (F277W, F356W, F444W) is used at the selection band. Aperture photometry is measured in both 0.2 and 0.32 arcsecond apertures. A set of science images PSF homogenised to the F444W band are also produced using empirical PSF models generated following the technique used in \citet{Weaver2024}. We find the profile of the PSF model is similar in shape but slightly broader than theoretical models generated by {\tt WebbPSF} \citep{Perrin2012,Perrin2014}.  In the following subsections we describe how we select and measure the galaxies using different criteria and later how the results depend on these.

\subsection{Photometry and Photo-Z procedure}

Photometric redshifts for our sources are conducted with the use of the {\tt EAZY-py} SED fitting tool \citep{Brammer2008}. Here, we employ the use of the default FSPS templates as well as Sets 1 and 4 generated in the work by \citep{Larson2023}. These additional templates provide galaxy spectra which are younger, bluer and have stronger emission lines compared to the default template set on its own. EAZY is run using the \citet{Madau1995} treatment for absorption by neutral Hydrogen, adds dust attenuation up to 3.5 magnitudes from \citet{Calzetti2000} and employs no galaxy luminosity prior. EAZY is run twice with upper redshift limits of 25 and 6 in order to conduct a direct comparison between the best fitting high-z and low-z solutions. For our photo-z's, we quote the peak probability redshift as the primary solution and the 16th/84th percentiles as the error. We also investigate how our results would change when using different size apertures.

\begin{figure}%
    \centering
    \hspace{-0.4cm}\includegraphics[width=9cm]{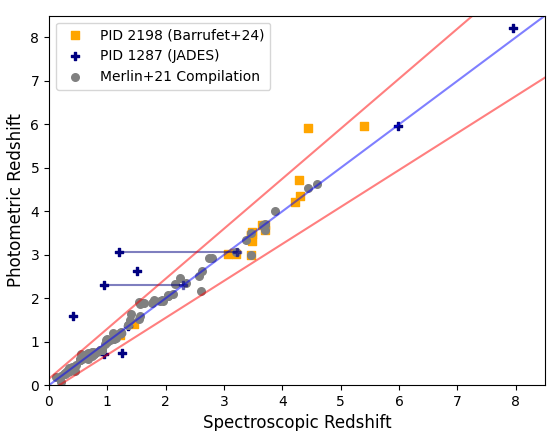} %
    \caption{A comparison of our photometric redshifts using the full JOF dataset against publicly available NIRSpec programmes covering the JOF field \citep{Barrufet2024,Carnani2024} and historical spectroscopic redshifts in the \citet{Merlin2021} compilation. Two JADES PRISM spectra obtain bimodal spec-z solutions which we indicate with a connecting line showing the photo-z prefers one solution over the other.}%
    \label{fig:specz}%
\end{figure}

To assess our photometric redshifts, we compare our outputs to 20 spectroscopic redshifts crossmatched from `v3' of the Dawn JWST Archive \citep[DJA:][]{Heintz2024}. All 20 sources originate from PID 2198 \citep[PI: L. Barruffet][]{Barrufet2024} and are luminous sources $23<m_{\rm F277W}<26$ between redshifts $1.2<z<5.4$ with the majority of targets between $3<z<4.5$ (See Figure \ref{fig:specz}).  We find that 19/20 sources have redshifts consistent with their spectroscopic results. The one outlier has spectroscopic redshift of $z=4.4358$ and an uncertain photometric redshift of $z=5.9^{+0.0}_{-3.8}$ featuring 3 strong peaks in its probability density function at $z=$5.9, 4.2 and 2.1. More spectra of this sub-field was taken as part of the JADES programme (under PID 1287, PI: K. Isaak) in January 2024. We have reduced the PRISM spectra from this programme following the default NIRSpec settings of the {\tt msaexp} tool \footnote{\url{https://zenodo.org/records/7579050}}, which is similar to the methods used by the DJA. For our purposes we only show 13 conservative spec-z results with 2 or more emission lines identified with by-eye vetting. We also include a compilation of historic spectroscopic redshifts from \citet{Merlin2021}, which features 122 cross-matches with 3D-HST \citep{Momcheva2015}, ACES \citep{Cooper2012}, \citep{Morris2015}, \citet{Ravikumar2007}, \citet{Szokoly2004}, VANDELS \citep{Garilli2021}, \citet{Balestra2010}, VUDS \citep{LeFevre2015}, VVDS \citep{LeFevre2013} and \citet{Vanzella2008}. A visual comparison is shown in Figure \ref{fig:specz}.

\subsection{High-z Selection Criteria}

High redshift galaxies at $z>7.5$ are selected following the criteria and are similar to those used in the construction of the EPOCHS sample \citep{Adams2024,Conseliece2024,Austin2024}. We summarise the primary cuts below:

\begin{enumerate}
    \item $\leq3 \sigma$ detections in all bands blueward of the expected Lyman break. 
    \item $\geq 5\sigma$ detection in the first 2 bands directly redward of the Lyman break, and $\geq 2\sigma$ detection in all other redward bands, excluding medium bands. If the galaxy appears only in the long wavelength NIRCam photometry (i.e. a F200W or higher dropout), we increase the requirement to a 7$\sigma$ detection to minimise the potential for selecting red camera artefacts which can appear to look like noisy $16<z<18$ sources.
    \item $\int^{1.10\times z_{phot}}_{0.90\times z_{phot}} \ P(z) \ dz \ \geq \ 0.6 $ to ensure the majority of the redshift PDF is located within the primary peak.
    \item $\chi^2_{red} < 3 (6)$ for best-fitting SED to be classed as robust (good).
    \item $\delta \chi^2 \geq$ 4 between high-z and low-z EAZY \ runs (where maximum redshift is set to 6). This ensures that the high-z solution is much more statisically probable. 
    \item If the 50\% encircled flux radius ({\tt FLUX\_RADIUS} parameter in SExtractor) is consistent with the size of the PSF in the F444W band, then we require that $\delta \chi^2 \geq$ 4 between the best-fitting high-z galaxy solution and the best-fitting brown dwarf template \citep{Marley2021,Harvey2024} which can look like high-z galaxies \citep{holwerda2024}.
    \item 50\% encircled flux radius is $\geq$1.5 pixels in the long-wavelength wideband NIRCam photometry (F277W, F356W, F444W). This avoids detecting hot pixels in the LW detectors as F200W dropouts. 
\end{enumerate}

\subsection{The initial full sample of high-z candidates}

\begin{figure}%
    \centering
    \hspace{-0.4cm}\includegraphics[width=9cm]{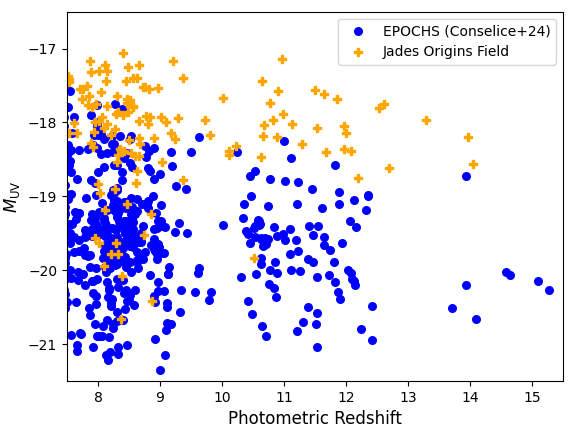} %
    \caption{A compariosn of the redshfits and $M_{\rm UV}$ values obtained for the full 0.2as selected JOF sample vs the EPOCHS sample \citep{Conseliece2024} which compiled a number of wider area, but shallower surveys together. Here, we see the JOF field enables intrinsically fainter sources to be identified.}%
    \label{fig:MUV}%
\end{figure}

Utilising the full set of NIRCam data available in this field and conducting the above photo-z selection procedure, we obtain 62 galaxies which we class as robust at $z>7.5$ using the full JOF dataset. The distribution of sources is as follows, with a full table of high-z candidates presented in Table \ref{tab:fulllist}: 30 sources with $7.5<z<8.5$, 18 sources with $8.5<z<9.5$, 9 sources with $9.5<z<11.5$, 4 sources with $11.5<z<13.5$ and 1 source with $z>13.5$. Redshifts and $M_{UV}$ values for the sample are shown in Figure \ref{fig:MUV}. Since we only utilise JWST data in our selections, our sample is incomplete at $z<8.0$ due to the demand for a complete break to be observed and the lack of data bluewards of 0.9 microns. While HST data is present in this field, it is significantly shallower (mag 28.3-29.1 in F606W) and non-uniform \citep{Whitaker2019}. Since ultra-deep HST data can not always be a guarantee for such JWST fields, we focus on the use of just JWST photometry for the purpose of this study.  Below we discuss some of the details of our selected galaxies and comparisons to previous similar studies on this field.

\subsubsection{A robust candidate galaxy at $z\sim14.7$}

\begin{figure}%
    \centering
    \includegraphics[width=9cm]{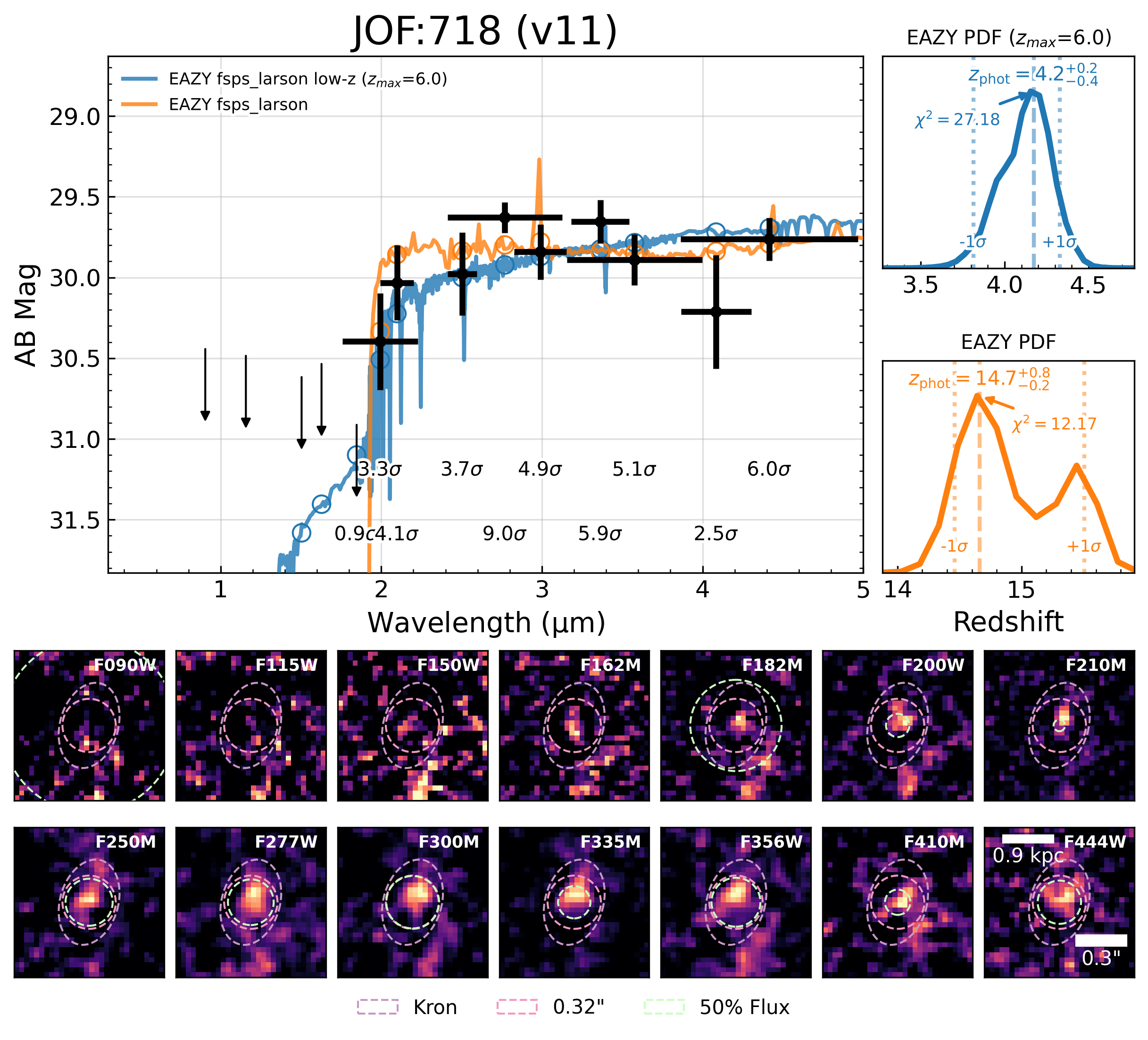} %
    \caption{The SED fit for JADES+53.10762-27.86013 (known as JOF:718 in the our catalogues), the highest redshift robust candidate. Displayed are the EAZY fits to 0.32as photometry for the high-z solution, in orange, and low-z solution, in blue. The 0.2as photometry displays a similar SED but lower redshift of $14.0^{+0.2}_{-0.2}$.}%
    \label{fig:highz15}%
\end{figure}

The highest redshift, robust galaxy in our catalogues that passes visual inspection is JADES+53.10762-27.86013, which is a faint ($m_{\rm AB}\sim29.8-30$) galaxy selected at $z=14.66$ (See Figure \ref{fig:highz15}) with a $\delta\chi$ of 15 relative to a low redshift solution at $z=4.2$. This agrees with the $\delta\chi^2$ expectations set out in the JOF survey paper \citep{Eisenstein2024}. Key for determining this is the slightly blue slope measured in the detected bands and the lack of line emission in the F250M band, which covers H$\beta$ and O[III] at $z=4.2$. This object is also identified in both JADES photo-z selection papers covering the field, \citet{Hainline2024} and \citet{Robertson2024}, as a robust candidate with $z=14.44$ and $z=14.63$ respectively. NIRSpec spectroscopy was taken of this source, but due its faint luminosity and the large slit losses in the observations, no definitive redshift was obtained \citep{Carnani2024}.

\begin{table*}
\caption{The 15 high-z galaxy candidates in the JADES Origins Field presented in \citet{Robertson2024} We split the sources by their original classification as Robust ($z>11.5$), Contributing (some $P(z)>11.5$) and Secondary ($z>11.5$ but failing some selection criteria). Alongside the original JADES teams findings, we present our equivalent photometric redshift estimations and classifications. `Faint' describes sources that are $<5\sigma$ detected in the first two bands redwards of the Lyman Break. Our use of 0.32as apertures does mean the faintest sources are lost for this reason. `Blended' sources are flagged as potentially having rogue light from a close neighbour within the aperture. 'Masked' sources are typically near module edges or stellar diffraction spikes. Objects with \dots values did not obtain a successful SED fit.}
\begin{tabular}{llllll}
Name / Coordinates                   & JADES Class  & JADES z & Our z (0.32as)     &   Our z (0.2as)          & EPOCHS Comments     \\
JADES+53.09731-27.84714 & Robust       & 11.53   & $9.0^{+1.6}_{-0.5}$ & $2.4^{+8.6}_{-0.0}$  & Faint, Broad PDF, Balmer Break (0.2as)        \\
JADES+53.02618-27.88716 & Robust       & 11.56   & $11.4^{+0.3}_{-0.4}$ & $11.1^{+0.2}_{-0.2}$ & Robust       \\
JADES+53.04017-27.87603 & Robust       & 12.10   & $11.2^{+0.5}_{-0.3}$ & $11.5^{+0.2}_{-0.3}$ & Faint (0.32), Robust (0.2)        \\
JADES+53.03547-27.90037 & Robust       & 12.38   & $16.3^{+0.3}_{-0.4}$ & $14.0^{+0.2}_{-0.3}$ & Faint (0.32), Robust (0.2)        \\
JADES+53.06475-27.89024 & Robust       & 12.93   & $12.9^{+0.3}_{-0.1}$ & $12.7^{+0.1}_{-0.1}$ & High $\chi^2$ (0.32), Robust (0.2)       \\
JADES+53.02868-27.89301 & Robust       & 13.52   & $12.3^{+0.3}_{-0.2}$ & $12.2^{+0.1}_{-0.1}$ & Robust       \\
JADES+53.07557-27.87268 & Robust       & 14.38   & $16.9^{+2.2}_{-0.5}$ & $14.4^{+1.4}_{-0.5}$ & Faint        \\
JADES+53.08294-27.85563 & Robust       & 14.39   & $3.6^{+0.1}_{-0.1}$ & $14.2^{+0.1}_{-0.2}$ & Balmer Break (0.32), Blended, $3\sigma$ blue (0.2) (Spec-z 14.18) \\
JADES+53.10762-27.86013 & Robust       & 14.63   & $14.7^{+0.8}_{-0.2}$ & $14.0^{+0.2}_{-0.2}$ & Robust       \\
JADES+53.07248-27.85535 & Contributors & 10.61   & $11.9^{+0.1}_{-0.7}$ & $11.0^{+0.2}_{-0.3}$ & Robust      \\
JADES+53.03139-27.87219 & Contributors & 10.76   & $11.0^{+0.4}_{-0.6}$ & $10.7^{+0.2}_{-0.3}$ & Faint        \\
JADES+53.09292-27.84607 & Contributors & 11.05   & $0.2^{+8.2}_{-0.1}$ & $10.4^{+0.3}_{-0.5}$ & Faint (0.32), 1.6 micron bump (0.32), fail $\delta\chi^2$ (0.2), Masked        \\
JADES+53.06857-27.85093 & Contributors & 11.17   & \dots & \dots & Blended, Masked        \\
JADES+53.08468-27.86666 & Secondary    & 12.90   & $3.5^{+10.3}_{-0.2}$ & $12.5^{+0.8}_{-0.1}$ & Balmer Break (0.32), Robust (0.2) \\
JADES+53.07385-27.86072 & Secondary    & 13.06   & $1.7^{+6.6}_{-0.5}$ & $1.6^{+1.1}_{-0.6}$ & Balmer Break, Faint 
\end{tabular}
\end{table*}

\subsection{Comparisons to the sample selected in Robertson+24}

The official JOF team released their initial sample of high-z galaxies in \citet{Robertson2024}. This consisted of nine robust galaxies at $z>11.5$, including four robust galaxies with redshift PDF's extending above $z>11.5$ and a further two less confident galaxies at $z>11.5$. We find cross matches to all 15 objects in our catalogues with typical matching separations less than a pixel. Only 4 of these 15 galaxies enter our robust sample. The primary reason for non-selections are that these galaxies fall just below our $5\sigma$ selection criteria, this is because we initially utilise 0.32 arcsecond diameter apertures while \citet{Robertson2024} uses 0.2 arcsecond diameter apertures. As a result, their faintest sources are typically 3-4$\sigma$ detected in our work, though we find the majority still have high-z primary solutions.

For a full breakdown of our findings for the sources presented in \citet{Robertson2024}, we find that: 4/9 sources in their main sample are selected in our work, 0/4 of their lower redshift sources are selected and 0/2 of their dubious sources are selected. In total, four sources are identified in our work to be at $z<4$ and better fit by Balmer breaks than Lyman-breaks. These include both of the sources described as dubious in the work by \citet{Robertson2024}. The additional two include one of their lower-z candidates and the spectroscopically confirmed JADES-GS-z14-0 \citep[$z=14.18$][]{Carnani2024,Schows2025,Carnani2025}, which is blended with a $z=3$ neighbour resulting in $3\sigma$ detections in the bluest bands. We find one source is better fit at $z\sim9$ as opposed to $z\sim11$, however this source has a redshift PDF with a large positive wing and so it is not in significant large tension with the initial \citet{Robertson2024} estimations.

The small area of the JOF field provides some limitations to our work. Firstly, the impact of source blending and artefacts can normally be accounted for when employing the use of large survey areas by removing masked areas/segmentation maps from any area/volume calculations. But for small areas, cosmic variance and small number counts combined with poor luck can result in e.g. one of two-to-three $z\sim14$ candidates being lost to blending within a masked area of 10\%, resulting in underestimations of number density. Whilst the precise aperture location for JADES-GS-z14-0 could be shifted to reduce the flux from the neighbour, the application of this to larger area programmes would become increasingly impractical. Secondly, the small area of the JOF will limit the range of potential SED shapes that are explored. Subsequently, rare or clustered subpopulations that are more or less likely to be mis-modelled in SED fitting could be under- or overrepresented.

\subsection{Differences when using 0.32 or 0.2 arcsecond apertures}

Various studies targeting high redshift samples have primarily made use of two different sized circular apertures when conducting initial flux measurements. These have been 0.3-0.32 arcsecond diameters (amounting to 5-6 pixels in radius) or 0.2 arcsecond diameters (amounting to 3.3 pixels in radius). There are potential advantages and disadvantages to using each. The use of larger apertures is slightly more forgiving for the precision of the WCS alignment between photometric bands and it encloses more flux from the PSF ($\sim70-80\%$), requiring smaller aperture corrections. The smaller aperture size can enable higher SNR measurements to be made but can be more greatly influenced by bad pixels, correlated noise, minor image misalignments and it requires larger corrections for flux outside of the apertures. At a radius of 0.1 arcsec, the PSF for JWST's reddest bands has a very steep slope for its encircled flux verses radius, meaning precise PSF models are required as small changes to the FWHM will result in larger changes in the aperture correction verses the use of a larger aperture.

To assess if there are significant differences in sample selection as a consequence of this choice, we repeat our selection procedure on a catalogue using 0.2 arcsecond apertures. This catalogue selects 130 sources at $z>7.5$ compared to the 62 sources that the 0.32as catalogue did. Of these sources, 52 are consistent between the two samples. for the 10 sources lost, half (5) switch to dominant low-z solutions due to $2.5\sigma+$ detections in F090W whilst 4 sources have the $\delta\chi^2$ reduced below constraints and 1 source scatters below the $z>7.5$ cut.

Among the faint sources added to the high-z sample are an additional four galaxies from the \citet{Robertson2024} sample. Non-selected galaxies from that work with high-z solutions typically only just fail our detection criteria (e.g. 4.5$\sigma$ in rest-frame UV). The spectroscopically confirmed JADES-GS-z14-0 now obtains a high-z solution but remains non-selected on account of contamination from the neighbouring source producing a 3.1$\sigma$ detection in F150W. The source JADES+53.10762-27.86013 (Figure \ref{fig:highz15}) remains the most robust ultra-high-z source in the sample and it is selected with both choices of aperture size. With the smaller aperture size, the $\delta\chi^2$ to a low-z solution increases from 15 to 22 though the photo-z decreases from 14.7 to 14.0.

\section{Quantifying the Advantages of Medium Bands}

The JADES origins field is the the only blank field to reach AB magnitude depths of 30+ in such a variety of wide and medium photometric bands. It thus presents a great playground for observing the impacts that such a sampling of galaxy SED's provide. In this section, we explore how the addition of medium-width bands impacts the selection of high redshift candidates. We conduct this study both through the use of real data from the JOF field and with the use of mock catalogues generated by the JAGUAR semi-analytic model \citep{Williams2018}.

\subsection{Adding and Removing Bands from Real Data}

To begin, we repeat our photometric redshift and selection procedures from Section 3 but with more restricted combinations of photometric bands. These runs are designed to replicate the filter choices of other JWST deep field observations and so include the use of no medium-width bands (similar to e.g. GLASS \& NGDEEP), the use of just F410M (similar to e.g. CEERS \& PEARLS) and the use of both F410M and F335M (e.g. JADES). We respectively name these catalogues C-0, C-1 and C-2 to indicate the number of medium-width bands used. Our fiducial catalogue containing the full JOF dataset is named C-ALL. In this section, we examine the changes to the selected high-z sample (both sources which are lost from the fiducial sample and added to the fiducial sample) in order to examine how variable high-z selections can be and what the reasons are for this variability.

Our original selection using the full dataset (C-ALL) identified 62 galaxies at $z>7.5$. When applying our selections to the three catalogues with fewer medium bands (C-0, C-1, C-2), each respectively identified 68, 69, 71 candidates. Of these objects, 49, 53, 53 objects are consistent with the fiducial sample (C-ALL), showing that sample overlap with C-ALL is relatively high (79-86\%). The largest increase in sample overlap occurs when a single medium-band is added to the broad-band-only dataset. Sources which are identified as confident in C-ALL but omitted when fewer bands are used are largely consistent between the different band combinations too (e.g. $82\%$ of sources selected in C-ALL but omitted in C-2 were also omitted in C-1 and C-0).   

We begin our analysis by examining the 13 unique objects that were selected by the C-ALL catalogue yet omitted in the C-0 catalogue. Nine of these sources are low SNR, with F200W detections at $<8\sigma$. The remaining four are between 8$\sigma$ and 12$\sigma$. Additionally, most of these sources (10) are at the lower end of the redshift range that we consider ($z<8.6$) where we expect completeness to decrease due to the absence of HST data below $1\mu$m in this study. For the three higher-z galaxies at $z>10$ (all fainter than $8\sigma$), in C-0 we find one galaxy switches to a low-z dominant solution and two fail $\delta\chi$ cuts to the next best low-z solution. In these cases, the medium-width bands in the blue module show a smooth UV slope just before the break, indicating that the break is not a consequence of emission line contamination in the broad-bands.

\begin{figure*}%
    \centering
    \subfloat[\centering Example source better fit as a Balmer break with strong lines.]{{\includegraphics[width=7cm]{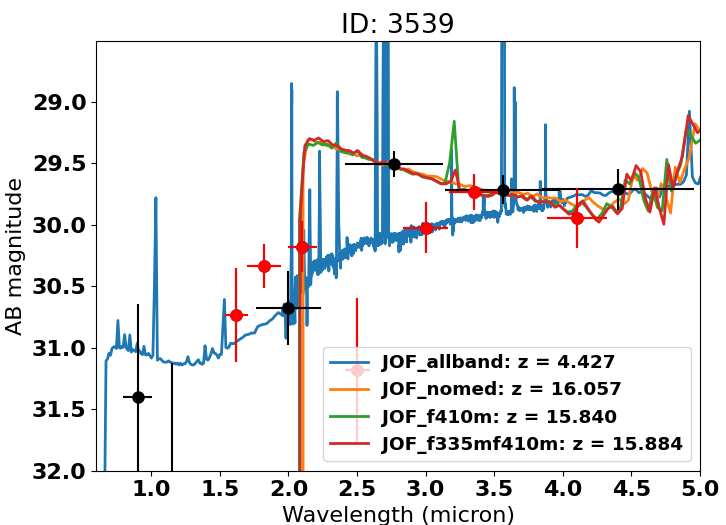} }}%
    \qquad
    \subfloat[\centering Example of a source better fit by a Balmer break with weak/no lines.]{{\includegraphics[width=7.1cm]{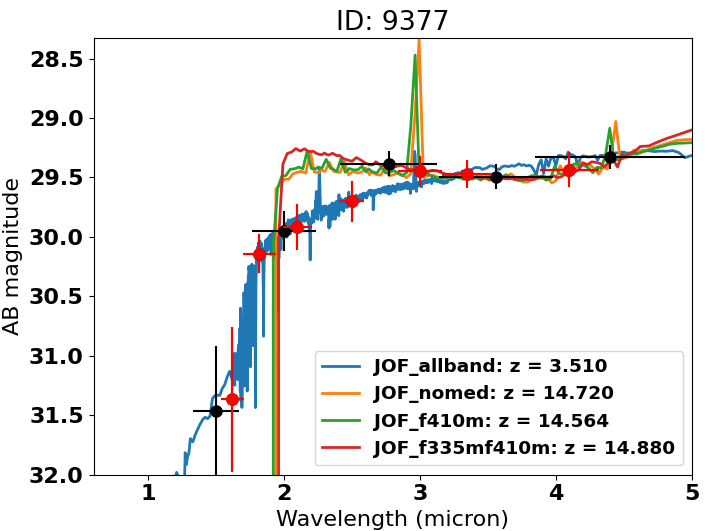} }}%
    \qquad
    \subfloat[\centering Example source which is better fit by a 1.6 micron bump.]{{\includegraphics[width=7.1cm]{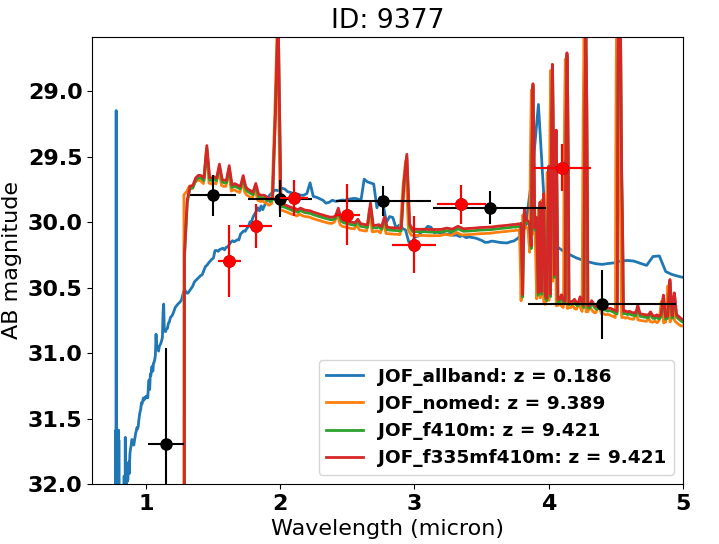} }}%
    \qquad
    \subfloat[\centering Example source which is poorly fit when blue medium-width bands are added in.]{{\includegraphics[width=7.1cm]{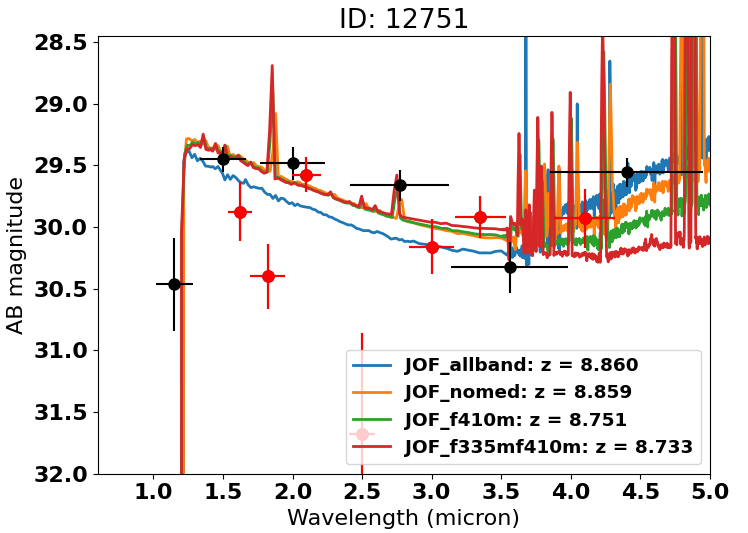} }}%
    \caption{The SED fits for four example JOF sources that represent the primary categories of galaxies which are selected in catalogues that employ lower numbers of photometric bands, but become non-selected when all available information is utilised. Photometric data points coloured in black are broad-width bands while those in red are medium-width.}
    \label{fig:lostsources}%
\end{figure*}

We next examine sources that are added to the high-z sample when fewer bands are used. For the C-0, C-1 and C-2 catalogues we find that 19, 16, 18 new sources are selected compared to the C-ALL sample. These extra sources are largely consistent between these catalogues, with 13/16 and 14/18 sources in the C-1 and C-2 catalogues appearing in the C-0 catalogue. There are a total of 22 unique sources which get selected as high-z when fewer photometric bands are available. We find that the sources that are selected outside of the C-ALL catalogue are generally faint, with 16/22 sources below $8\sigma$ in the expected rest-frame UV.

When examining these new sources in the C-ALL catalogue, we find 12/22 sources have primary photo-z solutions that are low-z. These sources can be broken down into 3 categories; 2/12 sources are better fit with 1.6$\mu$m bumps with all the medium-width photometry available, 2/12 sources are better fit by a Balmer break with no strong emission lines and 8/12 sources are better fit by a Balmer break with strong emission lines. We display an example of each in Figure \ref{fig:lostsources}. The strong line emitter category includes three sources exhibiting SED shapes similar to that found in \citet{ArrabalHaro2023} which followed up a $z=16.4$ candidate \citep[see also][]{Donnan2022}. The use of medium-width bands bluewards of 3 microns is found to be essential to identifying sharp colour differences indicative of emission line presence.

We find that the remaining 10/22 sources lost in C-ALL retain high-z primary solutions but do not meet $\delta\chi^2$ cuts relative to low-z solutions or fall below the $z>7.5$ selection. All but one of these galaxies had high-z solution of $7.5<z<9.5$. Three sources are confident $z\sim8$ sources that scatter below the $z>7.5$ cuts for the sample selection. The remainder all contain poorly fit blue medium-width bands (F162M, F182M, F210M) which show colours of up to 1mag and could be indicative of poorly modelled emission lines. The one source at $z\sim12$ has F182M and F210M photometry showing a smoother break than would otherwise be expected from a Lyman break, reducing its confidence relative to the Balmer break alternative.

In brief, sources that are selected as robust, high-z galaxies in one combination of photometric bands, but not in another, fall within two broad categories. 1) Objects where bluer medium-width bands (<3$\mu$m) confidently identify or rule-out a low-z solution; 2) Objects where bluer medium-width band observations can not be well described by the template sets used, resulting in poorer quality fits. In cases where confident low-z solutions can be attributed, the most common source of confusion is Balmer break sources with strong emission lines, which represent 2/3rds of potential contaminants.

\subsection{Results when repeating the experiment on a simulated catalogue}

\begin{figure*}%
    \centering\vspace{-0.3cm}
    \includegraphics[width=0.9\textwidth]{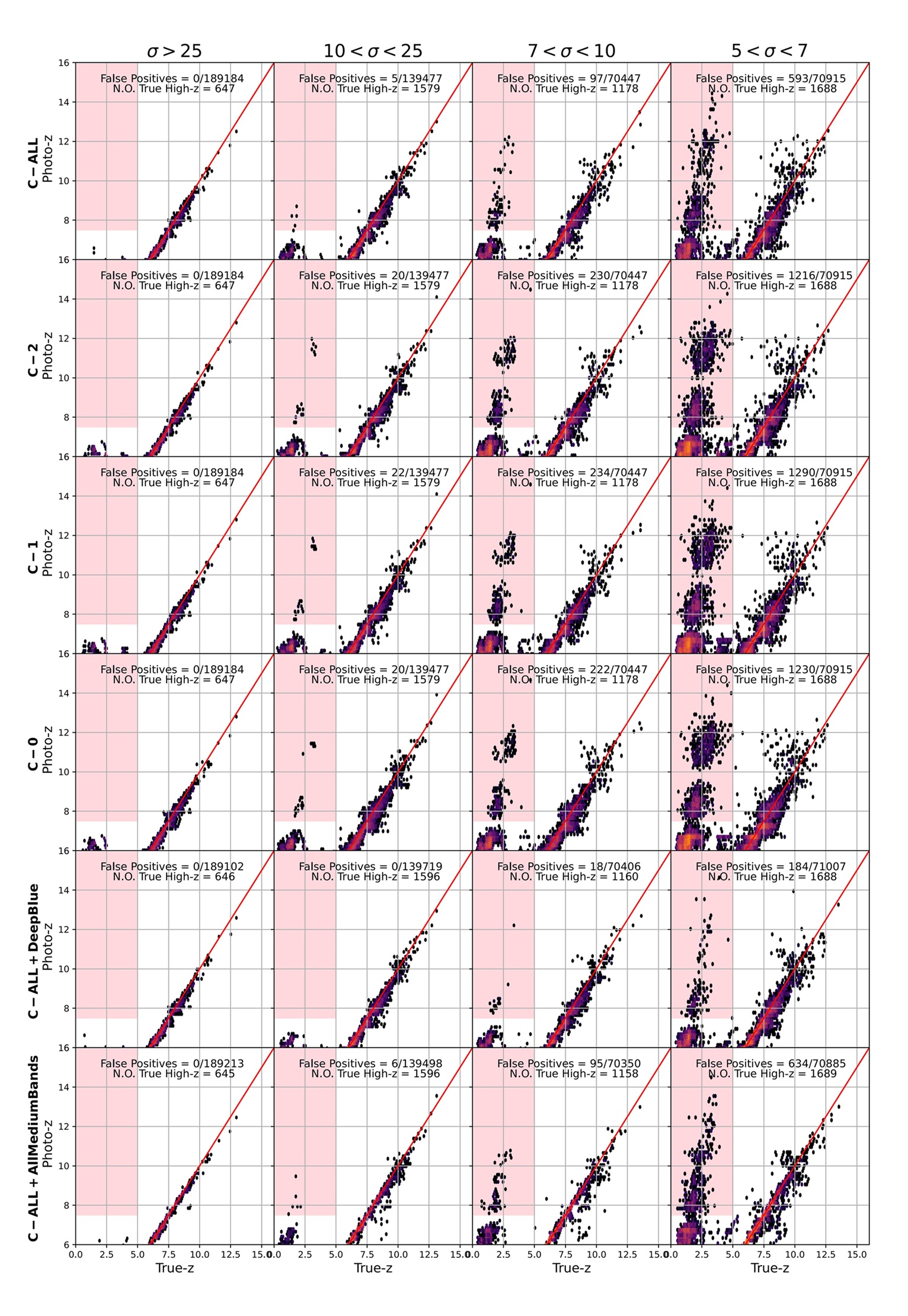} %
    \vspace{-1.0cm}
    \caption{A break down of the photo-z performance on our simulated JAGUAR catalogues. Brighter objects (defined in blocks of sigma confidence in F200W) are grouped to the left and fainter objects grouped to the right. Results for the various filter combinations are shown vertically, with the catalogue name provided on the y-axis label. False positives indicate the number of $z<5$ galaxies with a fit photo-z greater than 7.5 (shaded region). Brighter objects have greater photo-z performance, in addition, more photometric bands and deeper blue photometry improves outlier rates and scatter. Crucially, the inclusion of the JOFs medium-width bands around 1.5-3 microns greatly decreases the cloud of low-z sources identified at very high-z. Displayed are raw photo-z's, not a final high-z selection which is explored in Figure \ref{fig:JAG-Complete}, $\chi^2$ cuts on low-z solutions and other quality cuts greatly reduce the cloud of contaminants. Further improvements would benefit from additional broadband time at 0.9-2 microns to rule out Balmer breaks.}%
    \label{fig:jag-photz}%
\end{figure*}

To contextualise the results from observations, we employ the use of the JAGUAR semi-analytic model and the mock photometry that was generated from it \citep{Williams2018}. This particular SAM was used due to its performance in replicating galaxy colours in JWST photometric bands \citep{Wilkins2023}. We use three different realisations of the JAGUAR simulation to provide over 300 square arcminutes of simulated sky area, a factor of 30 greater than probed by the JADES Origins Field. We use such a large area in order to produce a large statistical sample of high-z galaxies and assess if similar behaviours witnessed in the observations are replicated. The catalogues have a magnitude cut applied 1 magnitude fainter than the $5\sigma$ limits of the JOF field. This results in approximately 1 million sources in total across all redshifts.

Multiple simulated catalogues are generated using different combinations of filters as trialled in the previous section. We scatter the photometry of the simulated sources based on the average depths in each photometric band measured for the JOF field. The catalogue is then formatted in the same way as our observed catalogue and the mock observations run though our EAZY pipeline. This process is repeated 5 times for each of the different filter combinations we use to assess the sensitivity of results to the random scatter applied to photometry. We find that completeness and contamination rates scatter by around 4 percentage points between different realisations of our catalogues. To quantify the impact the addition of medium-width photometric filters has in a high-z context, we focus on the completeness of high-z objects selected and the contamination rate of low-z ($z<6$) sources entering the sample. 

We also add in some additional hypothetical situations to explore avenues for further improving photo-z performance. In one run (deep\_blue), we increase the depth of the bluest JOF bands (F090W, F115W) by 0.4 mags (equivalent to 2X exposure time) to make their depths more comparable to the rest of the wavelength range (AB mag 30.2). In a second run (all\_band) we add in the remaining medium-width bands of F140M, F360M, F430M, F460M and F480M. These are set to have the same depth as their neighbouring medium-width band, so 29.9 for F140M and 29.8 for the red bands.

We present the full result of simulating photometric redshifts in Figure \ref{fig:jag-photz}. We find that the inclusion of more medium-width bands leads to reductions in outlier rate and scatter. Outlier rate and scatter also increase towards fainter luminosities and a cloud of contaminant low-z galaxies at $z_{\rm true}\sim3$ and $z_{\rm phot}\sim11$ also increases in density. Adding medium-width bands between 2-3 microns reduces this cloud, but adding depth to the blue broad-bands is most affective at reducing the number of objects lying in this undesirable region.

\begin{figure*}%
    \centering
    \subfloat[\centering Selection Completeness \& Contamination rates for high-redshift sources ($8<z<10$) in the JAGUAR simulation]{{\includegraphics[width=8.5cm]{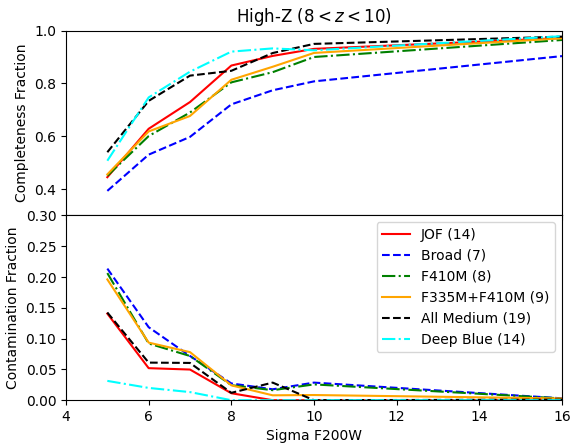} }}%
    \qquad
    \subfloat[\centering Selection Completeness \& Contamination rates for ultra-high-redshift sources ($10<z<14$) in the JAGUAR simulation]{{\includegraphics[width=8.5cm]{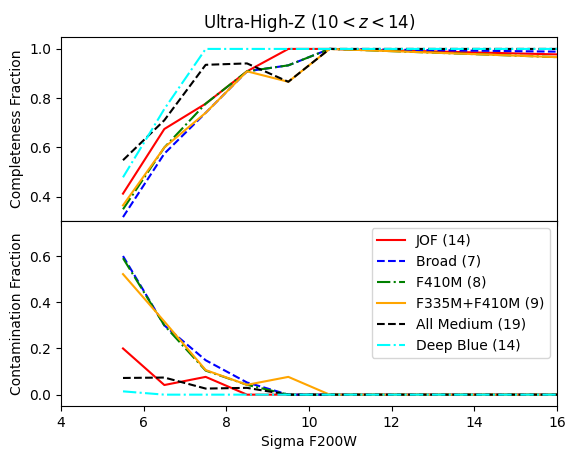} }}%
    \caption{Completeness and Contamination fractions as measured by running simulated JAGUAR catalogs through our selection pipeline using the observational set up of the Jades Origins Field. The top component of each figure is completeness, while the lower component is contamination. Each line describes the results from one observational set up. The red solid line describes the findings using the fiducial Jades Origins Fields combination of filters and depths achieved within those filters. The blue dotted line shows the results when all medium bands in the JOF field are omitted from use, leaving only broad-width bands. The green dash-dotted line shows results where the F410M medium-band is used in conjunction with broad bands. The solid yellow line further adds the F335M medium band. The final two lines show hypothetical survey expansions to the JOF field, 'All Medium' adds all remaining medium-width bands to the JOF field to similar depths while 'Deep Blue' doubles the exposure time (0.4 mags) in bands bluer than 1.5 microns. With increasing numbers of medium bands, completeness is observed to increase while contamination decreases. The legend indicates which bands were used and the total number of bands. Lines typically scatter by up to 4 percentage points when repeated 5 times with random elements re-rolled, with larger scatter in the 5-6$\sigma$ regime in the ultra-high-z bin.}%
    \label{fig:JAG-Complete}%
\end{figure*}

Our selection criteria is designed to identify sources at $z>7.5$, however there are a large number of sources around this boundary which scatter above and below this selection limit as a consequence of the use of photo-z's. To simplify discussion of completeness and contamination, we consider sources with a true redshift of $z>8$ in the following subsections. We also break up this discussion into two redshift bins; $8<z<10$ and $z>10$. In the lower redshift bin, the presence of the Balmer break and rest-optical emission lines in the reddest bands may result in different behaviours when adding medium-width bands when compared to higher redshifts, where the available UV emission lines are typically weaker and photometry generally probes the UV slope. We present the results of this process in Figure \ref{fig:JAG-Complete} and discuss them in more detail in the following subsections.

\subsubsection{High-z Sample Completeness}

No single selection criteria is entirely perfect when working with noisy data. Inevitably, a balance is struck between completeness and contamination with every effort made to maximise the former and minimise the latter. In this work, we examine the number of true sources at $z>8$ in the JAGUAR catalogs that are recovered by our selection processes when various combinations of medium-width bands are used.

Beginning with the redshift bin of $8<z<10$, we find that completeness (Number of sources at $z>8$ selected divided by the total number of $z>8$ sources) is generally high when considering sources at $10\sigma+$. Without any medium-band information, this completeness is over 80\% at the $10\sigma$ mark, while including even a single medium band can substantially raise this to 90\%. As would likely be expected, increasing the number of medium-width bands leads to an increase in completeness, though the largest benefit comes from the inclusion of a single medium-width band and subsequent bands are less impactful. This improvement with a single medium-width band is noted in the previous subsection, where the biggest jump in the JOF sample overlap was when only F410M was added to the catalogues. 

To better understand the benefits that the single F410M medium band provides, we explore the reasons why completeness increases by $\sim10\%$ in the $8<z<10$ bin when it is included. Examining the full sample of 215 of these sources, we find that all have high-z primary solutions, but the dominant population are those with true redshifts of $8<z<9$ but have a photo-z of $7.0<z_{\rm phot}<7.5$, resulting in their non-selection with our $z>7.5$ criteria. These sources have strong O[III]/H$\beta$ emission boosting F444W fluxes. Without medium band information, the strong emission lines are sometimes placed at the bluest end of the F444W filter providing a photo-z of $z\sim7$. The addition of F410M reveals an absence of emission lines at the bluest side of F444W and the F300M to F335M colour also provides a more precise Balmer break position. Such sources that have systematically underestimated photo-z's account for $77\%$ of the sources contained within this completeness increase. For the remaining sources they fail our $\delta\chi^2$ constraints when no medium-width bands are available. From a sample completeness perspective, these results thus support the general `deep-field' design of including a single medium-width band and F410M presents one of the better balances of sensitivity and longest wavelength (to study e.g. rest-optical emissions lines or the Balmer break up to $z\sim10$) as has been employed by many early survey designs as a balance of cost verses effectiveness.

Examining higher redshifts at $z>10$, completeness is very high ($97\%+$) for sources detected to $10\sigma+$, however the number of these sources present in the simulation a these luminosities is very low. As observed for the $8<z<10$ sources, completeness turns over and falls off below $10\sigma$. The use of 0, 1, 2, or even 7 medium-width bands does not provide a significant difference in the completeness curve, due to the lack of any strong, defining features present in the rest frame UV. Medium bands primarily aid in increasing or decreasing the $\delta\chi^2$ constraint relative to the next-best low-z solution. It is only in our hypothetical scenarios that completeness is improved, with either the inclusion of the reddest medium bands (providing some constraints on the Balmer break at $z\sim10.5$) or greater depths in the blue broad bands leading to notable improvements in the 5-8$\sigma$ regime.

\subsubsection{High-z Sample Contamination}

Contamination is when members of an unwanted sub-population enter a sample selection. In our case, contaminant galaxies are those which are of significantly lower redshift ($z<6$) than the target sample of $z>7.5$. The contaminating fraction is defined as the number of these low-z sources divided by the total number of sources (contaminant plus true) selected. It should be noted that JAGUAR was produced in the pre-JWST era. Comparing its intrinsic UV LF to observations across cosmic history \citep[e.g.][]{Bouwens2021}, JAGUAR number densities typically are slightly lower but within a factor of $\sim2$. However, at very high redshifts beyond previous HST limits ($z>11.5$), JAGUAR number densities are around factor of 20 below new observations \citep[e.g.][]{Adams2024,Donnan2024,Whitler2025}. Consequently, the ratio of potential low-z contaminants and high-z galaxies differs from reality and a JAGUAR-derived contamination rate would be too high. To compensate, correctly identified galaxies at $z>11.5$ are up-weighted by a factor of 10 in our contamination calculations.

In all of our simulated runs, we find that contamination is generally low for the JADES Origins Field. Even without the use of medium-width bands, contamination is zero for sources detected to $15\sigma+$ and begins to rise at increasing rates as the $5\sigma$ level is approached. The primary reason is with such high significance in the detections, the measured depth of the Lyman break discounts the potential Balmer break alternative in the bluer bands.

The inclusion of only 1 or 2 medium-width bands is found to make minimal difference to the total contamination rates, this is because these bands probe narrow wavelength ranges and so there are limited redshift ranges where medium-width bands have the potential to exhibit notably different fluxes (due to the presence of atomic lines) compared to their neighbouring wide- or medium-width bands. With the inclusion of a diverse range of medium-width bands, such as the 7 total medium bands used in the JOF fields, there is a notable reduction in contamination rates. Crucially, the Jades Origins Field employs the use of generally bluer medium-width bands (1.6-3.35$\mu m$) which probe the potential locations of rest-frame Balmer and Oxygen emission lines for the low-z Balmer break contaminants.

\begin{figure*}%
    \centering
    \subfloat[\centering JAGUAR galaxy R1-191915, $z=3.41$, $\log(M_\odot)=6.85$, $M_{\rm UV}=-15.2$, $sSFR=-8.2$]{{\includegraphics[width=7cm]{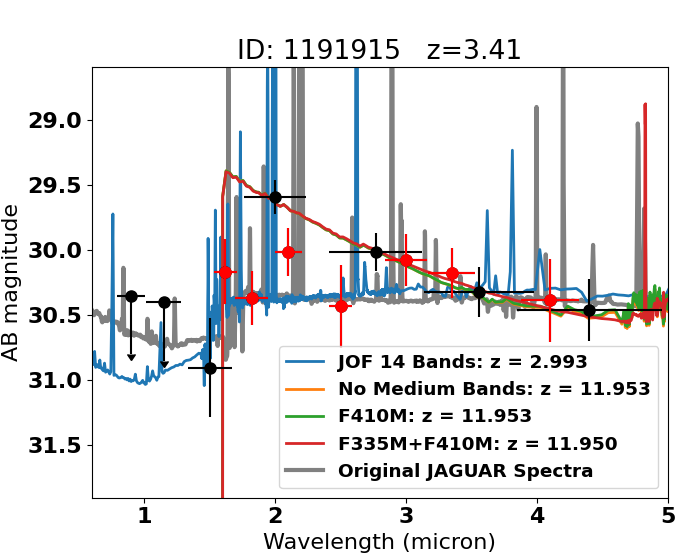} }}%
    \qquad
    \subfloat[\centering JAGUAR galaxy R1-171152, $z=2.97$, $\log(M_\odot)=7.03$, $M_{\rm UV}=-14.9$, $sSFR=-8.6$]{{\includegraphics[width=7.1cm]{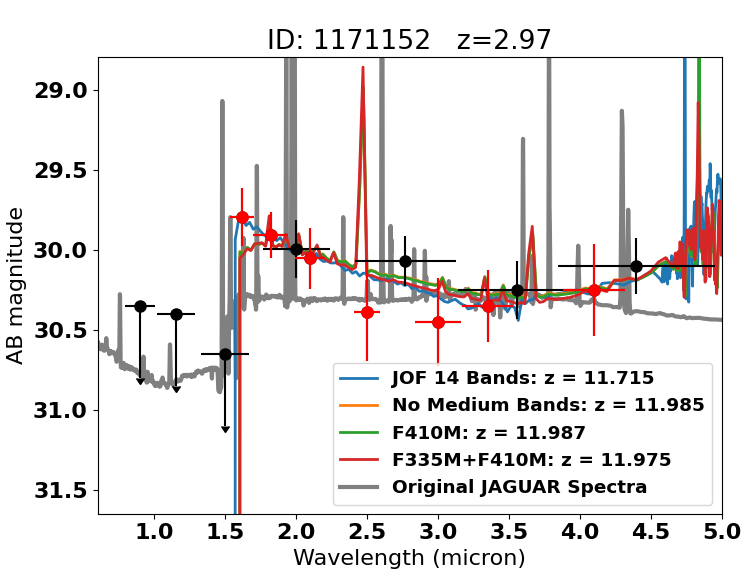} }}%
    \caption{The SED fits applied to mock photometry of two example simulated galaxies from the JAGUAR Semi-Analytic Model which are low-z but selected as high-z with NIRCam photometry. In all plots, the grey line shows the original JAGUAR spectra, the blue line shows the EAZY SED fit using of all 14 JOF filters and the yellow, green, red lines show results from using fewer filters. Photometric data points coloured in black are broad-width bands while those in red are medium-width, all photometric data points have had random scatter applied based on JOF field depths. Both sources show instances where strong emission lines (H$\beta$, [OIII] and H$\alpha$) can mimic blue high-z galaxies in broad-band colours. The first example (left) shows a case where the inclusion of many medium bands alleviates degeneracy between high-z and low-z solutions. The second case (right) an example is shown where the inclusion of many medium bands still results in the selection of a contaminant low-z galaxy as high-z.}%
    \label{fig:simbalmer}%
\end{figure*}

Contaminant galaxies are typically misidentified Balmer break sources at $1.5<z<4.5$. These misidentifications are driven by shallower bands probing the break regime (F090W and F115W, which are 0.3-0.5 mags shallower than the other broad-width bands) and the convenient positioning of strong emission lines within the broadband filters can replicate a blue SED slope of a high-z galaxy. Within Figure \ref{fig:simbalmer} we display two examples of low mass $z\sim3$ galaxies exhibiting broad-band colours that replicate blue high-z galaxies (see also the known real-world example of Balmer lines masquerading a Lyman break in \citet{Donnan2022,ArrabalHaro2023}).

We finally examine our two hypothetical expansions to the JOF survey design. First we explore the inclusion of the remaining 5 medium-width photometric bands (F140M, F360M, F430M, F460M, F480M). When conducting  our completeness and contamination simulations, we find that contamination rates do not change compared to the current survey design. This is because the majority of these bands are at the extreme red end of NIRCam's wavelength coverage. For the contaminating $1.5<z<4$ population, emission lines are typically weak within this wavelength regime, limiting the information that medium-width bands provide. Completeness does increase for fainter sources, particularly for sources with prominent Balmer breaks at 4-5 microns. For the second hypothetical case, we increase the depths of the two bluest bands (F090W, F115W) by 0.4 mags to better match the depth of the other bands. We find this approach to be highly impactful on both completeness and contamination rates. This is because the deeper bands probing the break regime improve the limits placed on a non-detection and rule out Balmer break solutions more confidently. 

Our findings indicate that while medium bands can be useful, from a very-high-z perspective it is important to employ those that would be located around the anticipated low-z Balmer line alternatives to rule out contaminants. In addition, the sensitivity profile of NIRCam has resulted in a number of surveys which have F090W and F115W depths that are notably shallower than the F200W and F277W filters used to select the rest-frame UV of ultra-high-z candidates. By placing additional time into these filters, NIRCam is capable of obtaining simultaneous observations in the desirable medium-band in the 2.5-4 micron regime in the red, enabling the contaminant population to be tackled from both angles. Future survey designs, or extensions to current programmes, with a high-z emphasis may with to consider dedicating more time to the above in order to maximise reliable samples near their detection limits.

\begin{figure*}%
    \centering
    \subfloat[\centering Original SED of Object ID 5009 in our fiducial JOF catalogue]{{\includegraphics[width=8.5cm]{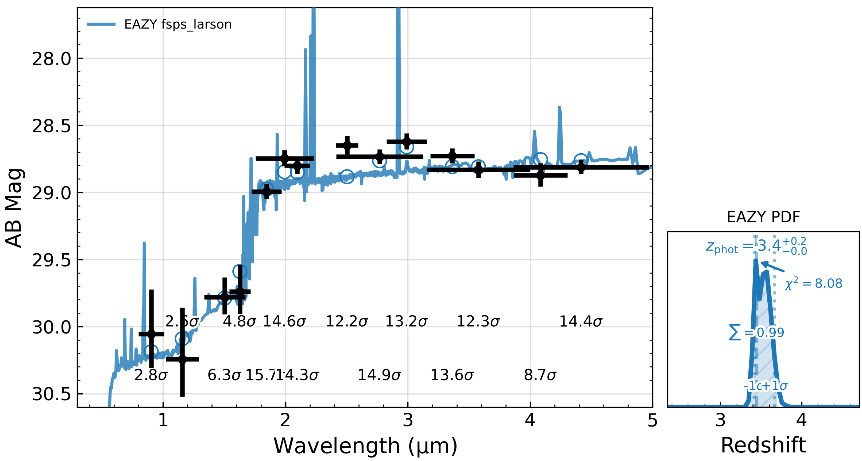} }}%
    \qquad
    \subfloat[\centering The same object observed in intentionally degraded imaging]{{\includegraphics[width=8.5cm]{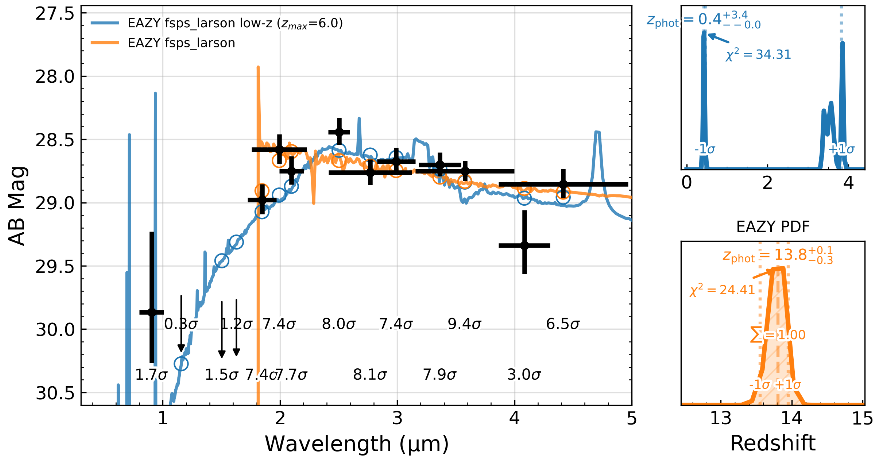} }}%
    \caption{An example of a source identified in the original JOF imaging as a Balmer break candidate at $z=3.4$ but is selected as a robust high-z candidate when the imaging is intentionally degraded by $\sim0.7$ mags. Blue lines show the best fitting $z<6$ solution while the orange line shows the best fitting $6<z<25$ solution. For the original SED fit (left) no valid high-z solution was identified. For the degraded observations (right), mild detections in the blue are removed and medium band excesses as a result of emission lines become more insignificant relative to the continuum detection, resulting in a dominant high-z solution.}%
    \label{fig:fakenoise}%
\end{figure*}

\section{The impact of depth on Photo-z selections \& contamination}

The JADES Origins Field not only provides a vast array of JWST filters, but also exquisite depth. In this section, we experiment with artificially degrading our imaging to match the filter and depth combinations of other JWST surveys in order to assess if high-z selection criteria behave the way it is expected, or if certain contaminant populations enter our samples. Within this experiment, we degrade the depths of our images by factors of [2,4,8]. This process is conducted through the addition of random gaussian noise which is scaled based on the original distribution of background pixel values. Background pixels are identified with the use of the segmentation map generated by SExtractor which is then further dilated by 25 pixels to ensure the faint outskirts of resolved sources are excluded. The associated error and weight maps are scaled accordingly based on the degradation factors used. These images are then rerun through our full catalogue generation and sample selection pipeline. The goal is to examine how previously high S/N sources with 14 photometric bands are now handled when fewer bands and lower S/N is available.

\subsection{Local Depth calculations}

The above process adds a purely Gaussian noise to the data, however the original noise in the images is not perfectly Gaussian. It is affected by the presence of artefacts in the images, contributions from undetected sources and correlated noise. Consequently, while the standard deviation of the pixel distribution increases by factors of e.g. 2, the local depths measured on the resultant images do not degrade by the $\sim0.7$mags expected but by a smaller degree, particularly at small degradation levels where other factors continue to dominate. Our degradation of factors of 2, 4, 8 result in average local depths which are 0.35, 0.77, 1.40 mags shallower on average. This shows that as the Gaussian noise becomes dominant over e.g. the correlated noise, the expected behaviour is reached.

\subsection{Changes in Photometric Redshifts \& Selections for High-z Sources}

We begin our investigation by assessing which objects meet our full selection criteria after all photometric bands are degraded by the same level. We identify 46, 21, 12 sources when the images are degraded by the three steps described above. The number of consistent objects between the samples generated from degraded imaging was 34, 17, 9, meaning that with each degradation step, sources were both lost and gained from the sample selection. For the remainder of this section we explore what these sources are and why their classification as robust high-z galaxies (or not) changed.

We examine the sources lost from the final sample when images are degraded. Due to the shape of the high-z luminosity function, there are more galaxies at low SNR than at high SNR, and so as the noise levels are increased, large numbers of galaxies are lost from the sample due to the constraint of requiring $5\sigma$ detection's in the rest-frame UV. 16 galaxies are lost from the sample when the images are degraded by 0.35 mags, with 14/16 retaining initial high-z solutions. A total of 10 sources are lost due to detection limits in the rest-frame UV. For the remaining six sources, two switch to dominant low-z solutions and the other four now have too significant of a secondary solution at low-z. All six sources lost due to increasing low-z confidence have detections of less than $7\sigma$ in the degraded imaging. A similar trend is found when comparing the original catalogue to those degraded by 0.77 and 1.4 mags. Under the assumption that the photo-z is correct when using the full depth available, the completeness of high-z sources in the 5-7$\sigma$ regime in each of our degraded images is around 56-73\%, broadly in line with completeness expectations from our JAGUAR simulations and slightly higher than the 40\% completeness found when comparing our 0.32 and 0.2as aperture catalogues.

While many sources are lost as images are degraded in quality, there are a select few new sources that enter the sample. In total, there are 12, 3, 3 sources in each of the iterative degradations which are selected as high-z in the shallower imaging but ruled out as high-z when deeper imaging is employed. We find that there are 3,0,2 instances where the new source has no cross match in the deeper catalogue, we find these instances to be cases where in the deeper catalogue, faint and extended emission from a large resolved neighbour has blended these sources in the segmentation map. For the imaging degraded by 0.35 mags, the remaining 9 sources added to the catalogue are either classed as low redshift sources (6 objects), detected in the blue (1 object) or have poor SED statistics (2 objects). In the catalogue degraded by 0.77 mags, all 3 sources are low-z with the deeper imaging as is the 1 other object in the 1.4mag degraded catalogue. 

The vast majority of sources which are selected as high-z sources in shallow catalogues but found to be confidently low-z in deeper catalogues are within the 5-8.5$\sigma$ regime in the respective shallow catalogue. Using the numbers of contaminants found, under the assumption that the photo-z using the full depth imaging is correct, the contamination rate in the 5-8.5$\sigma$ regime ranges from 20-38\% between catalogues. There is a single instance of a strongly detected source $10\sigma+$ switching from low-z in deep catalogues to high-z in shallower catalogues. This object is JADES-GS-z14-0 \citep{Carnani2024,Schows2025,Carnani2025}. It is found that the increasing background results in this source's successful selection as the faint extended emission from its neighbour is lost to noise, creating the expected $z\sim14$ non-detections from a Lyman dropout.

The contamination rate as measured directly from real degraded sources is slightly higher than initially anticipated from the JAGUAR simulation work conducted in the previous sections. However, this estimation (along with the completeness value) is based on a small number of sources and highly subject to the distribution of colours and number density of potential contaminant sources. The key result here is that Balmer-Lyman confusion does take place in the 5-8$\sigma$ regime on a regular enough basis for it to be a notable concern. This is exacerbated by the relative lack of depth in the bluest bands used in deep field programmes, with 0.3+ mag differences between the depths in F090W and F115W compared to F200W a regular occurance in deep JWST surveys. We display a full example of such a source in Figure \ref{fig:fakenoise}.

\section{Measuring the faint-end of the UV Luminosity Function with the JOF}

Our findings indicate that completeness is high and contamination low for sources detected at $8\sigma+$ in the rest-frame ultraviolet. In the context of studying galaxies as a population (e.g. by measuring the Ultraviolet Luminosity Function), the primary impact of an incorrect determination of completeness or contamination would be constrained to the faintest $\sim0.5$ magnitudes probed. Other studies of the UV LF in recent literature have largely focussed on the completeness estimation and report similar findings to our own, with generally high ($80+\%$) completeness which rapidly falls off over a 0.3-0.5 magnitude range \citep[See][for examples]{Robertson2024,Finkelstein2024}. with the largest potential ramifications for the measurement of the faint-end slope `$\alpha$'.

To observe the potential variability in the measurement of the UV LF, we measure the UV LF with the JOF field using the various sub-catalogues of photometric redshifts that we have produced throughout this work (different aperture sizes, different numbers of medium bands). We begin by constructing a UV LF using the base catalogue of all 14 photometric bands and 0.32as aperture diameters. We measure the UV LF using the 1/Vmax method in three redshift bins, $8.5<z<9.5$, $9.5<z<11.5$ and $11.5<z<13.5$, using an unmasked area of 8.7 square arcminutes. We apply the completeness correction based on the $M_{\rm UV}$ and redshift of each source using the JAGUAR simulations previously described and whose method is described in \citet{Adams2024}. We employ a cut where no galaxies are used if their assigned completeness is below 30\% or the contamination rate greater than 30\%. The value of $M_{\rm UV}$ used for each source is measured by placing a 100 angstrom top hat filter between 1450-1550 angstroms in the rest-frame SED obtained during the EAZY photo-z process. These SED's are scaled up by the difference between the aperture corrected magnitude and the total magnitude from SExtractor's MAG\_AUTO parameter in the nearest JWST filter to the rest-frame 1500 angstroms in order to account for light omitted due to the partly resolved nature of high-z galaxies.

Since the JOF field is limited to a single NIRCam pointing, cosmic variance is a significant potential source of systematic bias in our measurements. To broadly estimate the impact of Cosmic variance on number counts, we utilise the cosmology calculator from \citet{Trapp2020}, finding cosmic variance errors of the order 30\% for the three redshift bins. This error source is added in quadrature to the Poisson error, which we calculate using the Poisson confidence interval based on the $\chi^2$ distribution $I = [0.5\chi^2_{2N,a/2}, 0.5\chi^2_{2(N+1),1-a/2}]$ since our number counts ($N$) are small in our bins \citep{Ulm1990}.

\begin{figure*}%
    \centering
    \subfloat[\centering Multi-Catalogue $z=9$ UV LF]{{\includegraphics[width=8.5cm]{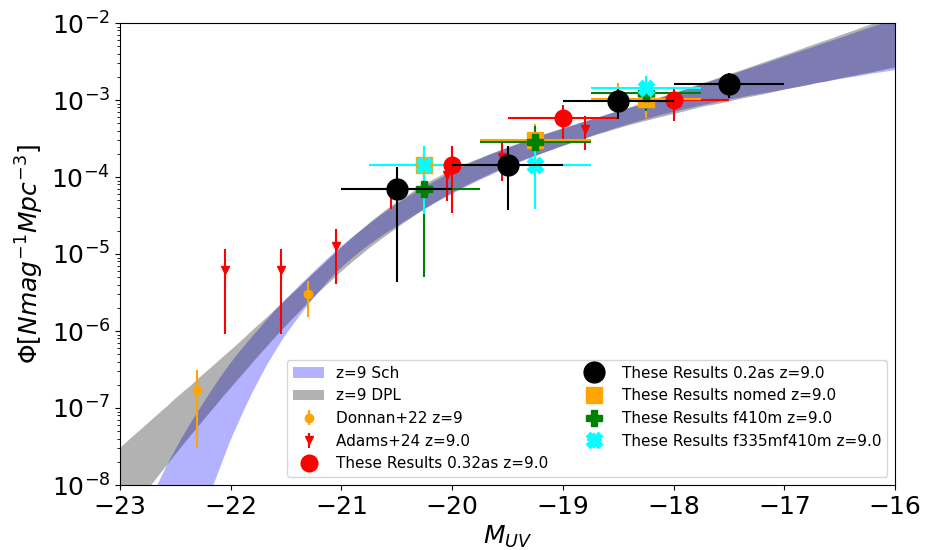} }}%
    \subfloat[\centering Fiducial $z=9$ UV LF]{{\includegraphics[width=8.5cm]{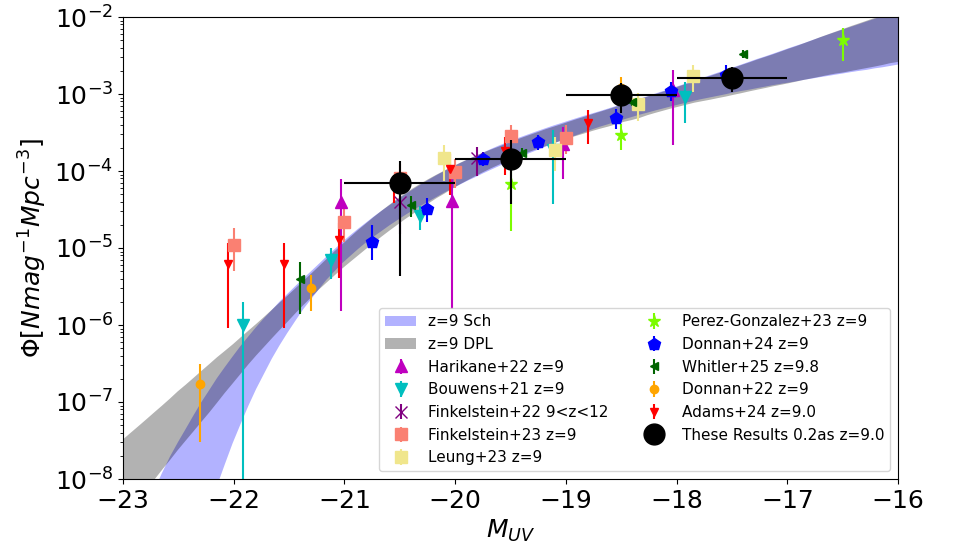} }}%
        \qquad
    \subfloat[\centering Multi-Catalogue $z=10.5$ UV LF]{{\includegraphics[width=8.5cm]{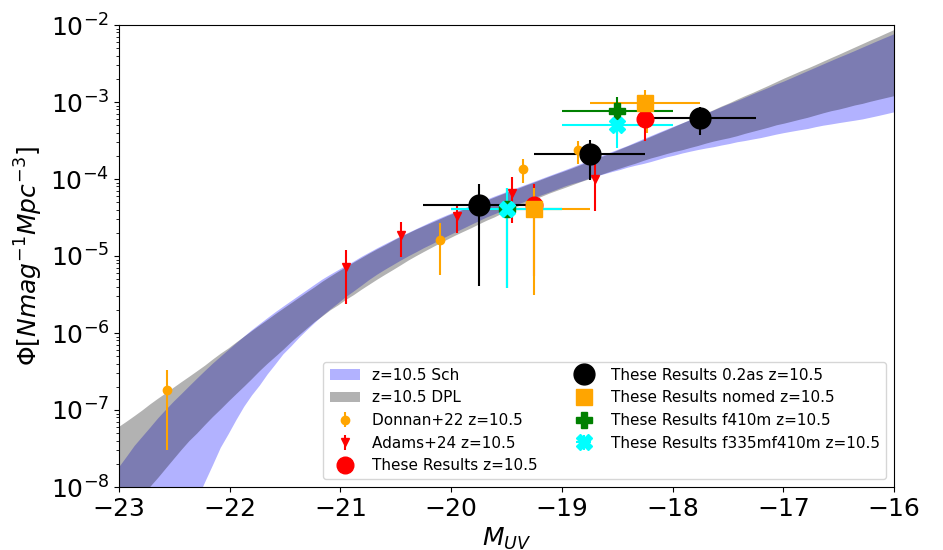} }}%
    \subfloat[\centering Fiducial $z=10.5$ UV LF]{{\includegraphics[width=8.5cm]{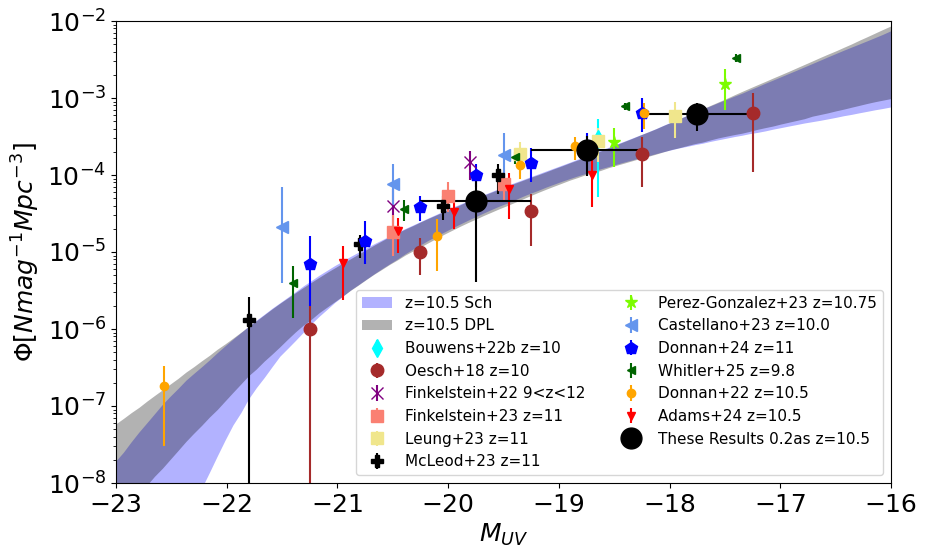} }}%
        \qquad
    \subfloat[\centering Multi-Catalogue $z=12.5$ UV LF]{{\includegraphics[width=8.5cm]{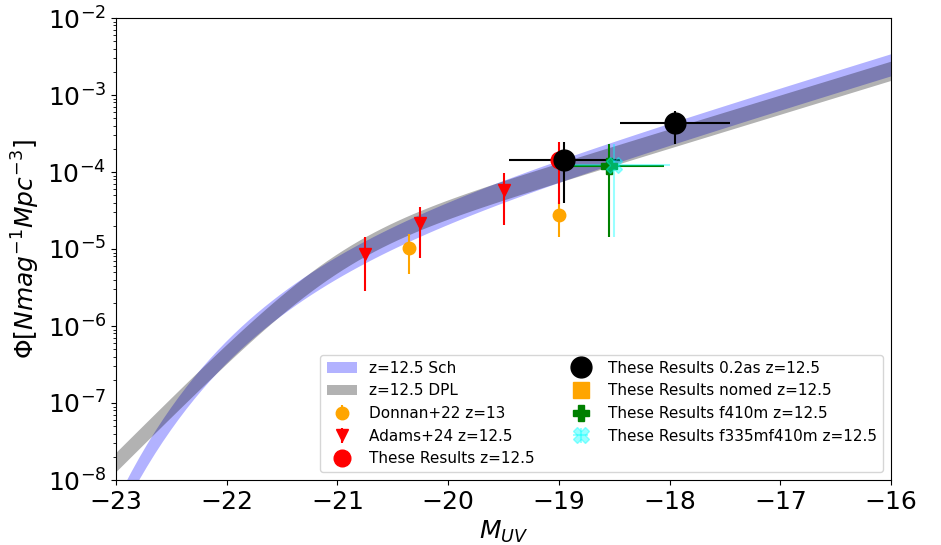} }}%
    \subfloat[\centering Fiducial $z=12.5$ UV LF]{{\includegraphics[width=8.5cm]{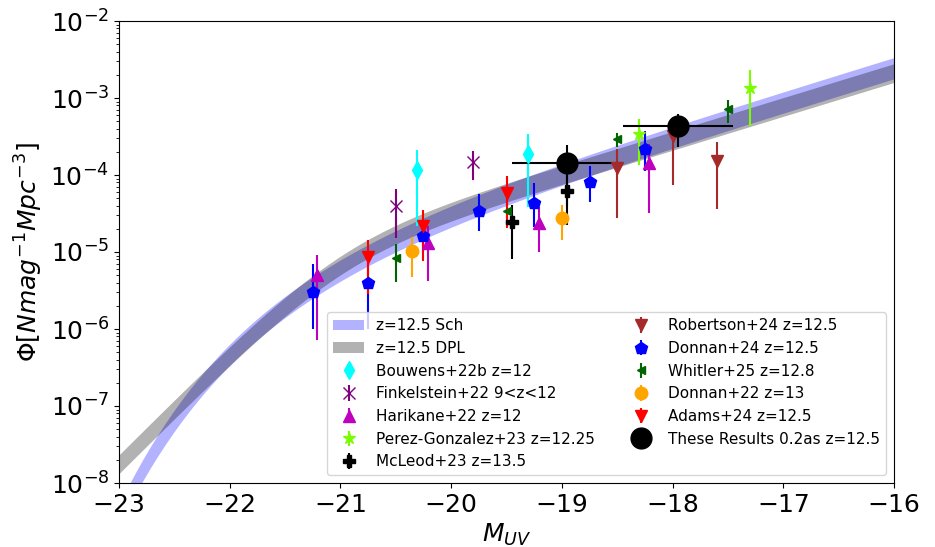} }}%
    \caption{The measured faint end of the UV Luminosity function within the redshift bins of $8.5<z<9.5$, $9.5<z<11.5$ and $11.5<z<13.5$. In the left column, we show the results from using either the 0.32as aperture catalogue, the 0.2as aperture catalogue or removing medium bands from the process. Various interpretations are found produce consistent results with the anticipated behaviour of the 0.2as catalogue extending the luminosity range faintwards. Shaded regions indicate the $1\sigma$ confidence interval on the best fit Schecter or DPL funciton to the fiducial UV LF (0.2as apertures plus the use of \citet{Adams2024} and bright VISTA data points from \citet{Donnan2022}). In the right column, we show our fiducial result alongside the range of previous observations using HST and JWST. These include: \citep{Oesch2018,Harikane2023,Bouwens2021,bouwens2022,Donnan2022,Finkelstein2022,Finkelstein2024,Leung2023,PerezGonzalez2023,Adams2024,McLeod2024,Donnan2024,Robertson2024}}.%
    \label{fig:UVLFs}%
\end{figure*}

The results of this process are presented in Figure \ref{fig:UVLFs}. The deep imaging provided by the JOF field enables galaxies to be identified down to $M_{\rm UV}\sim-17.00$ at $z\sim9$ and $M_{\rm UV}\sim-17.50$ at $z\sim12.5$. Compared to \citet{Adams2024}, which compiled 6 JWST surveys together over 180 square arcminutes of area, the JOF field expands the faint end coverage by 1mag. We conduct our parametric fits to the UV LF using these new JOF results, the \citet{Adams2024} results and, where possible, constraints from the ground-based UltraVISTA survey from \citet{Donnan2022}. Our observations are consistent with the high number density of galaxies found by other JWST surveys that have pushed the depth frontier at $m>30$, including the NGDEEP \citep{Leung2023} and MIDIS surveys \citep{PerezGonzalez2023} and the JADES teams own UV LF measurements \citep{Robertson2024,Whitler2025}.

\begin{table}
\caption{The measured rest-frame UV LF and its error margin in the redshift bins $8.5<z<9.5$, $9.5<z<11.5$ and $11.5<z<13.5$. Column 1 shows the bin in absolute UV magnitude ($\lambda_{\rm rest}=1500$\AA), widths are set to 1 magnitude. Column 2 shows the number density of objects and column 3 shows the uncertainties in the number density, which consist of the Poisson error summed in quadrature with the derived cosmic variance from \citet{Trapp2020}.}
\centering
\begin{tabular}{lcc}
\hline
$M_{\rm UV}$     & $\Phi (10^{-5})$                        & $\delta\Phi (10^{-5})$                  \\
$[\textrm{mag}]$ & $[\textrm{mag}^{-1} \textrm{Mpc}^{-3}]$ & $[\textrm{mag}^{-1} \textrm{Mpc}^{-3}]$ \\ \hline
$z=9$          &          &            \\
-20.00          &  14.28        &  10.85          \\
-19.00          &  58.40        &  28.52          \\
-18.00          &  101.2        &  47.9          \\ \hline
$z=10.5$          &          &            \\
-19.25          &  45.33        &  39.87          \\
-18.25          &  50.60        &  29.38          \\ \hline
$z=12.5$          &          &            \\
-19.00          &  14.30        &  10.50          \\ \hline \hline
$z=9$ 0.2as          &          &            \\
-20.50          &  7.03        &  6.59          \\
-19.50          &  14.38        &  10.65          \\
-18.50          &  96.84        &  40.32          \\
-17.50          &  163.03        &  58.41         \\ \hline
$z=10.5$ 0.2as         &          &            \\
-19.75          &  4.53        &  4.13          \\
-18.75          &  20.82        &  11.20          \\
-17.75          &  61.49        &  24.17          \\ \hline
$z=12.5$ 0.2as         &          &            \\
-19.00          &  14.30        &  10.50          \\
-18.00          &  42.90        &  20.00          \\ \hline
\end{tabular} \label{Tab:Points}
\end{table}

\begin{table*}
\caption{The best-fit parameters for the final Schechter (Sch) and Double Power Law (DPL) functional forms for both the Levenberg–Marquard (LM) approach and the MCMC approach when applied to our deeper 0.2as catalogue with all available bands. For the MCMC, we show the median and standard error obtained from the posterior. The individual MCMC sample with highest probability closely matches the best fitting parameters of the LM methodology. Values presented with an asterisk indicate the parameter is fixed to this value in the fitting procedure.}
\centering
\begin{tabular}{ll|llll}
Redshift & Method   & $\log10(\Phi)$ & $M^*$ & $\alpha$ & $\beta$ \\ \hline
9        & Sch/LM   & $-3.52 \pm 0.34$    & $-20.00\pm 0.34$    & $-1.81 \pm 0.31$      & --      \\
9        & Sch/MCMC & $-3.64^{+0.36}_{-0.61}$    & $-20.09^{+0.39}_{-0.59}$    & $-1.89^{+0.38}_{-0.36}$ & --      \\
9        & DPL/LM  & $-4.21 \pm 0.38$    & $-20.72\pm 0.41$    & $-2.11 \pm 0.24$      & $-6.0 \pm 2.1$      \\
9        & DPL/MCMC & $-4.01^{+0.42}_{-0.42}$ & $-20.41^{+0.61}_{-0.38}$ & $-2.04^{+0.34}_{-0.28}$     & $-4.81^{+0.74}_{-0.77}$     \\ \hline
10.5        & Sch/LM   & $-4.82 \pm 0.49$    & $-21.09 \pm 0.50$    & $-2.16 \pm 0.30$ & --      \\
10.5       & Sch/MCMC & $-4.71^{+0.59}_{-0.82}$    & $-20.89^{+0.77}_{-0.79}$ & $-2.13^{+0.46}_{-0.38}$       & --     \\
10.5        & DPL/LM   & $-5.34\pm1.02$    & $-21.52\pm1.34$    & $-2.36\pm0.30$      & $-5.4\pm4.2$      \\
10.5        & DPL/MCMC & $-4.94^{+0.68}_{-0.74}$    & $-20.92^{+0.93}_{-0.80}$   &  $-2.28^{+0.44}_{-0.34}$     & $-4.48^{+0.94}_{-1.00}$     \\
10.5        & Sch/LM   & $-4.74 \pm 0.20$    & $-21.01\pm 0.28$    & $-2.1^*$ & --      \\
10.5        & Sch/MCMC & $-4.68^{+0.28}_{-0.23}$    & $-20.89^{+0.48}_{-0.32}$ & $-2.1^*$       &  --    \\
10.5        & DPL/LM   & $-4.75 \pm 0.37$    & $-20.82 \pm 0.60$    & $-2.1^*$      & $-4.3 \pm 0.9$      \\
10.5        & DPL/MCMC & $-4.74^{+0.40}_{-0.31}$    & $-20.73^{+0.66}_{-0.51}$   &  $-2.1^*$     & $-4.56^{+0.86}_{-0.93}$     \\\hline
12.5        & Sch/LM   & $-4.71 \pm 0.11$    & $-21.0^*$    & $-2.1^*$ & --      \\
12.5        & Sch/MCMC & $-4.76 ^{+0.12}_{-0.18}$    & $-21.0^*$ & $-2.1^*$       & --     \\
12.5        & DPL/LM   & $-4.84 \pm 0.10$    & $-21.0^*$    & $-2.1^*$      & $-4.6^*$      \\
12.5        & DPL/MCMC & $-4.87^{+0.11}_{-0.14}$    & $-21.0^*$   &  $-2.1^*$     & $-4.6^*$     \\\hline
\end{tabular} \label{tab:fits}
\end{table*}

\subsection{Fitting the UV LF}

To parameterise the UV LF and describe any potential evolution, it has been common process to fit the UV LF of galaxies with either a Schechter \citep[SCH: power law with an exponential decline][]{Schechter1976} or Double Power Law (DPL) parameterisation. These both have a normalisation parameter ($\Phi^*$), a characteristic luminosity/magnitude where the shape changes ($M^*$) and a faint-end slope ($\alpha$). The DPL functional form has an additional bright-end slope ($\beta$) in replacement to the exponential decay in the Schechter form, adding more flexibility when describing brighter galaxies. 

We elect to fit the UV LF in two different ways. First we fit with a classical $\chi^2$ minimisation technique \citep[][]{Levenberg1944,Marquardt1963} and second we fit using a Markov-Chain-Monte-Carlo (MCMC) using {\tt emcee}. When conducting our fits, we combine the measurements of this study with those from \citet{Adams2024} which mimics our methods almost identically (which also contains VISTA based-data points at the bright-end from \citet{Donnan2022}). Each MCMC run is conducted with 500 walkers, burns-in for 5000 steps and conducts an additional 5000 steps. Due to the still relative lack of galaxies at these high redshifts and the degeneracies between parameters within these functional forms, we conduct fits with all parameters left free at $z=9$, at $z=10.5$ we experiment with leaving all variables free and with fixing the faint-end slope to $\alpha=-2.1$. At $z=12.5$, we fix the shape broadly to the $z=10.5$ result and only fit for the normalisation. The value of $\alpha=-2.1$ is chosen as when it is left free at $z=9$ and $z=10.5$, the best fit values are typically in the range of $-2.3<\alpha<-1.9$ in agreement with recent predictions from \citet{Donnan2025} and other observational studies using JWST and HST, though still with relatively large error margins.

Relative to results from \citet{Adams2024}, uncertainties in the faint-end slopes (when left free) have decreased by around $\sim40-50$\% thanks to the increased depth provided by the JOF field allowing us to identify intrinsically fainter galaxies ($\sim1$mag). 

\subsection{A consistent faint-end of the UV LF between photometric catalogues}

We repeat the above procedure for measuring the UV LF with four other versions of our photometric catalogue. The first three use the same 0.32as aperture data, but omit medium-width bands leaving only zero, F410M or F410M+F335M bands available, mimicking the bands used in similar deep field JWST surveys. We found in Section 4 that between catalogues that employ different combinations of medium-width bands, the number of selected sources at $z>8$ varies by only $\sim10$\% and that the anticipated completeness also varies by of order 10-15\%. Therefore differences between UV LF's measured by the different catalogues may be expected to be consistent.

We find that for the $z=9$ and $z=10.5$ UV LF, the majority of galaxies used are consistent between the C-ALL, C-0, C-1 and C-2 catalogues, though up to 20\% of the galaxies occupying the faintest bin can change. These galaxies fall into two categories; 1) they either have a photo-z close to the redshift bin boundary which changes slightly between catalogues (e.g. $z_p=8.6$ in the $8.5<z<9.5$ bin) scattering them in/out or 2) they are among the faintest in the sample and are no longer selected due to changes in $\chi^2$ confidence relative to low-z solutions in different catalogues. For the $z=12.5$ redshift bin, the C-ALL catalogue employs the use of 2 galaxies, while C-2 and C-1 use 1 galaxy and C-0 uses no galaxies. The scatter in redshift is larger in this regime, with two $z\sim12$ galaxies in C-ALL now $z\sim11$ or $z\sim14$ in C-0. Additionally, completeness and contamination are also sensitive. For C-ALL, the average completeness is 83\% for the galaxies used while it drops to 40\% for C-2 and C-1, with this factor of two in completeness making up some of the difference in measured number density obtained from using 1 or 2 galaxies.

The four catalogue iterations (C-ALL, C-0, C-1, C-2) all produce consistent UV LFs when errors from poisson statistics and cosmic variance are considered, though these error sources are large. A wider area medium-band study would be required to elaborate further. Our findings agree with earlier JWST-based work that pushed the $m>30$ boundary and found high number densities of moderately luminous ($-19<M_{\rm UV}<-18$) galaxies both within this very field \citep{Robertson2024} and in other fields \citep{Leung2023,PerezGonzalez2023}.

When using a different aperture size, we also observe minor changes to the measured UV LF that match expectations. The deeper coverage provided by the use of smaller apertures enables intrinsically fainter galaxies ($\sim0.5$ mags) to be identified while the brighter galaxies found in the wider aperture catalogue have the majority recovered with similar redshifts and properties. We provide the final UV LF fits to the 0.2as version of the UV LF on account of its constraining power on the faint-end slope in Table \ref{tab:fits} and provide the points in Table \ref{Tab:Points}.

\subsubsection{Testing the impact of Damped Lyman Alpha systems}

Most photo-z studies in the early Universe make use of the Lyman break formed by the attenuation of UV flux from the neutral IGM \citep{Madau1995,Inoue2014}. At very high redshifts, the Lyman break is essentially a step function. However, NIRSpec observations have shown a prevalence of damped Lyman alpha (DLA) systems in the early Universe. DLA's arise from neutral gas in close proximity to a galaxy and can soften the shape of the Lyman break. This can result in systematic offsets in photometric redshifts as reported in \citet{asada2024}. We experiment with the implementation of the DLA attenuation described in \citet{asada2024} in our photo-z process. Since the JAGUAR SAM doesn't include such attenuation, we focus on the application to the real JOF catalogues.

We find that including DLA features in our template fitting results in systematic shifts towards lower-z for our $z>7.5$ sample. The magnitude of these shifts is an average of 0.08 in redshift between $9<z<11$ and 0.18 between ($11<z<13$). Using our fiducial catalogue, we find that the UV LF does not change significantly when using these slightly modified redshifts due ot the low numbers and broad bins. These redshift biases of up to 0.18 at $z\sim12$ will likely be more impactful when working with larger, more statistical samples of galaxies from e.g. multi-survey studies.

\subsection{Comparisons to Other Observational Studies}

In this subsection we examine our binned number densities and best fitting Schechter/DPL parameters with those of other JWST-based observations undertaken in the past couple of years. At $z\sim9$, our new number densities at $-20<M_{\rm UV}<-17.5$ are in close agreement to observations from most other observational studies \citep{Finkelstein2024,Leung2023,PerezGonzalez2023,Donnan2022,Donnan2024,Adams2024,Whitler2025}, showing a general consensus across multiple different survey fields. The only exception is the GLASS field which may be slightly overdense at the bright end \citep{Castellano2022}.

At $z=10.5$, the data points from \citet{Adams2024} are on the lower-end of the number densities measured from multiple JWST-based studies though are generally consistent within $1\sigma$ with the exception of the point at $M_{\rm UV}=-18.75$ which is over $1\sigma$ below the works of \citet{Donnan2024}, \citet{Leung2023} and \citet{PerezGonzalez2023}. On the other hand, the fainter data points added by the inclusion of the JOF at $M_{\rm UV}=-18.75$ and $M_{\rm UV}=-17.75$ are in very close agreement to the observations from those same studies. This could indicate a potential overestimation of sample completeness at the faint-end of \citet{Adams2024}, though it agrees with pre-JWST observations from \citet{Oesch2018}, or it could be driven by the impacts of cosmic variance. Most studies reaching this regime primarily make use of ultra-deep blank fields (NGDEEP, MIDIS, JOF etc.) which are all located the wider GOODS-S region. The inclusion of more ultra-deep or gravitationally lensed fields will be required to fully assess the impact of cosmic variance at the very faint end.

At $z=12.5$, our observations again agree with the other surveys reaching $m>30$, now including \citet{Robertson2024}, with the same caveat that these are all located within the wider GOODS-S field. Previous fits within \citet{Adams2024} resulted in a relatively large uncertainty in the normalisation $\log(\Phi^*)$ which has now been further constrained to the upper 50\%, showing that the UV LF evolves relatively slowly from $z=10.5$ to $z=12.5$. However, our fits provide galaxy number densities that are around $1\sigma$ (or $\sim0.15$dex) higher than the multi-survey study of \citet{Donnan2024}, further raising the question of potential cosmic variance induced systematics which will require more independent deep field programmes to solve.

\subsection{The UV Luminosity Density}

\begin{figure*}%
    \centering
    \hspace{-0.4cm}\includegraphics[width=15cm]{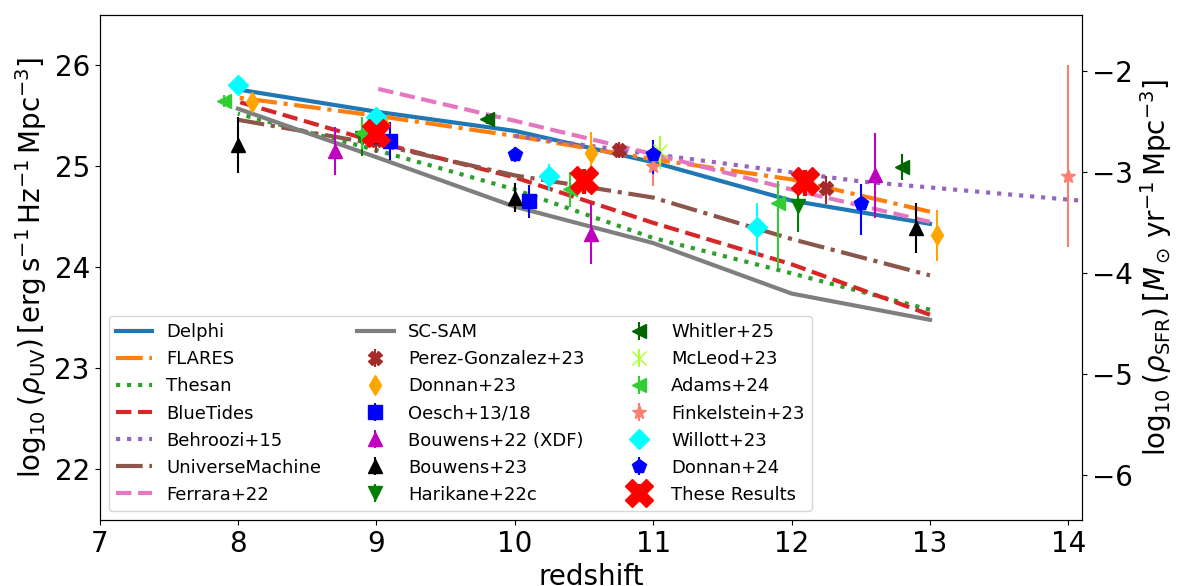} %
    \caption{The measured UV Luminosity density and star formation rate density at $8<z<14$ from a compilations of HST and JWST-based studies plotted over the predictions from a selection of simulations and semi-analytical models. Observations include \citet{Oesch2013,Oesch2018,Donnan2022,bouwens2022,PerezGonzalez2023,Donnan2024,Harikane2023,McLeod2024,Finkelstein2024,Willott2024,Adams2024}. Predictions include those from Delphi \citep{dayal2022}, FLARES \citep{Wilkins2023}, Thesan \citep{Kannan2022}, BlueTides \citep{Marshall2022}, SCSAM \citet{yung2024}, \citet{Behroozi2015}, UniverseMachine \citep{Behroozi2019}, \citet{Ferrara2022}. At $8<z<9$ observations are in close agreement with each other, while at $z\sim10-11$ the spread increases to span the full range of predictions before re-concentrating at $z>12$ around higher SFRD.}%
    \label{fig:lumden}%
\end{figure*}

Using our fiducial UV LF measurement, we calculate the UV Luminosity density down to $M_{\rm UV}<-17$ by integrating 1000 randomly drawn samples from the MCMC chains. This is then converted into an estimated star formation rate density (SFRD) replicating the procedure of \citet{Harikane2023} and \citet{Adams2024}. Here, we use the relations presented in \citet{Madau2014}, obtaining a conversion factor $k=1.15\times10^{-28} [M_\odot {\rm yr}^{-1}/({\rm erg}\, {\rm s}^{-1} {\rm Hz}^{-1})]$. The results of this process are presented in Figure \ref{fig:lumden}. The key result here is that the high number density of observed galaxies at $-19<M_{\rm UV}<-18$ galaxies at $z>10$ have constrained the previously uncertain UV Luminosity Density at $z\sim12$ to the upper bound of our previous measurements in \citet{Adams2024}, resulting in a greater tension with a selection of models which predict relatively low numbers of high-z galaxies.

When fitting a trivial log-linear evolution \citep[e.g.][]{Donnan2022,McLeod2024,Donnan2024} to our updated UV luminosity density measurements (including the $z=8$ result from \citet{Adams2024}, we find a slope of $-0.23\pm0.04$ and normalisation of $27.51\pm0.31$. The steepness of this evolution is shallower than initially measured in \citet{Adams2024}, almost identical to the value measured by \citet{McLeod2024} but steeper than found in \citet{Donnan2024}. When compared to theoretical predictions, there is a general bimodality in the star formation rate density expected, with a selection of simulations and analytic prescriptions predicting low star formation rates (SCSAM \citet{yung2024}, Thesan \citet{Kannan2022}, UniverseMachine \citet{Behroozi2019} in addition to \citet{Harikane2022}) and a selection predicting higher (FLARES \citet{Wilkins2023}, Delphi \citet{dayal2022}, \citet{Ferrara2022} in addition to \citet{Behroozi2015}). At $z=10$, our observations more closely match to those predicting lower SFRD, while at $z=12$ the opposite is the case, with observations more closely matching predictions of high SFRD. Further increased volume and sightlines probing fainter luminosities (e.g. the complete JADES survey and lensing clusters from e.g. PEARLS, CANUCS and GLIMPSE) will ultimately be required in order fully constrain the shape of the UV LF at $z>10$ and $-19<M_{\rm UV}<-16$.

\section{Conclusions}

In this paper, we use the recently conducted JADES Origins Field survey, in parallel with the JAGUAR semi-analytic simulation, to conduct an assessment of the completeness and contamination rates when photometrically selecting the highest redshift galaxies at $z>8$ with JWSTs NIRCam. We explore the advantage that the addition of medium-with photometric filters provides and intentionally degrade imaging to observe how selections differ as SNR values decrease.  
 We also measure the resulting UV luminosity function using our methods to determine the best measurement possible at these depths in the JOF and how other realisations of the galaxy population would affect this. Following these experiments, we identify the follow key conclusions:

\begin{enumerate}
    \item Even without the use of any medium-width bands, sample completeness is generally high (80\%+ at 10$\sigma$+). As the number of medium-width bands increases, so does sample completeness alongside a decrease in contamination. Medium width bands are most effective in decreasing contamination when they probe central NIRCam wavelengths (2-3.5 microns) where they cover the expected locations of emission lines for contaminant low-z sources (Balmer and Oxygen lines).
    \item A drop off in completeness rates and an increase in contamination rates begin to occur when probing the luminosity regime where the rest-frame UV is less than $8\sigma$. This is caused by an increasing degeneracy between Lyman and Balmer break solutions, exacerbated by the generally shallower blue photometric bands (F090W and F115W) common in deep field surveys due to NIRCam's sensitivity profile. This enables sources to be detected in the anticipated rest-frame UV at depths often fainter than the $5\sigma$ limits of the neighbouring bands probing the Lyman break.
    \item Degrading JOF images results in changes to the selected high-z sample broadly following the expectations of our simulation work. Completeness falls off in the 5-7$\sigma$ regime as previously high-z sources are now degenerate with Balmer break solutions. In addition, small numbers of low-z Balmer break sources with minor blue detections in the original imaging can be selected as high-z sources when images are degraded to be shallower. Completeness in the faint regime is measured from degraded imaging to be around 56-73\%, broadly agreeing with our simulated estimates, while contamination is slightly higher than expected at 20-38\%, through it is measured using a small number of sources.
    \item Instances of selecting contaminant sources can be alleviated through the use of medium-width bands positioned where low-z emission lines are to be expected and through an increase in observing time spent on the bluest photometric bands. The modular design of NIRCam is advantageous in this regard, since doubling the time spent in the bluest bands can be conducted in parallel to obtaining some of the redder medium-width bands. In the event that early JWST surveys are expanded upon or new fields contributed, we recommend the inclusion of additional time in the blue broad-bands relative to the red broad-bands in order to maximise the utility of the depths the red broad-width bands can achieve.
    \item Our fiducial measurement of the faint end of the UV LF at redshifts 8.5 to 13.5 agree with previous estimates from early JWST data, evidencing high numbers of moderately faint galaxies ($M_{\rm UV}\sim-18.0$) at this early time. The estimated star formation rate density of the Universe at $z=12$ is higher than previously measured in \citet{Adams2024} and closer to other early JWST-based measurements. However, there is the caveat of cosmic variance, as all other surveys reaching these depths have been concentrated within the wider GOODS-South region \citep[e.g. NGDEEP and MIDIS][]{Leung2023,PerezGonzalez2023}.
    \item When reproducing UV LF measurements using different catalogues that employ different combinations of photometric bands, we find no significant difference in the obtained results. Indicating that an adequate understanding of sample completeness and contamination can enable large, statistical studies to be reliably conducted with broad-band dominated surveys.
\end{enumerate}

We have shown that medium-width photometric bands can improve the reliability of photo-z's for the faintest high-z targets in the 5-8$\sigma$ regime. When studying and following up the deepest fields at $m>30$, they may prove crucial to identifying targets before significant time investment from spectroscopy or other follow-up work is committed. There is concern is that faint, dusty and/or low-mass Balmer break sources can mimic the colours of Lyman break galaxies in many early JWST survey designs \citep[see also discussion in][]{Gandolfi2025}. There are now several medium-band programmes already planned, underway or completed (e.g. JEMS, MegaScience, Technicolour, JUMPS) which will continue to pave the path forwards and open avenues to study extreme emission line systems, high-z drop outs and further stress-test photo-z pipelines.

It is also worth stressing that development continues to improve high-z selections. This work displays the expected reliability and issues facing currently employed systems. New templates of galaxy SEDs are continuously being tested \citep[e.g.][]{Larson2023,Clausen2024}, understanding of IGM attenuation is improving \citep[e.g.][]{asada2024} and the application of informed priors can be used to more conservatively examine fainter sources \citep[e.g.][]{Donnan2024}. Increasing spectroscopic samples of populations which can masquerade as high-z galaxies (be it galactic brown dwarfs, dusty line emitters or Balmer break galaxies) or objects occupying unexplored parameter spaces \citep[e.g. RUBIES ][]{degraaff2024} will be pivotal to refining high redshift selections and SED template sets. With large spectroscopic programmes like JADES, and now with CAPERS underway, a more statistical analysis of SED modelling comparing to real spectra will soon be enabled. CAPERS in particular will explore high-z sources from multiple, independently produced input catalogues which will be key to revealing why different groups select different sources and reveal more insights into the properties of contaminant sources \citep[see e.g.][]{Castellano2025}.

\section*{Acknowledgements}
We acknowledge support from the ERC Advanced Investigator Grant EPOCHS (788113), as well as two studentships from the STFC. JAAT acknowledges support from the Simons Foundation and JWST program 3215. Support for program 3215 was provided by NASA through a grant from the Space Telescope Science Institute, which is operated by the Association of Universities for Research in Astronomy, Inc., under NASA contract NAS 5-03127.

This work is based on observations made with the NASA/ESA \textit{Hubble Space Telescope} (HST) and NASA/ESA/CSA \textit{James Webb Space Telescope} (JWST) obtained from the \texttt{Mikulski Archive for Space Telescopes} (\texttt{MAST}) at the \textit{Space Telescope Science Institute} (STScI), which is operated by the Association of Universities for Research in Astronomy, Inc., under NASA contract NAS 5-03127 for JWST, and NAS 5–26555 for HST. The authors thank all involved with the construction and operation of JWST, without whom this work would not be possible. The authors also thank the team who designed and proposed for the JADES (PID 1280,1287) and JADES Origins Field (PID 3215) observing programmes, whose data was used in this study. Elements of this work also used observational products from CANUCS, GLASS, CEERS, FRESCO, JEMS, NGDEEP, PEARLS and MegaScience JWST surveys. We thank all of those involved in these collaborations for their efforts in designing and carrying out these observations.

This work makes use of {\tt astropy} \citep{Astropy2013,Astropy2018,Astropy2022}, {\tt matplotlib} \citep{Hunter2007}, {\tt reproject}, {\tt DrizzlePac} \citep{Hoffmann2021}, {\tt SciPy} \citep{2020SciPy-NMeth}, {\tt msaexp} \url{https://zenodo.org/records/7579050} and {\tt photutils} \citep{larry_bradley_2022_6825092} and the {\tt galfind} photometric toolbox \url{https://github.com/duncanaustin98/galfind}. The authors thank Anthony Holloway and Sotirios Sanidas for providing their expertise in high performance computing and other IT support throughout this work. We also thank Elizabeth Stanway, Stephen Wilkins and Joseph Caruana for their helpful discussions. Some of the data products presented herein were retrieved from the Dawn JWST Archive (DJA). DJA is an initiative of the Cosmic Dawn Center (DAWN), which is funded by the Danish National Research Foundation under grant DNRF140.

\section*{Data Availability}

This work makes use of raw data from the JADES GTO survey programme ID's 1280, 1287 and the Jades Origins Field GO programme 3215. All raw data was obtained from the MAST portal operated by STScI, a DOI linking to the data is provided here: \href{https://archive.stsci.edu/doi/resolve/resolve.html?doi=10.17909/a2e3-bj58}{DOI 10.17909/a2e3-bj58}

Templates for the Wisp artefact which were generated in this work utilised JWST STAGE 2 calibrated products from programmes: 1180 \citep[JADES, PI: D. Eisenstein][]{Eisenstein2023JADES,Bunker2024,Rieke2023,Eisenstein2023DR2,Hainline2024}, 1208 \citep[CANUCS, PI: C. Willott;][]{Willott2024}, 1210 (JADES, PI: N. Luetzgendorf), 1324 \citep[GLASS, PI: T. Treu;][]{Treu2022,Mascia2024}, 1345 \citep[CEERS, PI: S. Finkelstein;][]{Bagley2023,Finkelstein2024}, 1895 \citep[FRESCO, PI: P. Oesch;][]{Oesch2023}, 1963 \citep[JEMS, PI: C. Williams;][]{Williams2023}, 2079 \citep[NGDEEP, PI: S. Finkelstein;][]{Bagley2024}, 2738 \citep[PEARLS, PI: R. Windhorst;][]{Windhorst2023}, 3215 \citep[JOF, PI: D. Eistenstein;][]{Eisenstein2024,Robertson2024}, 4111 \citep[MegaScience, PI: K. Suess;][]{Suess2024}. The final product is made available on the following cloud-based repository: \url{https://1drv.ms/f/s!AjXt-wkeMSXAgrlI7h5rQUe-Lx2rpA?e=IxpN6M}

If the reader is interested in any other products from this work (images, catalogues etc.) they are welcome to contact the lead author.



\bibliographystyle{mnras}
\bibliography{example,mnras_template} 

\begin{thebibliography}{}
\makeatletter
\relax
\def\mn@urlcharsother{\let\do\@makeother \do\$\do\&\do\#\do\^\do\_\do\%\do\~}
\def\mn@doi{\begingroup\mn@urlcharsother \@ifnextchar [ {\mn@doi@}
  {\mn@doi@[]}}
\def\mn@doi@[#1]#2{\def\@tempa{#1}\ifx\@tempa\@empty \href
  {http://dx.doi.org/#2} {doi:#2}\else \href {http://dx.doi.org/#2} {#1}\fi
  \endgroup}
\def\mn@eprint#1#2{\mn@eprint@#1:#2::\@nil}
\def\mn@eprint@arXiv#1{\href {http://arxiv.org/abs/#1} {{\tt arXiv:#1}}}
\def\mn@eprint@dblp#1{\href {http://dblp.uni-trier.de/rec/bibtex/#1.xml}
  {dblp:#1}}
\def\mn@eprint@#1:#2:#3:#4\@nil{\def\@tempa {#1}\def\@tempb {#2}\def\@tempc
  {#3}\ifx \@tempc \@empty \let \@tempc \@tempb \let \@tempb \@tempa \fi \ifx
  \@tempb \@empty \def\@tempb {arXiv}\fi \@ifundefined
  {mn@eprint@\@tempb}{\@tempb:\@tempc}{\expandafter \expandafter \csname
  mn@eprint@\@tempb\endcsname \expandafter{\@tempc}}}

\bibitem[\protect\citeauthoryear{{Adams} et~al.,}{{Adams}
  et~al.}{2024}]{Adams2024}
{Adams} N.~J.,  et~al., 2024, \mn@doi [\apj] {10.3847/1538-4357/ad2a7b}, \href
  {https://ui.adsabs.harvard.edu/abs/2024ApJ...965..169A} {965, 169}

\bibitem[\protect\citeauthoryear{{Arrabal Haro} et~al.,}{{Arrabal Haro}
  et~al.}{2023}]{ArrabalHaro2023}
{Arrabal Haro} P.,  et~al., 2023, \mn@doi [\nat] {10.1038/s41586-023-06521-7},
  \href {https://ui.adsabs.harvard.edu/abs/2023Natur.622..707A} {622, 707}

\bibitem[\protect\citeauthoryear{Asada et~al.,}{Asada et~al.}{2024}]{asada2024}
Asada Y.,  et~al., 2024, arXiv preprint arXiv:2410.21543

\bibitem[\protect\citeauthoryear{{Astropy Collaboration} et~al.,}{{Astropy
  Collaboration} et~al.}{2013}]{Astropy2013}
{Astropy Collaboration} et~al., 2013, \mn@doi [\aap]
  {10.1051/0004-6361/201322068}, \href
  {https://ui.adsabs.harvard.edu/abs/2013A&A...558A..33A} {558, A33}

\bibitem[\protect\citeauthoryear{{Astropy Collaboration} et~al.,}{{Astropy
  Collaboration} et~al.}{2018}]{Astropy2018}
{Astropy Collaboration} et~al., 2018, \mn@doi [\aj] {10.3847/1538-3881/aabc4f},
  \href {https://ui.adsabs.harvard.edu/abs/2018AJ....156..123A} {156, 123}

\bibitem[\protect\citeauthoryear{{Astropy Collaboration} et~al.,}{{Astropy
  Collaboration} et~al.}{2022}]{Astropy2022}
{Astropy Collaboration} et~al., 2022, \mn@doi [\apj]
  {10.3847/1538-4357/ac7c74}, \href
  {https://ui.adsabs.harvard.edu/abs/2022ApJ...935..167A} {935, 167}

\bibitem[\protect\citeauthoryear{{Austin} et~al.,}{{Austin}
  et~al.}{2024}]{Austin2024}
{Austin} D.,  et~al., 2024, \mn@doi [arXiv e-prints]
  {10.48550/arXiv.2404.10751}, \href
  {https://ui.adsabs.harvard.edu/abs/2024arXiv240410751A} {p. arXiv:2404.10751}

\bibitem[\protect\citeauthoryear{{Bagley} et~al.,}{{Bagley}
  et~al.}{2023}]{Bagley2023}
{Bagley} M.~B.,  et~al., 2023, \mn@doi [\apjl] {10.3847/2041-8213/acbb08},
  \href {https://ui.adsabs.harvard.edu/abs/2023ApJ...946L..12B} {946, L12}

\bibitem[\protect\citeauthoryear{{Bagley} et~al.,}{{Bagley}
  et~al.}{2024}]{Bagley2024}
{Bagley} M.~B.,  et~al., 2024, \mn@doi [\apjl] {10.3847/2041-8213/ad2f31},
  \href {https://ui.adsabs.harvard.edu/abs/2024ApJ...965L...6B} {965, L6}

\bibitem[\protect\citeauthoryear{{Balestra} et~al.,}{{Balestra}
  et~al.}{2010}]{Balestra2010}
{Balestra} I.,  et~al., 2010, \mn@doi [\aap] {10.1051/0004-6361/200913626},
  \href {https://ui.adsabs.harvard.edu/abs/2010A&A...512A..12B} {512, A12}

\bibitem[\protect\citeauthoryear{{Barrufet} et~al.,}{{Barrufet}
  et~al.}{2024}]{Barrufet2024}
{Barrufet} L.,  et~al., 2024, \mn@doi [arXiv e-prints]
  {10.48550/arXiv.2404.08052}, \href
  {https://ui.adsabs.harvard.edu/abs/2024arXiv240408052B} {p. arXiv:2404.08052}

\bibitem[\protect\citeauthoryear{{Behroozi} \& {Silk}}{{Behroozi} \&
  {Silk}}{2015}]{Behroozi2015}
{Behroozi} P.~S.,  {Silk} J.,  2015, \mn@doi [\apj]
  {10.1088/0004-637X/799/1/32}, \href
  {https://ui.adsabs.harvard.edu/abs/2015ApJ...799...32B} {799, 32}

\bibitem[\protect\citeauthoryear{{Behroozi}, {Wechsler}, {Hearin}  \&
  {Conroy}}{{Behroozi} et~al.}{2019}]{Behroozi2019}
{Behroozi} P.,  {Wechsler} R.~H.,  {Hearin} A.~P.,   {Conroy} C.,  2019,
  \mn@doi [\mnras] {10.1093/mnras/stz1182}, \href
  {https://ui.adsabs.harvard.edu/abs/2019MNRAS.488.3143B} {488, 3143}

\bibitem[\protect\citeauthoryear{{Bouwens} et~al.,}{{Bouwens}
  et~al.}{2021}]{Bouwens2021}
{Bouwens} R.~J.,  et~al., 2021, \mn@doi [\aj] {10.3847/1538-3881/abf83e}, \href
  {https://ui.adsabs.harvard.edu/abs/2021AJ....162...47B} {162, 47}

\bibitem[\protect\citeauthoryear{{Bouwens} et~al.,}{{Bouwens}
  et~al.}{2022}]{bouwens2022}
{Bouwens} R.~J.,  et~al., 2022, \mn@doi [\apj] {10.3847/1538-4357/ac5a4a},
  \href {https://ui.adsabs.harvard.edu/abs/2022ApJ...931..160B} {931, 160}

\bibitem[\protect\citeauthoryear{Bradley et~al.,}{Bradley
  et~al.}{2022}]{larry_bradley_2022_6825092}
Bradley L.,  et~al., 2022, astropy/photutils: 1.5.0,
  \mn@doi{10.5281/zenodo.6825092}, \url
  {https://doi.org/10.5281/zenodo.6825092}

\bibitem[\protect\citeauthoryear{{Brammer}, {van Dokkum}  \& {Coppi}}{{Brammer}
  et~al.}{2008}]{Brammer2008}
{Brammer} G.~B.,  {van Dokkum} P.~G.,   {Coppi} P.,  2008, \mn@doi [\apj]
  {10.1086/591786}, \href
  {https://ui.adsabs.harvard.edu/abs/2008ApJ...686.1503B} {686, 1503}

\bibitem[\protect\citeauthoryear{{Bunker} et~al.,}{{Bunker}
  et~al.}{2024}]{Bunker2024}
{Bunker} A.~J.,  et~al., 2024, \mn@doi [\aap] {10.1051/0004-6361/202347094},
  \href {https://ui.adsabs.harvard.edu/abs/2024A&A...690A.288B} {690, A288}

\bibitem[\protect\citeauthoryear{{Calzetti}, {Armus}, {Bohlin}, {Kinney},
  {Koornneef}  \& {Storchi-Bergmann}}{{Calzetti} et~al.}{2000}]{Calzetti2000}
{Calzetti} D.,  {Armus} L.,  {Bohlin} R.~C.,  {Kinney} A.~L.,  {Koornneef} J.,
   {Storchi-Bergmann} T.,  2000, \mn@doi [\apj] {10.1086/308692}, \href
  {https://ui.adsabs.harvard.edu/abs/2000ApJ...533..682C} {533, 682}

\bibitem[\protect\citeauthoryear{{Carniani} et~al.,}{{Carniani}
  et~al.}{2024}]{Carnani2024}
{Carniani} S.,  et~al., 2024, \mn@doi [\nat] {10.1038/s41586-024-07860-9},
  \href {https://ui.adsabs.harvard.edu/abs/2024Natur.633..318C} {633, 318}

\bibitem[\protect\citeauthoryear{{Carniani} et~al.,}{{Carniani}
  et~al.}{2025}]{Carnani2025}
{Carniani} S.,  et~al., 2025, \mn@doi [\aap] {10.1051/0004-6361/202452451},
  \href {https://ui.adsabs.harvard.edu/abs/2025A&A...696A..87C} {696, A87}

\bibitem[\protect\citeauthoryear{{Castellano} et~al.,}{{Castellano}
  et~al.}{2022}]{Castellano2022}
{Castellano} M.,  et~al., 2022, \mn@doi [\apjl] {10.3847/2041-8213/ac94d0},
  \href {https://ui.adsabs.harvard.edu/abs/2022ApJ...938L..15C} {938, L15}

\bibitem[\protect\citeauthoryear{{Castellano} et~al.,}{{Castellano}
  et~al.}{2025}]{Castellano2025}
{Castellano} M.,  et~al., 2025, \mn@doi [arXiv e-prints]
  {10.48550/arXiv.2504.05893}, \href
  {https://ui.adsabs.harvard.edu/abs/2025arXiv250405893C} {p. arXiv:2504.05893}

\bibitem[\protect\citeauthoryear{{Clausen}, {Steinhardt}, {Shao}  \& {Senthil
  Kumar}}{{Clausen} et~al.}{2024}]{Clausen2024}
{Clausen} T.,  {Steinhardt} C.~L.,  {Shao} A.,   {Senthil Kumar} G.,  2024,
  \mn@doi [arXiv e-prints] {10.48550/arXiv.2412.01893}, \href
  {https://ui.adsabs.harvard.edu/abs/2024arXiv241201893C} {p. arXiv:2412.01893}

\bibitem[\protect\citeauthoryear{{Coe} et~al.,}{{Coe} et~al.}{2012}]{Coe2012}
{Coe} D.,  et~al., 2012, \mn@doi [\apj] {10.1088/0004-637X/757/1/22}, \href
  {https://ui.adsabs.harvard.edu/abs/2012ApJ...757...22C} {757, 22}

\bibitem[\protect\citeauthoryear{{Conselice} et~al.,}{{Conselice}
  et~al.}{2024}]{Conseliece2024}
{Conselice} C.~J.,  et~al., 2024, \mn@doi [arXiv e-prints]
  {10.48550/arXiv.2407.14973}, \href
  {https://ui.adsabs.harvard.edu/abs/2024arXiv240714973C} {p. arXiv:2407.14973}

\bibitem[\protect\citeauthoryear{{Cooper} et~al.,}{{Cooper}
  et~al.}{2012}]{Cooper2012}
{Cooper} M.~C.,  et~al., 2012, \mn@doi [\mnras]
  {10.1111/j.1365-2966.2012.21524.x}, \href
  {https://ui.adsabs.harvard.edu/abs/2012MNRAS.425.2116C} {425, 2116}

\bibitem[\protect\citeauthoryear{{Curtis-Lake} et~al.,}{{Curtis-Lake}
  et~al.}{2023}]{CurtisLake2022}
{Curtis-Lake} E.,  et~al., 2023, \mn@doi [Nature Astronomy]
  {10.1038/s41550-023-01918-w}, \href
  {https://ui.adsabs.harvard.edu/abs/2023NatAs...7..622C} {7, 622}

\bibitem[\protect\citeauthoryear{{D'Eugenio} et~al.,}{{D'Eugenio}
  et~al.}{2024}]{DEugenio2024}
{D'Eugenio} F.,  et~al., 2024, \mn@doi [arXiv e-prints]
  {10.48550/arXiv.2404.06531}, \href
  {https://ui.adsabs.harvard.edu/abs/2024arXiv240406531D} {p. arXiv:2404.06531}

\bibitem[\protect\citeauthoryear{{D'Silva} et~al.,}{{D'Silva}
  et~al.}{2025}]{DSilva2025}
{D'Silva} J. C.~J.,  et~al., 2025, \mn@doi [arXiv e-prints]
  {10.48550/arXiv.2503.03431}, \href
  {https://ui.adsabs.harvard.edu/abs/2025arXiv250303431D} {p. arXiv:2503.03431}

\bibitem[\protect\citeauthoryear{{Dayal} et~al.,}{{Dayal}
  et~al.}{2022}]{dayal2022}
{Dayal} P.,  et~al., 2022, \mn@doi [\mnras] {10.1093/mnras/stac537}, \href
  {https://ui.adsabs.harvard.edu/abs/2022MNRAS.512..989D} {512, 989}

\bibitem[\protect\citeauthoryear{{Donnan} et~al.,}{{Donnan}
  et~al.}{2023}]{Donnan2022}
{Donnan} C.~T.,  et~al., 2023, \mn@doi [\mnras] {10.1093/mnras/stac3472}, \href
  {https://ui.adsabs.harvard.edu/abs/2023MNRAS.518.6011D} {518, 6011}

\bibitem[\protect\citeauthoryear{{Donnan} et~al.,}{{Donnan}
  et~al.}{2024}]{Donnan2024}
{Donnan} C.~T.,  et~al., 2024, \mn@doi [\mnras] {10.1093/mnras/stae2037}, \href
  {https://ui.adsabs.harvard.edu/abs/2024MNRAS.533.3222D} {533, 3222}

\bibitem[\protect\citeauthoryear{{Donnan}, {Dunlop}, {McLure}, {McLeod}  \&
  {Cullen}}{{Donnan} et~al.}{2025}]{Donnan2025}
{Donnan} C.~T.,  {Dunlop} J.~S.,  {McLure} R.~J.,  {McLeod} D.~J.,   {Cullen}
  F.,  2025, arXiv e-prints, \href
  {https://ui.adsabs.harvard.edu/abs/2025arXiv250103217D} {p. arXiv:2501.03217}

\bibitem[\protect\citeauthoryear{{Eisenstein} et~al.,}{{Eisenstein}
  et~al.}{2023a}]{Eisenstein2023JADES}
{Eisenstein} D.~J.,  et~al., 2023a, \mn@doi [arXiv e-prints]
  {10.48550/arXiv.2306.02465}, \href
  {https://ui.adsabs.harvard.edu/abs/2023arXiv230602465E} {p. arXiv:2306.02465}

\bibitem[\protect\citeauthoryear{{Eisenstein} et~al.,}{{Eisenstein}
  et~al.}{2023b}]{Eisenstein2024}
{Eisenstein} D.~J.,  et~al., 2023b, \mn@doi [arXiv e-prints]
  {10.48550/arXiv.2310.12340}, \href
  {https://ui.adsabs.harvard.edu/abs/2023arXiv231012340E} {p. arXiv:2310.12340}

\bibitem[\protect\citeauthoryear{{Eisenstein} et~al.,}{{Eisenstein}
  et~al.}{2023c}]{Eisenstein2023DR2}
{Eisenstein} D.~J.,  et~al., 2023c, \mn@doi [arXiv e-prints]
  {10.48550/arXiv.2310.12340}, \href
  {https://ui.adsabs.harvard.edu/abs/2023arXiv231012340E} {p. arXiv:2310.12340}

\bibitem[\protect\citeauthoryear{{Ferrara}, {Pallottini}  \& {Dayal}}{{Ferrara}
  et~al.}{2023}]{Ferrara2022}
{Ferrara} A.,  {Pallottini} A.,   {Dayal} P.,  2023, \mn@doi [\mnras]
  {10.1093/mnras/stad1095}, \href
  {https://ui.adsabs.harvard.edu/abs/2023MNRAS.522.3986F} {522, 3986}

\bibitem[\protect\citeauthoryear{{Finkelstein} \& {Bagley}}{{Finkelstein} \&
  {Bagley}}{2022}]{Finkelstein2022}
{Finkelstein} S.~L.,  {Bagley} M.~B.,  2022, arXiv e-prints, \href
  {https://ui.adsabs.harvard.edu/abs/2022arXiv220702233F} {p. arXiv:2207.02233}

\bibitem[\protect\citeauthoryear{{Finkelstein} et~al.,}{{Finkelstein}
  et~al.}{2024}]{Finkelstein2024}
{Finkelstein} S.~L.,  et~al., 2024, \mn@doi [\apjl] {10.3847/2041-8213/ad4495},
  \href {https://ui.adsabs.harvard.edu/abs/2024ApJ...969L...2F} {969, L2}

\bibitem[\protect\citeauthoryear{{Gaia Collaboration} et~al.,}{{Gaia
  Collaboration} et~al.}{2023}]{GAIADR3}
{Gaia Collaboration} et~al., 2023, \mn@doi [\aap]
  {10.1051/0004-6361/202243940}, \href
  {https://ui.adsabs.harvard.edu/abs/2023A&A...674A...1G} {674, A1}

\bibitem[\protect\citeauthoryear{{Gandolfi} et~al.,}{{Gandolfi}
  et~al.}{2025}]{Gandolfi2025}
{Gandolfi} G.,  et~al., 2025, \mn@doi [arXiv e-prints]
  {10.48550/arXiv.2502.02637}, \href
  {https://ui.adsabs.harvard.edu/abs/2025arXiv250202637G} {p. arXiv:2502.02637}

\bibitem[\protect\citeauthoryear{{Garilli} et~al.,}{{Garilli}
  et~al.}{2021}]{Garilli2021}
{Garilli} B.,  et~al., 2021, \mn@doi [\aap] {10.1051/0004-6361/202040059},
  \href {https://ui.adsabs.harvard.edu/abs/2021A&A...647A.150G} {647, A150}

\bibitem[\protect\citeauthoryear{{Giardino} et~al.,}{{Giardino}
  et~al.}{2022}]{Giardino2022}
{Giardino} G.,  et~al., 2022, in {Coyle} L.~E.,  {Matsuura} S.,   {Perrin}
  M.~D.,  eds,  Society of Photo-Optical Instrumentation Engineers (SPIE)
  Conference Series Vol. 12180, Space Telescopes and Instrumentation 2022:
  Optical, Infrared, and Millimeter Wave. p. 121800X (\mn@eprint {arXiv}
  {2208.04876}), \mn@doi{10.1117/12.2628980}

\bibitem[\protect\citeauthoryear{{Hainline} et~al.,}{{Hainline}
  et~al.}{2020}]{Hainline2020}
{Hainline} K.~N.,  et~al., 2020, \mn@doi [\apj] {10.3847/1538-4357/ab7dc3},
  \href {https://ui.adsabs.harvard.edu/abs/2020ApJ...892..125H} {892, 125}

\bibitem[\protect\citeauthoryear{{Hainline} et~al.,}{{Hainline}
  et~al.}{2024}]{Hainline2024}
{Hainline} K.~N.,  et~al., 2024, \mn@doi [\apj] {10.3847/1538-4357/ad1ee4},
  \href {https://ui.adsabs.harvard.edu/abs/2024ApJ...964...71H} {964, 71}

\bibitem[\protect\citeauthoryear{{Harikane} et~al.,}{{Harikane}
  et~al.}{2022}]{Harikane2022}
{Harikane} Y.,  et~al., 2022, \mn@doi [\apj] {10.3847/1538-4357/ac53a9}, \href
  {https://ui.adsabs.harvard.edu/abs/2022ApJ...929....1H} {929, 1}

\bibitem[\protect\citeauthoryear{{Harikane} et~al.,}{{Harikane}
  et~al.}{2023}]{Harikane2023}
{Harikane} Y.,  et~al., 2023, \mn@doi [\apjs] {10.3847/1538-4365/acaaa9}, \href
  {https://ui.adsabs.harvard.edu/abs/2023ApJS..265....5H} {265, 5}

\bibitem[\protect\citeauthoryear{{Harvey} et~al.,}{{Harvey}
  et~al.}{2025}]{Harvey2024}
{Harvey} T.,  et~al., 2025, \mn@doi [\apj] {10.3847/1538-4357/ad8c29}, \href
  {https://ui.adsabs.harvard.edu/abs/2025ApJ...978...89H} {978, 89}

\bibitem[\protect\citeauthoryear{{Heintz} et~al.,}{{Heintz}
  et~al.}{2024}]{Heintz2024}
{Heintz} K.~E.,  et~al., 2024, \mn@doi [arXiv e-prints]
  {10.48550/arXiv.2404.02211}, \href
  {https://ui.adsabs.harvard.edu/abs/2024arXiv240402211H} {p. arXiv:2404.02211}

\bibitem[\protect\citeauthoryear{{Hoffmann}, {Mack}, {Avila}, {Martlin},
  {Cohen}  \& {Bajaj}}{{Hoffmann} et~al.}{2021}]{Hoffmann2021}
{Hoffmann} S.~L.,  {Mack} J.,  {Avila} R.,  {Martlin} C.,  {Cohen} Y.,
  {Bajaj} V.,  2021, in American Astronomical Society Meeting Abstracts. p.
  216.02

\bibitem[\protect\citeauthoryear{{Holwerda} et~al.,}{{Holwerda}
  et~al.}{2024}]{holwerda2024}
{Holwerda} B.~W.,  et~al., 2024, \mn@doi [\mnras] {10.1093/mnras/stae316},
  \href {https://ui.adsabs.harvard.edu/abs/2024MNRAS.529.1067H} {529, 1067}

\bibitem[\protect\citeauthoryear{Hunter}{Hunter}{2007}]{Hunter2007}
Hunter J.~D.,  2007, \mn@doi [Computing in Science \& Engineering]
  {10.1109/MCSE.2007.55}, 9, 90

\bibitem[\protect\citeauthoryear{{Inoue}, {Shimizu}, {Iwata}  \&
  {Tanaka}}{{Inoue} et~al.}{2014}]{Inoue2014}
{Inoue} A.~K.,  {Shimizu} I.,  {Iwata} I.,   {Tanaka} M.,  2014, \mn@doi
  [\mnras] {10.1093/mnras/stu936}, \href
  {https://ui.adsabs.harvard.edu/abs/2014MNRAS.442.1805I} {442, 1805}

\bibitem[\protect\citeauthoryear{{Kannan}, {Garaldi}, {Smith}, {Pakmor},
  {Springel}, {Vogelsberger}  \& {Hernquist}}{{Kannan}
  et~al.}{2022}]{Kannan2022}
{Kannan} R.,  {Garaldi} E.,  {Smith} A.,  {Pakmor} R.,  {Springel} V.,
  {Vogelsberger} M.,   {Hernquist} L.,  2022, \mn@doi [\mnras]
  {10.1093/mnras/stab3710}, \href
  {https://ui.adsabs.harvard.edu/abs/2022MNRAS.511.4005K} {511, 4005}

\bibitem[\protect\citeauthoryear{{Kauffmann} et~al.,}{{Kauffmann}
  et~al.}{2020}]{Kauffmann2020}
{Kauffmann} O.~B.,  et~al., 2020, \mn@doi [\aap] {10.1051/0004-6361/202037450},
  \href {https://ui.adsabs.harvard.edu/abs/2020A&A...640A..67K} {640, A67}

\bibitem[\protect\citeauthoryear{{Kokorev} et~al.,}{{Kokorev}
  et~al.}{2024}]{Kokorev2024}
{Kokorev} V.,  et~al., 2024, arXiv e-prints, \href
  {https://ui.adsabs.harvard.edu/abs/2024arXiv240109981K} {p. arXiv:2401.09981}

\bibitem[\protect\citeauthoryear{Labb{\'e} et~al.,}{Labb{\'e}
  et~al.}{2023}]{labbe2023}
Labb{\'e} I.,  et~al., 2023, \mn@doi [Nature] {10.1038/s41586-023-05786-2},
  616, 266

\bibitem[\protect\citeauthoryear{{Larson} et~al.,}{{Larson}
  et~al.}{2023}]{Larson2023}
{Larson} R.~L.,  et~al., 2023, \mn@doi [\apj] {10.3847/1538-4357/acfed4}, \href
  {https://ui.adsabs.harvard.edu/abs/2023ApJ...958..141L} {958, 141}

\bibitem[\protect\citeauthoryear{{Le F{\`e}vre} et~al.,}{{Le F{\`e}vre}
  et~al.}{2015}]{LeFevre2015}
{Le F{\`e}vre} O.,  et~al., 2015, \mn@doi [\aap] {10.1051/0004-6361/201423829},
  \href {https://ui.adsabs.harvard.edu/abs/2015A&A...576A..79L} {576, A79}

\bibitem[\protect\citeauthoryear{LeF\`{e}vre et~al.,}{LeF\`{e}vre
  et~al.}{2013}]{LeFevre2013}
LeF\`{e}vre O.,  et~al., 2013, \mn@doi [\aap] {10.1051/0004-6361/201322179},
  559, A14

\bibitem[\protect\citeauthoryear{{Leung} et~al.,}{{Leung}
  et~al.}{2023}]{Leung2023}
{Leung} G. C.~K.,  et~al., 2023, \mn@doi [\apjl] {10.3847/2041-8213/acf365},
  \href {https://ui.adsabs.harvard.edu/abs/2023ApJ...954L..46L} {954, L46}

\bibitem[\protect\citeauthoryear{Levenberg}{Levenberg}{1944}]{Levenberg1944}
Levenberg K.,  1944, Quarterly of Applied Mathematics, 2, 164

\bibitem[\protect\citeauthoryear{{Madau}}{{Madau}}{1995}]{Madau1995}
{Madau} P.,  1995, \mn@doi [\apj] {10.1086/175332}, \href
  {https://ui.adsabs.harvard.edu/abs/1995ApJ...441...18M} {441, 18}

\bibitem[\protect\citeauthoryear{{Madau} \& {Dickinson}}{{Madau} \&
  {Dickinson}}{2014}]{Madau2014}
{Madau} P.,  {Dickinson} M.,  2014, \mn@doi [\araa]
  {10.1146/annurev-astro-081811-125615}, \href
  {https://ui.adsabs.harvard.edu/abs/2014ARA&A..52..415M} {52, 415}

\bibitem[\protect\citeauthoryear{{Marley} et~al.,}{{Marley}
  et~al.}{2021}]{Marley2021}
{Marley} M.~S.,  et~al., 2021, \mn@doi [\apj] {10.3847/1538-4357/ac141d}, \href
  {https://ui.adsabs.harvard.edu/abs/2021ApJ...920...85M} {920, 85}

\bibitem[\protect\citeauthoryear{Marquardt}{Marquardt}{1963}]{Marquardt1963}
Marquardt D.~W.,  1963, Journal of the Society for Industrial and Applied
  Mathematics, 11, 431

\bibitem[\protect\citeauthoryear{{Marshall} et~al.,}{{Marshall}
  et~al.}{2022}]{Marshall2022}
{Marshall} M.~A.,  et~al., 2022, \mn@doi [\mnras] {10.1093/mnras/stac2111},
  \href {https://ui.adsabs.harvard.edu/abs/2022MNRAS.516.1047M} {516, 1047}

\bibitem[\protect\citeauthoryear{{Mascia} et~al.,}{{Mascia}
  et~al.}{2024}]{Mascia2024}
{Mascia} S.,  et~al., 2024, \mn@doi [\aap] {10.1051/0004-6361/202450493}, \href
  {https://ui.adsabs.harvard.edu/abs/2024A&A...690A...2M} {690, A2}

\bibitem[\protect\citeauthoryear{Matthee et~al.,}{Matthee
  et~al.}{2023}]{matthee2023little}
Matthee J.,  et~al., 2023, arXiv preprint arXiv:2306.05448

\bibitem[\protect\citeauthoryear{{McLeod} et~al.,}{{McLeod}
  et~al.}{2024}]{McLeod2024}
{McLeod} D.~J.,  et~al., 2024, \mn@doi [\mnras] {10.1093/mnras/stad3471}, \href
  {https://ui.adsabs.harvard.edu/abs/2024MNRAS.527.5004M} {527, 5004}

\bibitem[\protect\citeauthoryear{{Merlin} et~al.,}{{Merlin}
  et~al.}{2021}]{Merlin2021}
{Merlin} E.,  et~al., 2021, \mn@doi [\aap] {10.1051/0004-6361/202140310}, \href
  {https://ui.adsabs.harvard.edu/abs/2021A&A...649A..22M} {649, A22}

\bibitem[\protect\citeauthoryear{{Meyer}, {Barrufet}, {Boogaard}, {Naidu},
  {Oesch}  \& {Walter}}{{Meyer} et~al.}{2024}]{Meyer2024}
{Meyer} R.~A.,  {Barrufet} L.,  {Boogaard} L.~A.,  {Naidu} R.~P.,  {Oesch}
  P.~A.,   {Walter} F.,  2024, \mn@doi [\aap] {10.1051/0004-6361/202348306},
  \href {https://ui.adsabs.harvard.edu/abs/2024A&A...681L...3M} {681, L3}

\bibitem[\protect\citeauthoryear{{Momcheva} et~al.,}{{Momcheva}
  et~al.}{2016}]{Momcheva2015}
{Momcheva} I.~G.,  et~al., 2016, \mn@doi [\apjs] {10.3847/0067-0049/225/2/27},
  \href {https://ui.adsabs.harvard.edu/abs/2016ApJS..225...27M} {225, 27}

\bibitem[\protect\citeauthoryear{{Morris} et~al.,}{{Morris}
  et~al.}{2015}]{Morris2015}
{Morris} A.~M.,  et~al., 2015, \mn@doi [\aj] {10.1088/0004-6256/149/6/178},
  \href {https://ui.adsabs.harvard.edu/abs/2015AJ....149..178M} {149, 178}

\bibitem[\protect\citeauthoryear{{Oesch} et~al.,}{{Oesch}
  et~al.}{2013}]{Oesch2013}
{Oesch} P.~A.,  et~al., 2013, \mn@doi [\apj] {10.1088/0004-637X/773/1/75},
  \href {https://ui.adsabs.harvard.edu/abs/2013ApJ...773...75O} {773, 75}

\bibitem[\protect\citeauthoryear{{Oesch}, {Bouwens}, {Illingworth}, {Labb{\'e}}
   \& {Stefanon}}{{Oesch} et~al.}{2018}]{Oesch2018}
{Oesch} P.~A.,  {Bouwens} R.~J.,  {Illingworth} G.~D.,  {Labb{\'e}} I.,
  {Stefanon} M.,  2018, \mn@doi [\apj] {10.3847/1538-4357/aab03f}, \href
  {https://ui.adsabs.harvard.edu/abs/2018ApJ...855..105O} {855, 105}

\bibitem[\protect\citeauthoryear{{Oesch} et~al.,}{{Oesch}
  et~al.}{2023}]{Oesch2023}
{Oesch} P.~A.,  et~al., 2023, \mn@doi [\mnras] {10.1093/mnras/stad2411}, \href
  {https://ui.adsabs.harvard.edu/abs/2023MNRAS.525.2864O} {525, 2864}

\bibitem[\protect\citeauthoryear{{P{\'e}rez-Gonz{\'a}lez}
  et~al.,}{{P{\'e}rez-Gonz{\'a}lez} et~al.}{2023}]{PerezGonzalez2023}
{P{\'e}rez-Gonz{\'a}lez} P.~G.,  et~al., 2023, \mn@doi [\apjl]
  {10.3847/2041-8213/acd9d0}, \href
  {https://ui.adsabs.harvard.edu/abs/2023ApJ...951L...1P} {951, L1}

\bibitem[\protect\citeauthoryear{{Perrin}, {Soummer}, {Elliott}, {Lallo}  \&
  {Sivaramakrishnan}}{{Perrin} et~al.}{2012}]{Perrin2012}
{Perrin} M.~D.,  {Soummer} R.,  {Elliott} E.~M.,  {Lallo} M.~D.,
  {Sivaramakrishnan} A.,  2012, in {Clampin} M.~C.,  {Fazio} G.~G.,  {MacEwen}
  H.~A.,   {Oschmann} Jacobus~M. J.,  eds,  Society of Photo-Optical
  Instrumentation Engineers (SPIE) Conference Series Vol. 8442, Space
  Telescopes and Instrumentation 2012: Optical, Infrared, and Millimeter Wave.
  p. 84423D, \mn@doi{10.1117/12.925230}

\bibitem[\protect\citeauthoryear{{Perrin}, {Sivaramakrishnan}, {Lajoie},
  {Elliott}, {Pueyo}, {Ravindranath}  \& {Albert}}{{Perrin}
  et~al.}{2014}]{Perrin2014}
{Perrin} M.~D.,  {Sivaramakrishnan} A.,  {Lajoie} C.-P.,  {Elliott} E.,
  {Pueyo} L.,  {Ravindranath} S.,   {Albert} L.,  2014, in {Oschmann}
  Jacobus~M. J.,  {Clampin} M.,  {Fazio} G.~G.,   {MacEwen} H.~A.,  eds,
  Society of Photo-Optical Instrumentation Engineers (SPIE) Conference Series
  Vol. 9143, Space Telescopes and Instrumentation 2014: Optical, Infrared, and
  Millimeter Wave. p. 91433X, \mn@doi{10.1117/12.2056689}

\bibitem[\protect\citeauthoryear{{Price} et~al.,}{{Price}
  et~al.}{2024}]{Price2024}
{Price} S.~H.,  et~al., 2024, \mn@doi [arXiv e-prints]
  {10.48550/arXiv.2408.03920}, \href
  {https://ui.adsabs.harvard.edu/abs/2024arXiv240803920P} {p. arXiv:2408.03920}

\bibitem[\protect\citeauthoryear{{Ravikumar} et~al.,}{{Ravikumar}
  et~al.}{2007}]{Ravikumar2007}
{Ravikumar} C.~D.,  et~al., 2007, \mn@doi [\aap] {10.1051/0004-6361:20065358},
  \href {https://ui.adsabs.harvard.edu/abs/2007A&A...465.1099R} {465, 1099}

\bibitem[\protect\citeauthoryear{{Rieke} et~al.,}{{Rieke}
  et~al.}{2023}]{Rieke2023}
{Rieke} M.~J.,  et~al., 2023, \mn@doi [\apjs] {10.3847/1538-4365/acf44d}, \href
  {https://ui.adsabs.harvard.edu/abs/2023ApJS..269...16R} {269, 16}

\bibitem[\protect\citeauthoryear{{Robertson} et~al.,}{{Robertson}
  et~al.}{2024}]{Robertson2024}
{Robertson} B.,  et~al., 2024, \mn@doi [\apj] {10.3847/1538-4357/ad463d}, \href
  {https://ui.adsabs.harvard.edu/abs/2024ApJ...970...31R} {970, 31}

\bibitem[\protect\citeauthoryear{{Robotham}, {D'Silva}, {Windhorst}, {Jansen},
  {Summers}, {Driver}, {Wilmer}  \& {Bellstedt}}{{Robotham}
  et~al.}{2023}]{Robotham2023}
{Robotham} A.~S.~G.,  {D'Silva} J.~C.~J.,  {Windhorst} R.~A.,  {Jansen} R.~A.,
  {Summers} J.,  {Driver} S.~P.,  {Wilmer} C.~N.~A.,   {Bellstedt} S.,  2023,
  \mn@doi [\pasp] {10.1088/1538-3873/acea42}, \href
  {https://ui.adsabs.harvard.edu/abs/2023PASP..135h5003R} {135, 085003}

\bibitem[\protect\citeauthoryear{{Schechter}}{{Schechter}}{1976}]{Schechter1976}
{Schechter} P.,  1976, \mn@doi [\apj] {10.1086/154079}, \href
  {https://ui.adsabs.harvard.edu/abs/1976ApJ...203..297S} {203, 297}

\bibitem[\protect\citeauthoryear{{Schouws} et~al.,}{{Schouws}
  et~al.}{2025}]{Schows2025}
{Schouws} S.,  et~al., 2025, \mn@doi [arXiv e-prints]
  {10.48550/arXiv.2502.01610}, \href
  {https://ui.adsabs.harvard.edu/abs/2025arXiv250201610S} {p. arXiv:2502.01610}

\bibitem[\protect\citeauthoryear{{Suess} et~al.,}{{Suess}
  et~al.}{2024}]{Suess2024}
{Suess} K.~A.,  et~al., 2024, \mn@doi [\apj] {10.3847/1538-4357/ad75fe}, \href
  {https://ui.adsabs.harvard.edu/abs/2024ApJ...976..101S} {976, 101}

\bibitem[\protect\citeauthoryear{{Szokoly} et~al.,}{{Szokoly}
  et~al.}{2004}]{Szokoly2004}
{Szokoly} G.~P.,  et~al., 2004, \mn@doi [\apjs] {10.1086/424707}, \href
  {https://ui.adsabs.harvard.edu/abs/2004ApJS..155..271S} {155, 271}

\bibitem[\protect\citeauthoryear{{Trapp} \& {Furlanetto}}{{Trapp} \&
  {Furlanetto}}{2020}]{Trapp2020}
{Trapp} A.~C.,  {Furlanetto} S.~R.,  2020, \mn@doi [\mnras]
  {10.1093/mnras/staa2828}, \href
  {https://ui.adsabs.harvard.edu/abs/2020MNRAS.499.2401T} {499, 2401}

\bibitem[\protect\citeauthoryear{{Treu} et~al.,}{{Treu}
  et~al.}{2022}]{Treu2022}
{Treu} T.,  et~al., 2022, \mn@doi [\apj] {10.3847/1538-4357/ac8158}, \href
  {https://ui.adsabs.harvard.edu/abs/2022ApJ...935..110T} {935, 110}

\bibitem[\protect\citeauthoryear{Ulm}{Ulm}{1990}]{Ulm1990}
Ulm K.,  1990, \mn@doi [American Journal of Epidemiology]
  {10.1093/oxfordjournals.aje.a115507}, 131, 373

\bibitem[\protect\citeauthoryear{{Vanzella} et~al.,}{{Vanzella}
  et~al.}{2008}]{Vanzella2008}
{Vanzella} E.,  et~al., 2008, \mn@doi [\aap] {10.1051/0004-6361:20078332},
  \href {https://ui.adsabs.harvard.edu/abs/2008A&A...478...83V} {478, 83}

\bibitem[\protect\citeauthoryear{Virtanen et~al.,}{Virtanen
  et~al.}{2020}]{2020SciPy-NMeth}
Virtanen P.,  et~al., 2020, \mn@doi [Nature Methods]
  {10.1038/s41592-019-0686-2}, \href {https://rdcu.be/b08Wh} {17, 261}

\bibitem[\protect\citeauthoryear{{Wang} et~al.,}{{Wang}
  et~al.}{2023}]{Wang2023}
{Wang} B.,  et~al., 2023, \mn@doi [\apjl] {10.3847/2041-8213/acfe07}, \href
  {https://ui.adsabs.harvard.edu/abs/2023ApJ...957L..34W} {957, L34}

\bibitem[\protect\citeauthoryear{{Weaver} et~al.,}{{Weaver}
  et~al.}{2024}]{Weaver2024}
{Weaver} J.~R.,  et~al., 2024, \mn@doi [\apjs] {10.3847/1538-4365/ad07e0},
  \href {https://ui.adsabs.harvard.edu/abs/2024ApJS..270....7W} {270, 7}

\bibitem[\protect\citeauthoryear{{Whitaker} et~al.,}{{Whitaker}
  et~al.}{2019}]{Whitaker2019}
{Whitaker} K.~E.,  et~al., 2019, \mn@doi [\apjs] {10.3847/1538-4365/ab3853},
  \href {https://ui.adsabs.harvard.edu/abs/2019ApJS..244...16W} {244, 16}

\bibitem[\protect\citeauthoryear{{Whitler} et~al.,}{{Whitler}
  et~al.}{2025}]{Whitler2025}
{Whitler} L.,  et~al., 2025, \mn@doi [arXiv e-prints]
  {10.48550/arXiv.2501.00984}, \href
  {https://ui.adsabs.harvard.edu/abs/2025arXiv250100984W} {p. arXiv:2501.00984}

\bibitem[\protect\citeauthoryear{{Wilkins} et~al.,}{{Wilkins}
  et~al.}{2023}]{Wilkins2023}
{Wilkins} S.~M.,  et~al., 2023, \mn@doi [arXiv e-prints]
  {10.48550/arXiv.2311.08065}, \href
  {https://ui.adsabs.harvard.edu/abs/2023arXiv231108065W} {p. arXiv:2311.08065}

\bibitem[\protect\citeauthoryear{{Williams} et~al.,}{{Williams}
  et~al.}{2018}]{Williams2018}
{Williams} C.~C.,  et~al., 2018, \mn@doi [\apjs] {10.3847/1538-4365/aabcbb},
  \href {https://ui.adsabs.harvard.edu/abs/2018ApJS..236...33W} {236, 33}

\bibitem[\protect\citeauthoryear{{Williams} et~al.,}{{Williams}
  et~al.}{2023}]{Williams2023}
{Williams} C.~C.,  et~al., 2023, \mn@doi [\apjs] {10.3847/1538-4365/acf130},
  \href {https://ui.adsabs.harvard.edu/abs/2023ApJS..268...64W} {268, 64}

\bibitem[\protect\citeauthoryear{{Willott} et~al.,}{{Willott}
  et~al.}{2024}]{Willott2024}
{Willott} C.~J.,  et~al., 2024, \mn@doi [\apj] {10.3847/1538-4357/ad35bc},
  \href {https://ui.adsabs.harvard.edu/abs/2024ApJ...966...74W} {966, 74}

\bibitem[\protect\citeauthoryear{{Windhorst} et~al.,}{{Windhorst}
  et~al.}{2023}]{Windhorst2023}
{Windhorst} R.~A.,  et~al., 2023, \mn@doi [\aj] {10.3847/1538-3881/aca163},
  \href {https://ui.adsabs.harvard.edu/abs/2023AJ....165...13W} {165, 13}

\bibitem[\protect\citeauthoryear{{Yung}, {Somerville}, {Finkelstein}, {Wilkins}
   \& {Gardner}}{{Yung} et~al.}{2024}]{yung2024}
{Yung} L.~Y.~A.,  {Somerville} R.~S.,  {Finkelstein} S.~L.,  {Wilkins} S.~M.,
  {Gardner} J.~P.,  2024, \mn@doi [\mnras] {10.1093/mnras/stad3484}, \href
  {https://ui.adsabs.harvard.edu/abs/2024MNRAS.527.5929Y} {527, 5929}

\bibitem[\protect\citeauthoryear{{Zavala} et~al.,}{{Zavala}
  et~al.}{2023}]{Zavala2023}
{Zavala} J.~A.,  et~al., 2023, \mn@doi [\apjl] {10.3847/2041-8213/acacfe},
  \href {https://ui.adsabs.harvard.edu/abs/2023ApJ...943L...9Z} {943, L9}

\bibitem[\protect\citeauthoryear{de Graaff et~al.,}{de~Graaff
  et~al.}{2024}]{degraaff2024}
de Graaff A.,  et~al., 2024, arXiv preprint arXiv:2409.05948

\makeatother
\end{thebibliography}




\appendix

\section{The Impact of Template Choice}

In this appendix, we examine the use of an alternate template on the results obtained in our work. We repeat the above analysis with the \citet{Larson2023} templates swapped out for a template set described in the work of \citet{Hainline2024} (in this section, we refer to this as the `JADES' template set and the previous templates as the `Larson' template set), noting that this new set does not provide ultraviolet slopes or rest-optical emission line strengths that are as extreme.

We find that when all 14 JOF bands are used in our JAGUAR catalogues, that the JADES template set selects 5\% more sources than the Larson template set. Overall, 87\% of sources are selected with the use of either template set. For sources which are not cross matched, half of them are $<7\sigma$ detected, a quarter are $7<\sigma<10$ and a quarter are brighter. The bright sources are all sources lying near the $z\sim7.5$ selection border and subsequently scatter above/below depending on the template set. Similarly, as photometric bands are removed, both template sets select fewer overall objects but the JADES template set loses sources at a faster rate, going from 5\% more sources with 14 photometric bands to 12\% fewer sources when only 7 bands are used. Sample overlap between template sets remains relatively consistent at $\sim88\%$.

\begin{figure*}%
    \centering
    \subfloat[\centering The UV LF as measured in 0.32as apertures comparing the use of the Larson and JADES template sets.]{{\includegraphics[width=8cm]{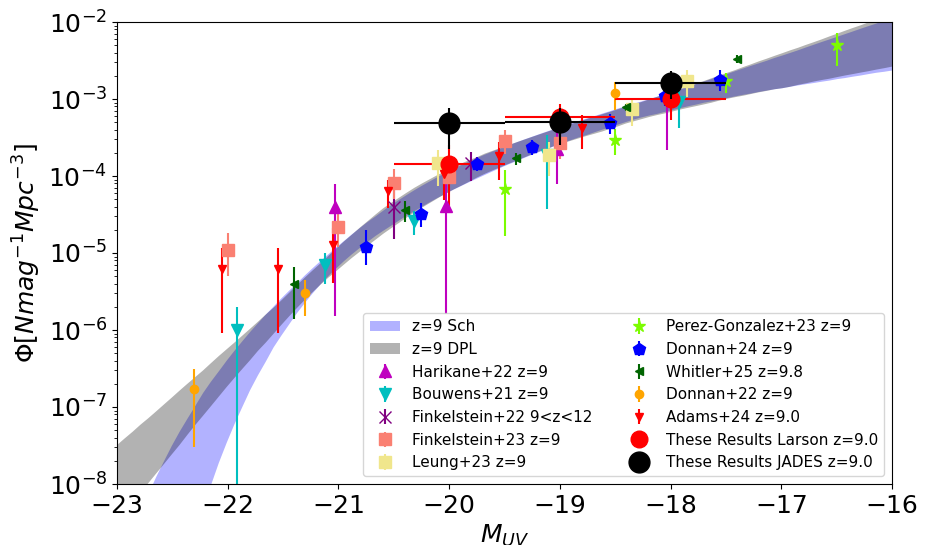} }}%
    \qquad
    \subfloat[\centering Completeness plot comparing JADES (faded lines) to the original Larson completenesses at $8<z<10$]{{\includegraphics[width=6.5cm]{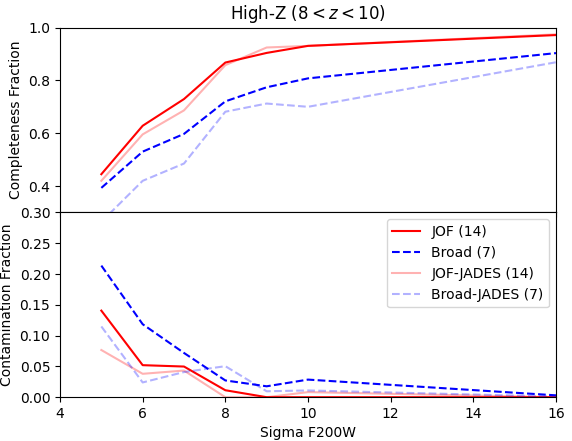} }}%
    \caption{The difference between the measured $z=9$ UV LF and its completeness/contamination curves when the SED template set is changed from \citet{Larson2023} to those used in \citet{Hainline2024}. Increased numbers of bright $z=8.5$ sources are driven by a systematic photo-z difference of 0.15 between the template sets.}%
    \label{fig:jadesLF}%
\end{figure*}

Following through to the completeness and contamination calculations. We find that the JADES template set has notably lower completeness than the Larson templates when the number of bands available is low (over 10 percentage points lower with only broad bands available). However, the contamination rates are also significantly lower (by factors of $\sim2$) in the faint regime of $<8\sigma$. For brighter galaxies at $z>>10\sigma$ both template sets converge to the same completeness levels. As more filters are added, the relative difference in the performance of the template sets decreases. The completeness and contamination in the $8<z<10$ bin is shown in Figure \ref{fig:jadesLF}.

We examine what the causes of these differences are in the two simulation-based samples:

\begin{itemize}
    \item \textbf{Selected by JADES and not by Larson:} 80\% of sources are better fit at redshifts $6<z<7.5$ with the Larson SEDs. This is caused by a systematic photo-z offset between the two template sets, with JADES-based photo-z's being higher by an average of 0.15 between $7.5<z<10$. This is driven by stronger Ly$\alpha$ in the JADES template set. Only 4\% of sources selected by JADES are fit by low-z Balmer breaks with Larson. The final 16\% of sources maintain high-z solutions but fail $\delta\chi^2$ constraints.
    \item \textbf{Selected by Larson and not by JADES:} 56\% of sources are better fit at redshifts $6<z<7.5$ with the JADES SEDs, 10\% are better fit by JADES templates as Balmer breaks and 34\% maintain high-z solutions but fail the $\delta\chi^2$ constraints.
\end{itemize}

When applying the JADES templates to the JOF catalogue, we find the catalogue using all available bands selects 66 sources (up from 62 with Larson templates) while with broad-width bands only we select 61 sources (down from 67 with Larson templates). 80\% of sources selected by the Larson templates appear in the JADES template-based selection. This reflects the general trend observed with the simulated catalogue above. 

Comparing the JOF selections that use all photometric bands:

\begin{itemize}
    \item \textbf{Selected by JADES and not by Larson:} There are 17 sources in this criteria, 7 are better fit by Larson templates as Balmer breaks, 9 are $7<z<9.5$ and one is a $z\sim16$ source. These either fail the $z>7.5$ cut or the $\delta\chi^2$ cut to the next best low-z solution with the Larson templates.
    \item \textbf{Selected by Larson and not by JADES:} There are 13 sources in this criteria, 10 are at $7<z<9.5$, and either scatter below the $z>7.5$ cut or just fail the $\delta\chi^2$ cuts for low-z solutions. Only three sources switch to a better fit Balmer break with the JADES templates (ID 1708, 5798, 6548 have $2.5<z<3.5$).
\end{itemize}

Conducting a by-eye examination, the Larson templates produce more frequent and stronger Balmer jumps to better fit solutions rejected by JADES templates, but there are some sources with very scattered photometry that are fit with extremely blue $\beta<-2.8$ UV slopes that JADES templates better fit with strong Balmer emission lines (Similar to those in Figure \ref{fig:simbalmer}).

\begin{figure}%
    \centering
    \includegraphics[width=9cm]{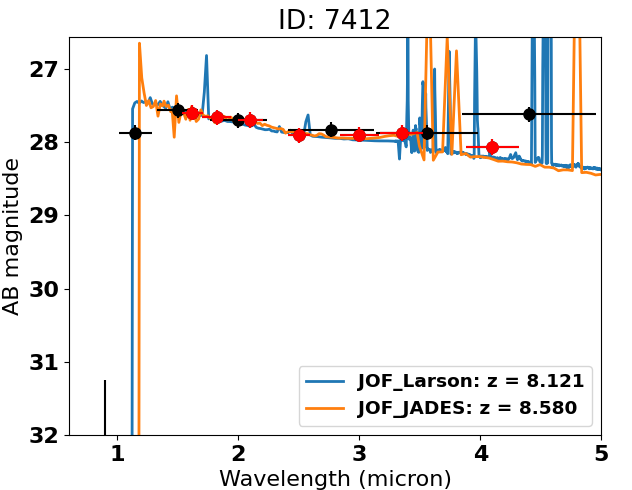} %
    \caption{Example galaxy SED at $z\sim8.5$ displaying the systematic redshift difference, driven by Ly$\alpha$, when the JADES template set \citep{Hainline2020} is used verses the Larson template set \citep{Larson2023}.}%
    \label{fig:jadesSED}%
\end{figure}

Finally, we test the JADES template-based selection in the UV LF, we find that our $z=10.5$ and $z=12.5$ UV LF's are unchanged within $1\sigma$ with the exception of the JADES templates selecting a $z=12.5$ source at $M_{\rm UV} = -23.2$ which has a dubiously red SED shape. This is largely because changes to the sample are occurring below the 30\% completeness or 30\% contamination cuts before the UV LF is measured. For the $z=9$ UV LF (which we display in Figure \ref{fig:jadesLF}), the narrower redshift bin, larger number of sources and $\sim0.15$ systematic shift increases the number densities measured by a factor of 3.5 at $M_{\rm UV} = -20$ whilst the faint end shifts by smaller amounts. All new, brighter sources have a photo-z with the JADES templates at $8.5<z<8.7$ whilst the Larson templates fit these at $7.9<z<8.2$ (an example SED is shown in Figure \ref{fig:jadesSED}) To solve this discrepancy more decisively, F430M, F460M and F480M measurements or spectroscopy would be required in order to constrain the location of the H$\beta$/[OIII] complex.

\section{Full list of high-z galaxy candidates}

In this appendix, we provide a full table of the high redshift galaxies selected in our `final' sample (catalogues using the full number of NIRCam bands available). We highlight which objects were previously identified in \citet{Robertson2024} as well as which objects are selected in only the 0.2as aperture version of our SED fits. Below the table, we show a selection of example SED fits across the full redshift range considered.

\clearpage
\onecolumn
\begin{longtable}{c|ll|lllllll|ll}
\caption{The full tabulated list of selected high-z targets in the JADES Origins Field. The first set of objects consist of those selected in the fiducial 0.32as aperture catalogue. The second set of objects show which objects are added when smaller 0.2as apertures are implemented. Shown are the best fit photometric redshift, photo-z at either side $1\sigma$ value, the total $M_{\rm UV}$, aperture corrected F277W magnitude and the next best-fitting low-z solution at $z<6$. Values of -1.0 indicate a failure to obtain a solution.} \\
EPOCHS ID & RA       & DEC       & zphot & zphot\_16 & zphot\_84 & $\chi^2$  & MUV    & F277W & $\sigma_{\rm F277W}$ & zphot$_{(z<6)}$ & $\chi^2$\_z6 \\ \hline
\textbf{0.32as} \\ \hline
10572  & 53.07346     & -27.85685    & 7.51     & 7.05         & 7.72         & 13.82         & -17.37    & 30.83            & 3.07         & 1.16     & 21.14         \\
10578  & 53.07332     & -27.85679    & 7.56     & 7.31         & 7.74         & 10.93         & -18.08    & 29.85            & 7.63         & 1.50     & 48.97         \\
9971   & 53.03078     & -27.88840    & 7.57     & 7.27         & 7.59         & 8.95          & -17.99    & 29.41            & 14.46        & 1.50     & 50.10         \\
288    & 53.07920     & -27.88182    & 7.65     & 7.60         & 7.72         & 14.71         & -17.96    & 29.63            & 8.66         & 0.29     & 28.06         \\
710    & 53.03646     & -27.91106    & 7.81     & 7.72         & 7.95         & 6.85          & -18.33    & 29.76            & 10.14        & 1.52     & 35.89         \\
6491   & 53.09007     & -27.85578    & 7.85     & 7.50         & 8.09         & 9.49          & -17.61    & 30.02            & 6.17         & 1.50     & 21.14         \\
1676   & 53.09919     & -27.86336    & 7.85     & 7.77         & 8.23         & 21.96         & -17.82    & 30.04            & 7.36         & 1.74     & 38.73         \\
2799   & 53.07720     & -27.87574    & 7.87     & 7.00         & 7.95         & 25.21         & -18.22    & 29.44            & 11.24        & 1.65     & 35.27         \\
9064   & 53.04792     & -27.87872    & 7.93     & 7.86         & 8.06         & 2.49          & -19.54    & 28.37            & 29.86        & 1.71     & 87.13         \\
11106  & 53.04431     & -27.87621    & 7.93     & 7.78         & 8.14         & 7.74          & -18.14    & 29.68            & 9.73         & 1.47     & 24.38         \\
9170   & 53.06998     & -27.86297    & 7.93     & 7.84         & 8.39         & 24.94         & -17.58    & 30.18            & 7.57         & 2.35     & 33.41         \\
7555   & 53.08745     & -27.85464    & 7.97     & 7.87         & 8.20         & 3.40          & -19.01    & 28.89            & 19.65        & 1.70     & 64.14         \\
4134   & 53.08222     & -27.86811    & 7.98     & 7.90         & 8.05         & 14.30         & -18.77    & 29.22            & 14.95        & 1.69     & 46.63         \\
5553   & 53.08741     & -27.86040    & 8.02     & 7.91         & 8.09         & 4.28          & -19.69    & 28.15            & 30.24        & 1.70     & 106.57        \\
7917   & 53.05373     & -27.87789    & 8.08     & 8.00         & 8.20         & 6.25          & -19.91    & 28.04            & 43.20        & 1.72     & 103.55        \\
6508   & 53.03580     & -27.89470    & 8.10     & 7.99         & 8.20         & 6.63          & -18.07    & 29.83            & 9.46         & 1.70     & 43.34         \\
7412   & 53.07052     & -27.86725    & 8.12     & 7.96         & 8.27         & 2.16          & -19.73    & 28.15            & 41.52        & 1.71     & 96.47         \\
14266  & 53.07027     & -27.84821    & 8.16     & 7.92         & 8.32         & 15.29         & -18.07    & 29.83            & 7.65         & 1.50     & 48.47         \\
6176   & 53.08678     & -27.85915    & 8.21     & 7.99         & 8.28         & 20.23         & -18.99    & 28.99            & 16.63        & 1.68     & 117.79        \\
5797   & 53.03769     & -27.89540    & 8.22     & 8.13         & 8.48         & 11.37         & -17.79    & 29.97            & 8.86         & 1.70     & 21.26         \\
5687   & 53.03334     & -27.89862    & 8.24     & 8.12         & 8.35         & 18.98         & -18.80    & 28.73            & 23.91        & 1.87     & 47.37         \\
885    & 53.08932     & -27.87269    & 8.26     & 8.18         & 8.35         & 5.55          & -19.63    & 28.26            & 40.38        & 1.70     & 129.98        \\
2162   & 53.10223     & -27.85926    & 8.29     & 8.18         & 8.44         & 2.41          & -19.93    & 28.28            & 35.07        & 1.70     & 104.73        \\
6111   & 53.08650     & -27.85920    & 8.32     & 8.22         & 8.40         & 20.12         & -19.82    & 27.85            & 50.58        & 1.75     & 65.44         \\
5376   & 53.08739     & -27.86032    & 8.32     & 8.19         & 8.40         & 7.34          & -20.61    & 27.72            & 45.61        & 1.94     & 87.57         \\
946    & 53.03658     & -27.91026    & 8.37     & 8.07         & 8.52         & 6.66          & -17.90    & 30.01            & 8.17         & 2.36     & 22.49         \\
3004   & 53.09827     & -27.85996    & 8.40     & 8.09         & 8.52         & 16.31         & -18.69    & 29.18            & 14.23        & 1.86     & 20.41         \\
3642   & 53.09238     & -27.86232    & 8.46     & 8.34         & 8.58         & 8.79          & -19.03    & 28.79            & 22.91        & 2.11     & 40.10         \\
3397   & 53.03386     & -27.90509    & 8.46     & 8.26         & 8.63         & 2.68          & -18.58    & 29.54            & 10.75        & 1.89     & 51.07         \\
4276   & 53.09083     & -27.86177    & 8.47     & 8.24         & 8.59         & 9.82          & -18.30    & 29.36            & 11.29        & 2.11     & 16.00         \\
9661   & 53.07849     & -27.85541    & 8.54     & 8.38         & 8.67         & 36.38         & -18.57    & 29.34            & 10.77        & 2.12     & 41.35         \\
10130  & 53.08105     & -27.85234    & 8.55     & 8.14         & 8.73         & 22.64         & -17.72    & 30.02            & 6.81         & 2.36     & 42.67         \\
1669   & 53.09686     & -27.86505    & 8.58     & 8.43         & 8.69         & 4.31          & -18.85    & 29.17            & 16.21        & 2.04     & 66.04         \\
9005   & 53.07037     & -27.86308    & 8.58     & 8.37         & 8.74         & 16.48         & -18.07    & 29.77            & 11.21        & 1.89     & 53.95         \\
3830   & 53.05999     & -27.88514    & 8.62     & 8.47         & 8.77         & 6.54          & -18.54    & 29.63            & 12.74        & 0.20     & 26.83         \\
9571   & 53.08794     & -27.84897    & 8.64     & 8.32         & 8.84         & 17.53         & -17.96    & 30.05            & 6.41         & 0.47     & 21.69         \\
15874  & 53.06717     & -27.84727    & 8.68     & 8.52         & 8.82         & 4.74          & -18.64    & 29.36            & 13.97        & 2.18     & 65.56         \\
15160  & 53.07205     & -27.84206    & 8.72     & 8.63         & 8.84         & 15.00         & -19.79    & 28.54            & 27.17        & 2.18     & 61.10         \\
2248   & 53.05415     & -27.89358    & 8.88     & 8.71         & 8.97         & 13.86         & -19.13    & 28.86            & 20.45        & 2.47     & 44.41         \\
15521  & 53.07213     & -27.84195    & 8.98     & 8.90         & 9.06         & 6.14          & -19.48    & 28.36            & 31.09        & 2.47     & 64.29         \\
4047   & 53.05833     & -27.88486    & 9.01     & 8.95         & 9.13         & 12.68         & -20.42    & 27.91            & 51.12        & 2.46     & 48.11         \\
14984  & 53.04118     & -27.86408    & 9.09     & 8.86         & 9.29         & 21.89         & -17.66    & 30.38            & 5.22         & 2.00     & 27.80         \\
6775   & 53.03080     & -27.89738    & 9.16     & 8.98         & 9.55         & 3.85          & -18.10    & 29.69            & 9.53         & 2.52     & 24.24         \\
9143   & 53.06774     & -27.86455    & 9.24     & 9.15         & 10.07        & 16.41         & -18.57    & 29.50            & 12.66        & 2.30     & 74.01         \\
4305   & 53.09749     & -27.85678    & 9.29     & 9.23         & 10.30        & 7.70          & -18.14    & 29.56            & 10.74        & 2.47     & 36.42         \\
4270   & 53.09748     & -27.85698    & 9.40     & 9.34         & 9.96         & 5.04          & -18.72    & 29.50            & 11.53        & 2.29     & 66.37         \\
2875   & 53.09872     & -27.86018    & 9.41     & 9.42         & 10.40        & 12.19         & -17.97    & 29.88            & 8.16         & 2.47     & 28.42         \\
14973  & 53.01911     & -27.87990    & 9.44     & 9.23         & 10.04        & 11.18         & -17.82    & 30.26            & 6.31         & 2.00     & 15.80         \\
10098  & 53.07608     & -27.85602    & 9.64     & 9.59         & 10.30        & 7.06          & -18.26    & 29.70            & 8.26         & 2.30     & 48.94         \\
9687   & 53.08537     & -27.85047    & 9.68     & 9.53         & 10.01        & 10.11         & -18.17    & 30.00            & 6.01         & 2.31     & 45.95         \\
13546  & 53.06020     & -27.85626    & 9.71     & 9.36         & 10.00        & 5.67          & -18.05    & 30.38            & 5.76         & 2.28     & 39.55         \\
9890   & 53.04680     & -27.87758    & 10.11    & 9.52         & 10.56        & 14.36         & -17.79    & 30.30            & 6.83         & 0.18     & 20.12         \\
14765  & 53.01654     & -27.88195    & 10.12    & 9.49         & 10.28        & 11.72         & -18.36    & 29.93            & 9.17         & 2.35     & 57.76         \\
5798   & 53.09909     & -27.85143    & 10.82    & 10.27        & 11.10        & 16.76         & -18.19    & 29.99            & 7.07         & 2.62     & 23.18         \\
1708   & 53.06006     & -27.89143    & 11.16    & 10.81        & 11.47        & 18.70         & -18.05    & 30.42            & 5.19         & 0.03     & 36.18         \\
6548   & 53.04767     & -27.88625    & 11.22    & 10.85        & 11.50        & 44.38         & -19.46    & 30.61            & 5.31         & 0.09     & 51.99         \\
11919*  & 53.02617     & -27.88716    & 11.41    & 11.05        & 11.69        & 14.09         & -18.04    & 30.39            & 6.40         & 0.06     & 33.34         \\
11424*  & 53.07248     & -27.85535    & 11.88    & 11.22        & 12.04        & 16.27         & -18.08    & 29.65            & 8.96         & 3.25     & 48.80         \\
3219   & 53.07804     & -27.87366    & 12.06    & 11.37        & 12.25        & 26.46         & -18.71    & 29.45            & 10.05        & 3.27     & 38.48         \\
8915*   & 53.02867     & -27.89301    & 12.30    & 12.14        & 12.56        & 24.43         & -18.85    & 29.83            & 9.25         & 3.27     & 50.71         \\
16010  & 53.05908     & -27.85070    & 12.77    & 12.67        & 13.87        & 15.83         & -19.64    & 30.04            & 7.34         & 3.34     & 23.95         \\
718*    & 53.10763     & -27.86013    & 14.66    & 14.46        & 15.51        & 11.98         & -18.58    & 29.95            & 8.96         & 4.17     & 27.18       \\ \hline
\textbf{0.2as} \\ \hline
1512   & 53.05454     & -27.89585    & 7.52     & 7.41      & 7.60      & 21.70         & -17.37    & 30.58            & 7.58         & 0.23     & 75.95         \\
11141  & 53.07165     & -27.85667    & 7.53     & 7.44      & 7.94      & 14.31         & -17.41    & 30.82            & 6.46         & 0.20     & 58.63         \\
386    & 53.08855     & -27.87444    & 7.53     & 7.28      & 7.56      & 18.42         & -17.45    & 30.61            & 7.13         & 1.43     & 50.60         \\
3532   & 53.04423     & -27.89679    & 7.66     & 7.61      & 8.04      & 22.45         & -18.14    & 30.33            & 8.77         & 0.24     & 30.58         \\
15041  & 53.05937     & -27.85099    & 7.71     & 7.69      & 8.07      & 24.79         & -17.55    & 30.31            & 7.81         & 1.40     & 51.76         \\
11939  & 53.07852     & -27.84957    & 7.74     & 7.40      & 7.89      & 10.22         & -17.68    & 30.28            & 9.04         & 1.66     & 57.95         \\
1197   & 53.10860     & -27.85810    & 7.75     & 7.71      & 8.12      & 30.55         & -17.54    & 30.95            & 5.40         & 0.47     & 45.28         \\
8936   & 53.07009     & -27.86345    & 7.84     & 6.78      & 8.04      & 20.41         & -17.77    & 30.90            & 5.80         & 1.92     & 24.46         \\
11322  & 53.09278     & -27.84107    & 7.86     & 7.37      & 7.93      & 32.64         & -17.64    & 30.74            & 6.06         & 1.50     & 84.06         \\
11754  & 53.08363     & -27.84632    & 7.88     & 7.73      & 8.10      & 25.07         & -17.16    & 30.52            & 6.44         & 1.74     & 42.40         \\
13696  & 53.07229     & -27.84706    & 7.88     & 7.82      & 7.95      & 19.69         & -17.31    & 31.26            & 2.62         & 2.19     & 58.33         \\
11370  & 53.06216     & -27.86298    & 7.90     & 7.74      & 8.05      & 29.45         & -17.79    & 30.94            & 5.26         & 0.09     & 35.56         \\
11772  & 53.09057     & -27.84137    & 8.01     & 7.81      & 8.28      & 12.43         & -17.30    & 30.96            & 5.09         & 2.01     & 18.27         \\
4155   & 53.03524     & -27.90195    & 8.05     & 7.93      & 8.12      & 28.91         & -17.59    & 30.90            & 5.93         & 0.12     & 52.68         \\
15769  & 53.07395     & -27.84454    & 8.07     & 7.80      & 8.19      & 38.43         & -17.93    & 30.60            & 8.35         & 1.50     & 80.36         \\
11891  & 53.06338     & -27.86048    & 8.11     & 7.45      & 8.22      & 14.72         & -17.22    & 30.84            & 6.02         & 0.19     & 23.63         \\
2176   & 53.07836     & -27.87688    & 8.13     & 7.90      & 8.31      & 16.47         & -17.43    & 30.79            & 6.35         & 0.18     & 24.69         \\
14290  & 53.04504     & -27.86981    & 8.14     & 8.01      & 8.27      & 22.58         & -18.02    & 29.87            & 14.63        & 0.02     & 31.90         \\
2338   & 53.07664     & -27.87763    & 8.15     & 8.00      & 8.50      & 23.60         & -17.30    & 30.29            & 9.85         & 1.91     & 53.78         \\
12988  & 53.07259     & -27.85059    & 8.27     & 8.12      & 8.41      & 24.64         & -18.09    & 30.40            & 8.03         & 0.23     & 74.61         \\
5484   & 53.08417     & -27.86305    & 8.29     & 7.98      & 8.39      & 32.54         & -17.89    & 30.39            & 8.24         & 2.11     & 74.44         \\
4828   & 53.04772     & -27.89114    & 8.30     & 8.21      & 8.43      & 23.62         & -17.64    & 30.58            & 8.95         & 2.24     & 62.56         \\
594    & 53.04853     & -27.90285    & 8.31     & 8.14      & 8.40      & 28.78         & -18.53    & 30.72            & 8.27         & 2.13     & 48.20         \\
3898   & 53.10441     & -27.85306    & 8.32     & 8.26      & 8.38      & 30.26         & -18.44    & 29.86            & 14.27        & 0.02     & 128.35        \\
8793   & 53.08600     & -27.85233    & 8.32     & 8.12      & 8.43      & 11.76         & -17.86    & 30.33            & 8.82         & 2.17     & 29.33         \\
12077  & 53.09063     & -27.84029    & 8.36     & 8.14      & 8.50      & 21.13         & -18.36    & 30.31            & 9.03         & 2.35     & 25.74         \\
12488  & 53.07285     & -27.85193    & 8.38     & 8.27      & 8.52      & 21.73         & -18.40    & 30.45            & 7.68         & 1.90     & 36.79         \\
2403   & 53.09886     & -27.86163    & 8.41     & 8.27      & 8.79      & 8.97          & -17.06    & 31.21            & 4.61         & 2.10     & 35.92         \\
2073   & 53.07734     & -27.87797    & 8.44     & 8.22      & 8.65      & 22.32         & -17.43    & 30.96            & 5.00         & 1.85     & 29.85         \\
12179  & 53.08439     & -27.84471    & 8.47     & 8.31      & 8.60      & 14.49         & -17.68    & 30.62            & 6.79         & 0.23     & 46.55         \\
9825   & 53.09241     & -27.84489    & 8.47     & 8.31      & 8.58      & 31.58         & -17.74    & 29.84            & 17.31        & 2.11     & 36.73         \\
4514   & 53.03171     & -27.90348    & 8.48     & 8.21      & 8.93      & 20.98         & -17.25    & 30.84            & 6.28         & 2.27     & 32.92         \\
8785   & 53.09541     & -27.84554    & 8.49     & 8.32      & 8.71      & 24.76         & -17.86    & 30.39            & 8.34         & 2.99     & 38.31         \\
8123   & 53.03554     & -27.89055    & 8.50     & 8.37      & 8.66      & 10.92         & -18.23    & 30.10            & 13.04        & 2.11     & 20.00         \\
8712   & 53.07131     & -27.86311    & 8.52     & 8.31      & 8.65      & 15.86         & -17.70    & 30.37            & 9.13         & 2.13     & 20.28         \\
13158  & 53.04123     & -27.86910    & 8.56     & 8.43      & 8.65      & 21.64         & -17.77    & 30.14            & 12.49        & 0.18     & 31.66         \\
3995   & 53.05173     & -27.89058    & 8.56     & 8.29      & 8.67      & 23.78         & -17.40    & 30.30            & 8.18         & 0.16     & 30.44         \\
7232   & 53.04621     & -27.88523    & 8.59     & 8.40      & 8.81      & 18.80         & -18.46    & 30.57            & 8.25         & 0.20     & 30.29         \\
13492  & 53.03934     & -27.87225    & 8.63     & 8.45      & 8.81      & 27.65         & -17.40    & 30.85            & 6.65         & 2.35     & 76.33         \\
16322  & 53.03939     & -27.86765    & 8.66     & 8.54      & 8.80      & 15.56         & -17.98    & 30.17            & 11.13        & 0.02     & 27.79         \\
9360   & 53.06697     & -27.86450    & 8.66     & 8.56      & 8.83      & 15.90         & -17.22    & 30.41            & 10.09        & 0.29     & 25.79         \\
6640   & 53.03028     & -27.89825    & 8.68     & 8.51      & 8.88      & 12.32         & -18.29    & 30.60            & 7.24         & 2.28     & 17.26         \\
10659  & 53.07008     & -27.85872    & 8.74     & 8.66      & 8.83      & 27.95         & -19.52    & 29.52            & 21.59        & 0.15     & 32.46         \\
13359  & 53.07008     & -27.84823    & 8.75     & 8.59      & 9.01      & 16.57         & -17.52    & 30.60            & 7.29         & 2.62     & 26.83         \\
7535   & 53.06896     & -27.86831    & 8.77     & 8.61      & 8.96      & 23.97         & -17.91    & 30.81            & 6.37         & 0.21     & 36.09         \\
14107  & 53.03826     & -27.87207    & 8.83     & 8.65      & 8.98      & 39.45         & -17.54    & 30.84            & 7.34         & 0.11     & 56.33         \\
5411   & 53.04858     & -27.88875    & 8.84     & 8.62      & 9.14      & 15.93         & -17.93    & 31.01            & 4.72         & 1.89     & 29.04         \\
15649  & 53.02357     & -27.87983    & 8.91     & 8.70      & 9.01      & 15.97         & -18.21    & 30.58            & 11.77        & 2.29     & 26.92         \\
10670  & 53.05118     & -27.87224    & 9.01     & 8.86      & 9.21      & 13.25         & -17.93    & 30.23            & 13.02        & 2.31     & 21.03         \\
1491   & 53.09101     & -27.86974    & 9.17     & 9.00      & 9.82      & 41.15         & -18.14    & 30.62            & 8.15         & 2.99     & 52.41         \\
5717   & 53.09244     & -27.85635    & 9.21     & 8.96      & 9.74      & 4.63          & -17.17    & 30.85            & 5.20         & 2.57     & 18.35         \\
2471   & 53.08328     & -27.87218    & 9.24     & 9.12      & 9.87      & 14.12         & -18.22    & 30.17            & 9.48         & 2.56     & 20.08         \\
11756  & 53.09075     & -27.84136    & 9.38     & 9.26      & 9.96      & 14.23         & -17.40    & 30.53            & 7.48         & 2.52     & 32.25         \\
4805   & 53.04887     & -27.89045    & 9.72     & 9.51      & 10.16     & 6.90          & -17.96    & 30.33            & 10.11        & 2.31     & 52.71         \\
3336   & 53.05681     & -27.88898    & 10.02    & 9.53      & 10.25     & 16.49         & -17.67    & 30.76            & 7.09         & 2.30     & 43.09         \\
7099   & 53.09208     & -27.85278    & 10.22    & 9.40      & 10.35     & 8.85          & -18.32    & 30.07            & 10.20        & 2.32     & 48.26         \\
3599   & 53.05248     & -27.89100    & 10.52    & 10.36     & 10.78     & 11.50         & -19.84    & 30.20            & 9.19         & 0.22     & 21.20         \\
15074  & 53.07207     & -27.84174    & 10.63    & 10.39     & 10.85     & 11.88         & -18.47    & 30.91            & 6.22         & 2.32     & 81.79         \\
13294  & 53.02961     & -27.87890    & 10.65    & 10.22     & 11.00     & 9.82          & -17.44    & 31.06            & 5.18         & 0.19     & 22.89         \\
7361   & 53.09859     & -27.84729    & 10.67    & 10.49     & 10.90     & 7.89          & -18.04    & 30.00            & 12.38        & 2.72     & 22.79         \\
15323  & 53.03300     & -27.87110    & 10.78    & 10.35     & 11.11     & 11.04         & -17.74    & 30.73            & 6.57         & 0.18     & 18.45         \\
15870  & 53.04149     & -27.86602    & 10.88    & 10.73     & 11.09     & 48.65         & -18.20    & 30.24            & 11.58        & 0.13     & 60.15         \\
16055  & 53.02278     & -27.87873    & 10.90    & 10.61     & 11.17     & 13.74         & -17.58    & 31.09            & 6.42         & 0.02     & 17.78         \\
14618  & 53.02374     & -27.87805    & 10.96    & 10.63     & 11.48     & 10.87         & -17.15    & 31.36            & 5.06         & 2.63     & 26.86         \\
5303   & 53.07460     & -27.87039    & 11.29    & 11.05     & 11.47     & 29.84         & -18.29    & 30.49            & 8.72         & 0.19     & 36.06         \\
8658   & 53.09045     & -27.84970    & 11.50    & 11.30     & 11.66     & 29.46         & -17.57    & 30.91            & 5.02         & 3.02     & 37.09         \\
12327*  & 53.04017     & -27.87603    & 11.54    & 11.24     & 11.66     & 15.45         & -18.07    & 30.98            & 5.96         & 0.08     & 59.82         \\
11469  & 53.04222     & -27.87684    & 11.64    & 10.96     & 11.74     & 9.81          & -17.62    & 30.68            & 8.61         & 2.99     & 20.84         \\
2625   & 53.06360     & -27.88606    & 11.67    & 11.39     & 11.78     & 23.54         & -18.40    & 30.49            & 10.96        & 3.02     & 31.44         \\
4093   & 53.08716     & -27.86492    & 11.85    & 11.61     & 12.00     & 13.40         & -17.69    & 30.24            & 9.02         & 3.25     & 25.16         \\
11822  & 53.04744     & -27.87208    & 11.97    & 11.83     & 12.12     & 9.17          & -18.05    & 30.25            & 10.45        & 3.33     & 14.86         \\
4473   & 53.08357     & -27.86645    & 12.00    & 11.86     & 12.18     & 41.54         & -18.14    & 30.80            & 5.29         & 2.88     & 48.47         \\
12050  & 53.04683     & -27.87191    & 12.08    & 11.97     & 12.20     & 26.03         & -18.38    & 30.14            & 11.90        & 3.33     & 35.32         \\
4072*   & 53.08468     & -27.86666    & 12.53    & 12.42     & 13.36     & 10.02         & -17.80    & 30.33            & 9.61         & 3.55     & 26.41         \\
877    & 53.10468     & -27.86187    & 12.61    & 12.49     & 13.08     & 20.66         & -17.74    & 30.66            & 7.52         & 0.21     & 78.36         \\
918*    & 53.06475     & -27.89023    & 12.69    & 12.61     & 12.77     & 6.29          & -18.62    & 30.03            & 13.65        & 3.55     & 90.77         \\
14071  & 53.04293     & -27.87044    & 13.29    & 12.49     & 13.50     & 11.91         & -17.96    & 30.56            & 8.24         & 3.26     & 20.65         \\
4703*   & 53.03547     & -27.90036    & 13.96    & 13.70     & 14.23     & 20.33         & -18.20    & 30.55            & 7.71         & 0.62     & 29.62 
\label{tab:fulllist}
\end{longtable}
\footnotesize{Notes: * indicates object was selected in \citet{Robertson2024}}

\begin{figure*}%
    \centering
    \subfloat[\centering Example $z=8$ source, ID 7917]{{\includegraphics[width=7.5cm]{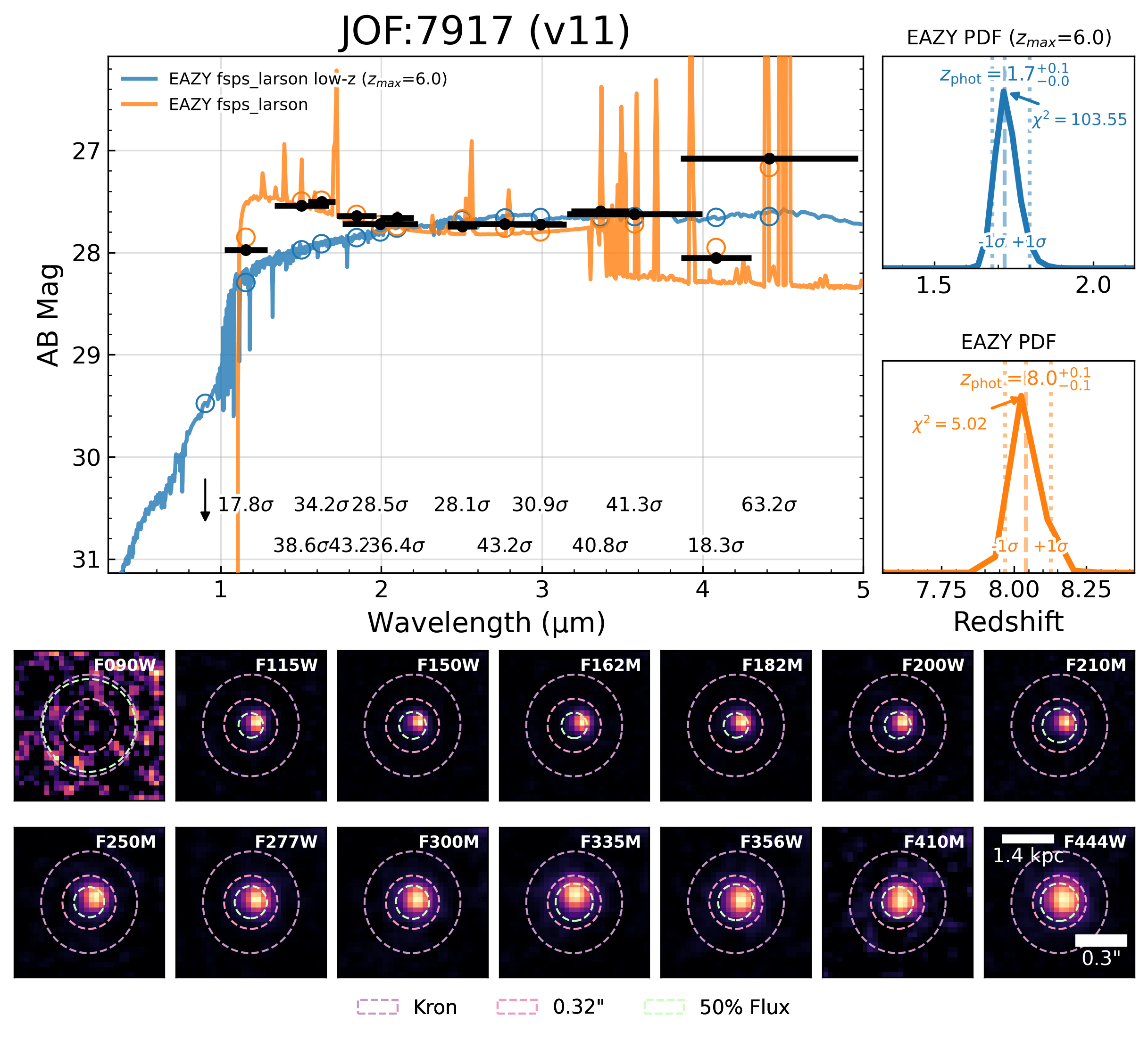} }}%
    \qquad
    \subfloat[\centering Example $z=8$ source, ID 946]{{\includegraphics[width=7.5cm]{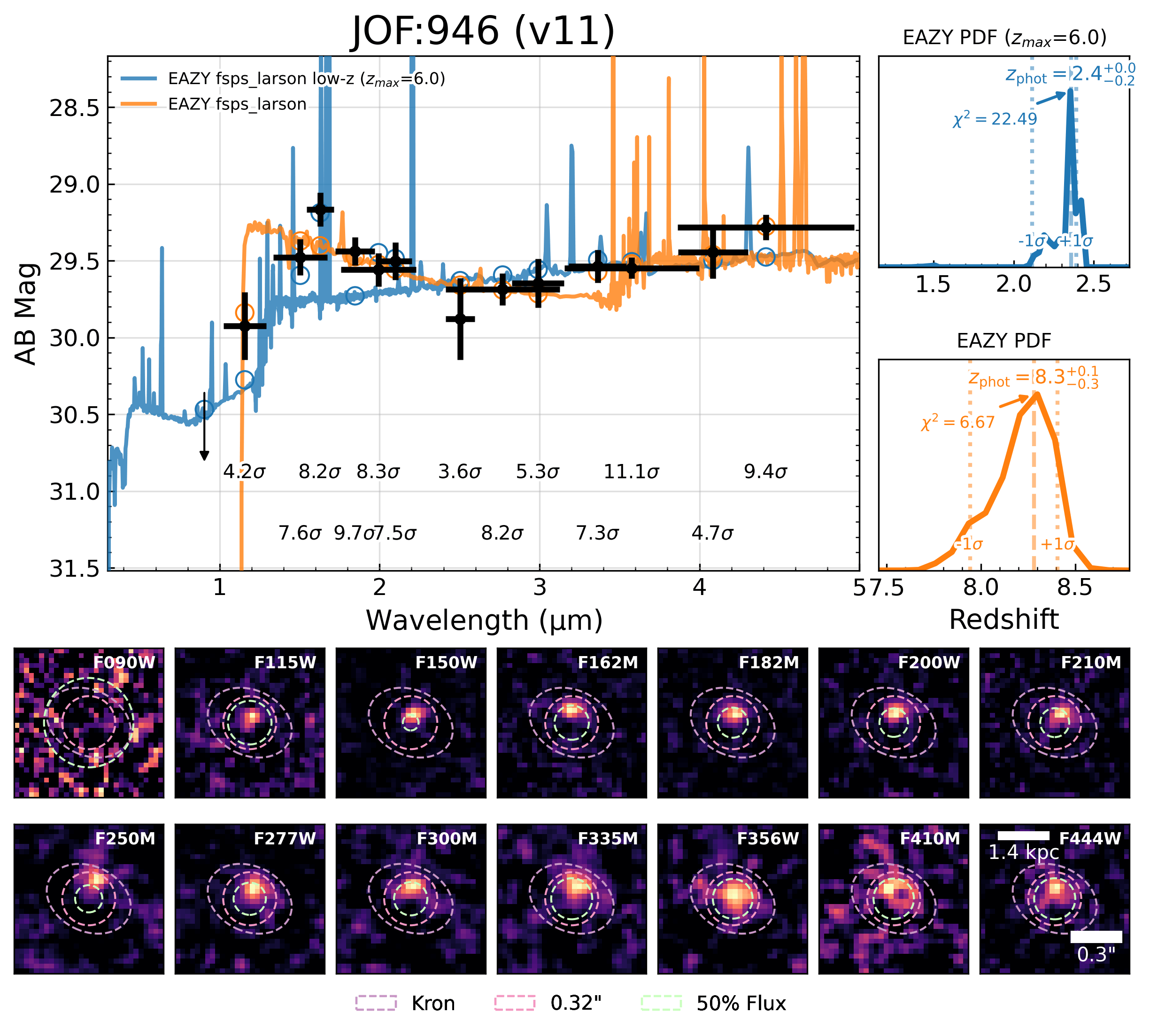} }}%
    \qquad
    \subfloat[\centering Example $z=9$ source, ID 4047]{{\includegraphics[width=7.5cm]{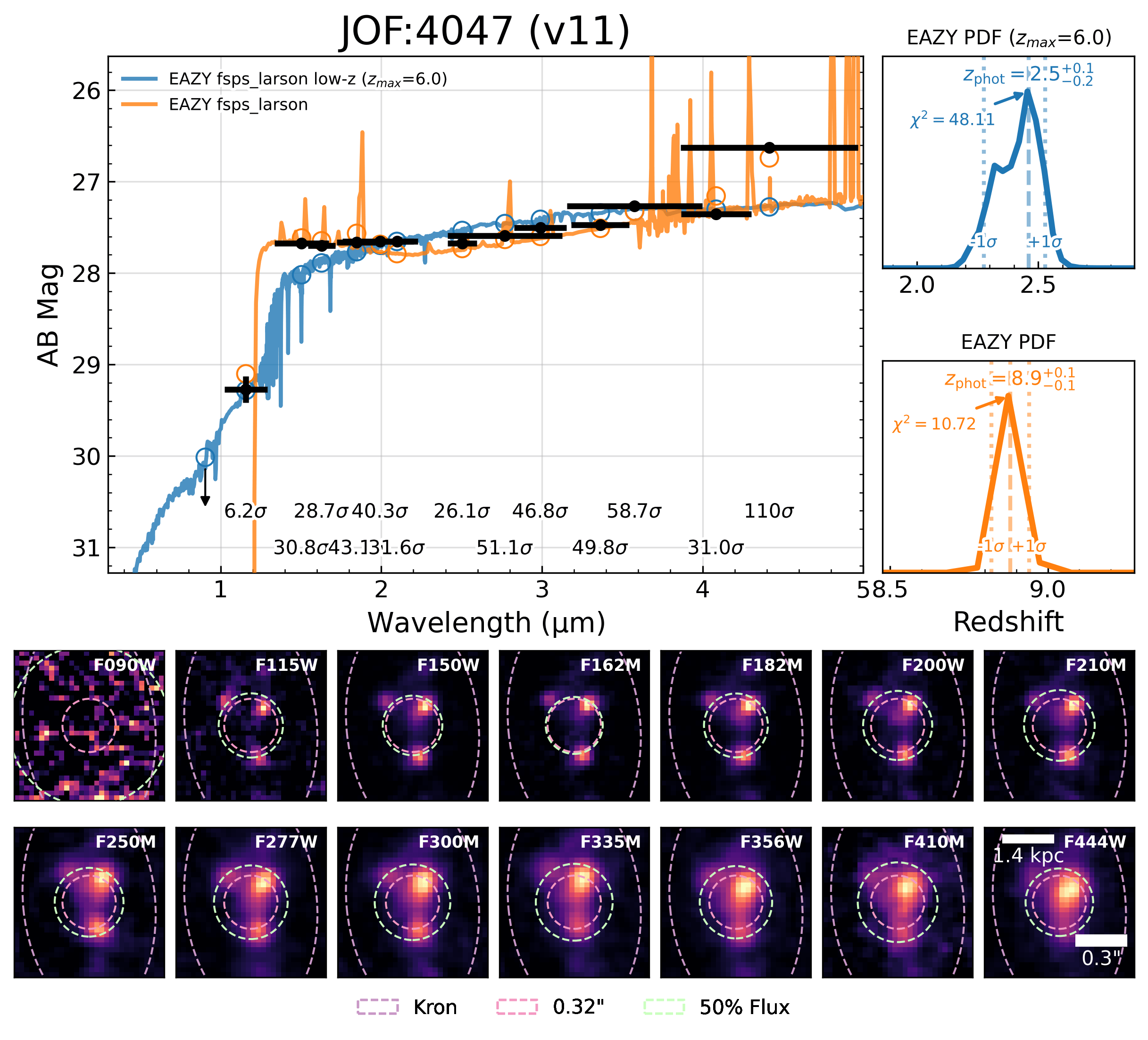} }}%
    \qquad
    \subfloat[\centering Example $z=9$ source, ID 1669]{{\includegraphics[width=7.5cm]{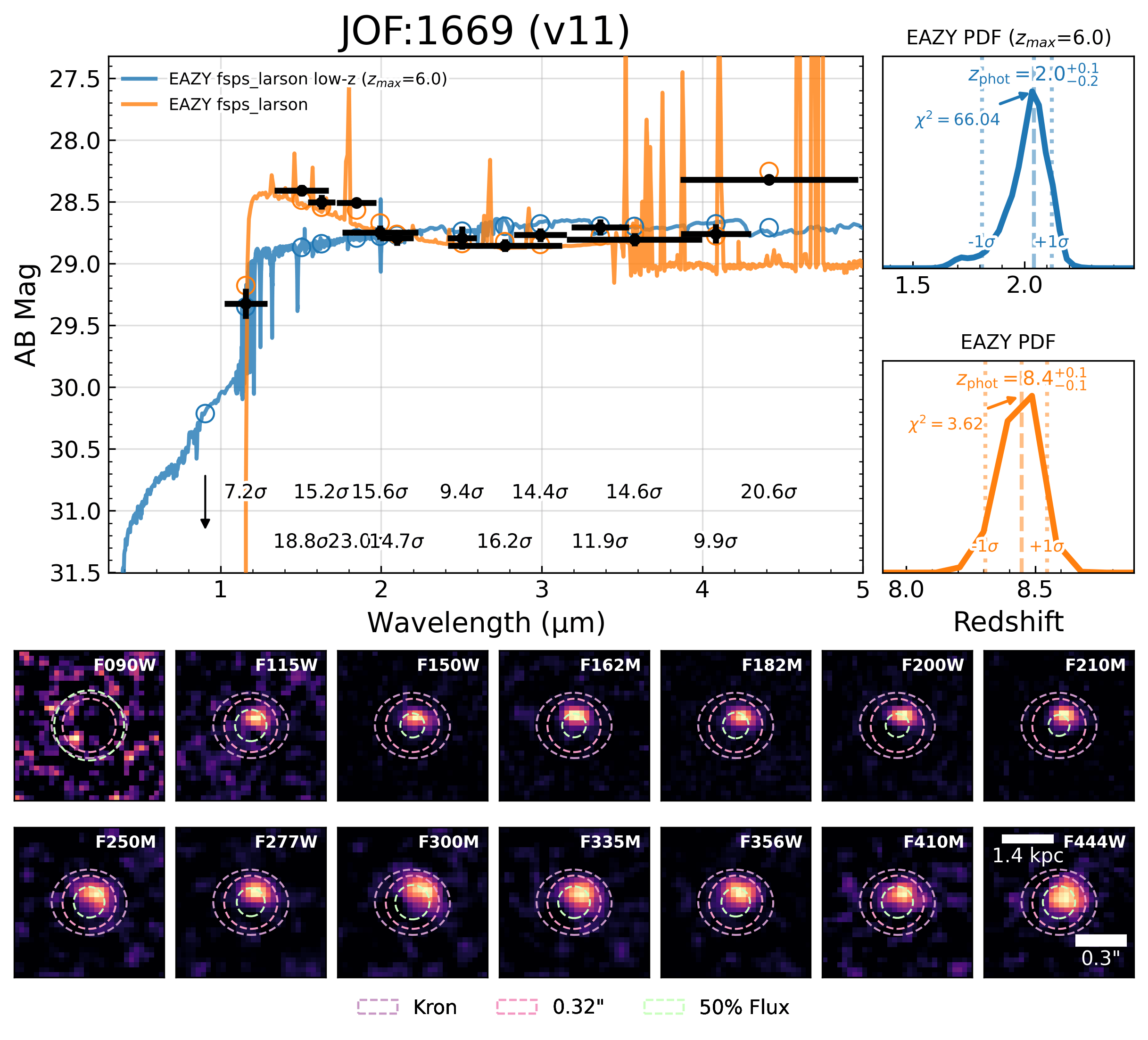} }}%
    \qquad
    \subfloat[\centering Example $z=10$ source, ID 10098]{{\includegraphics[width=7.5cm]{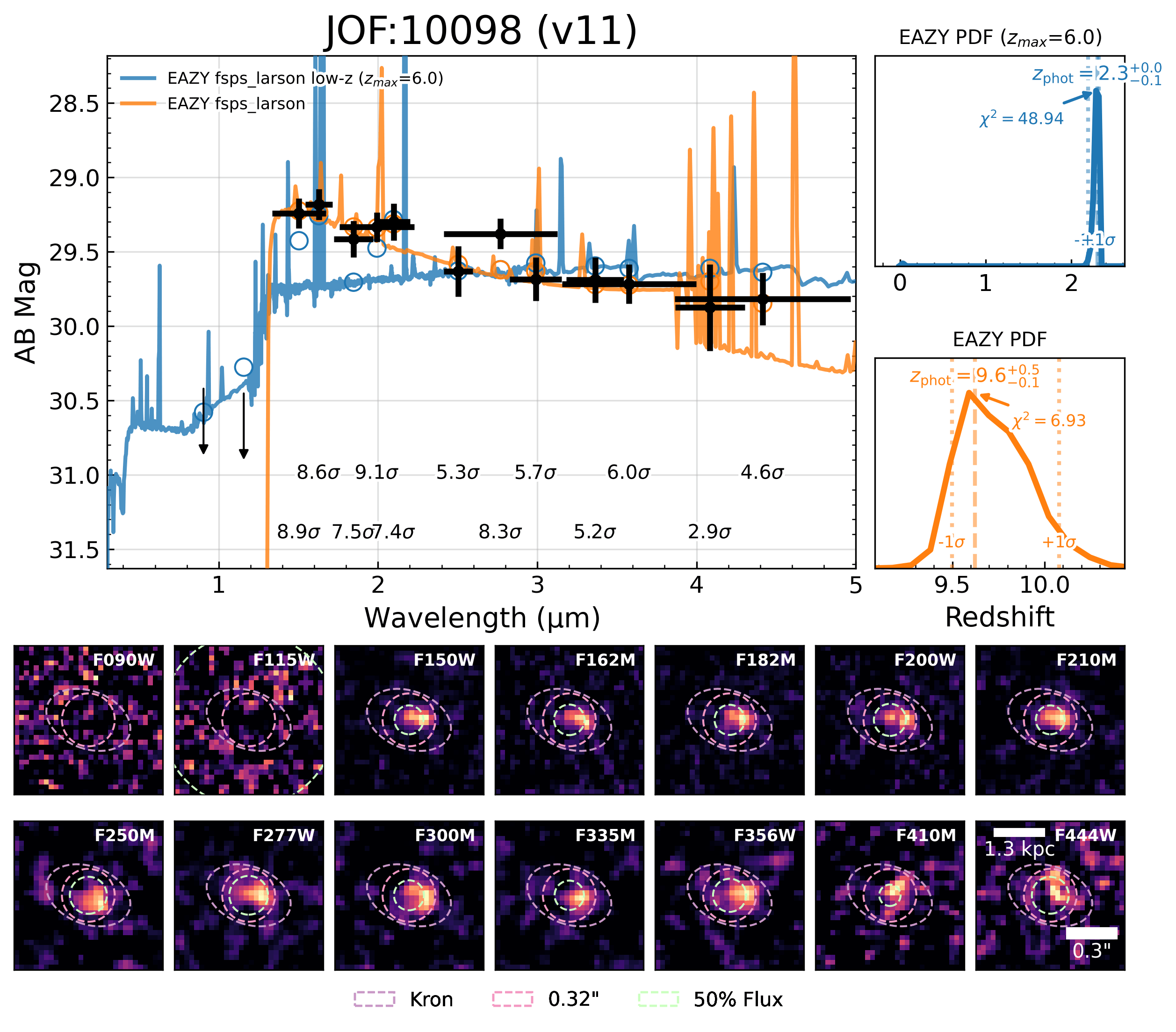} }}%
    \qquad
    \subfloat[\centering Example $z=10$ source, ID 9687]{{\includegraphics[width=7.5cm]{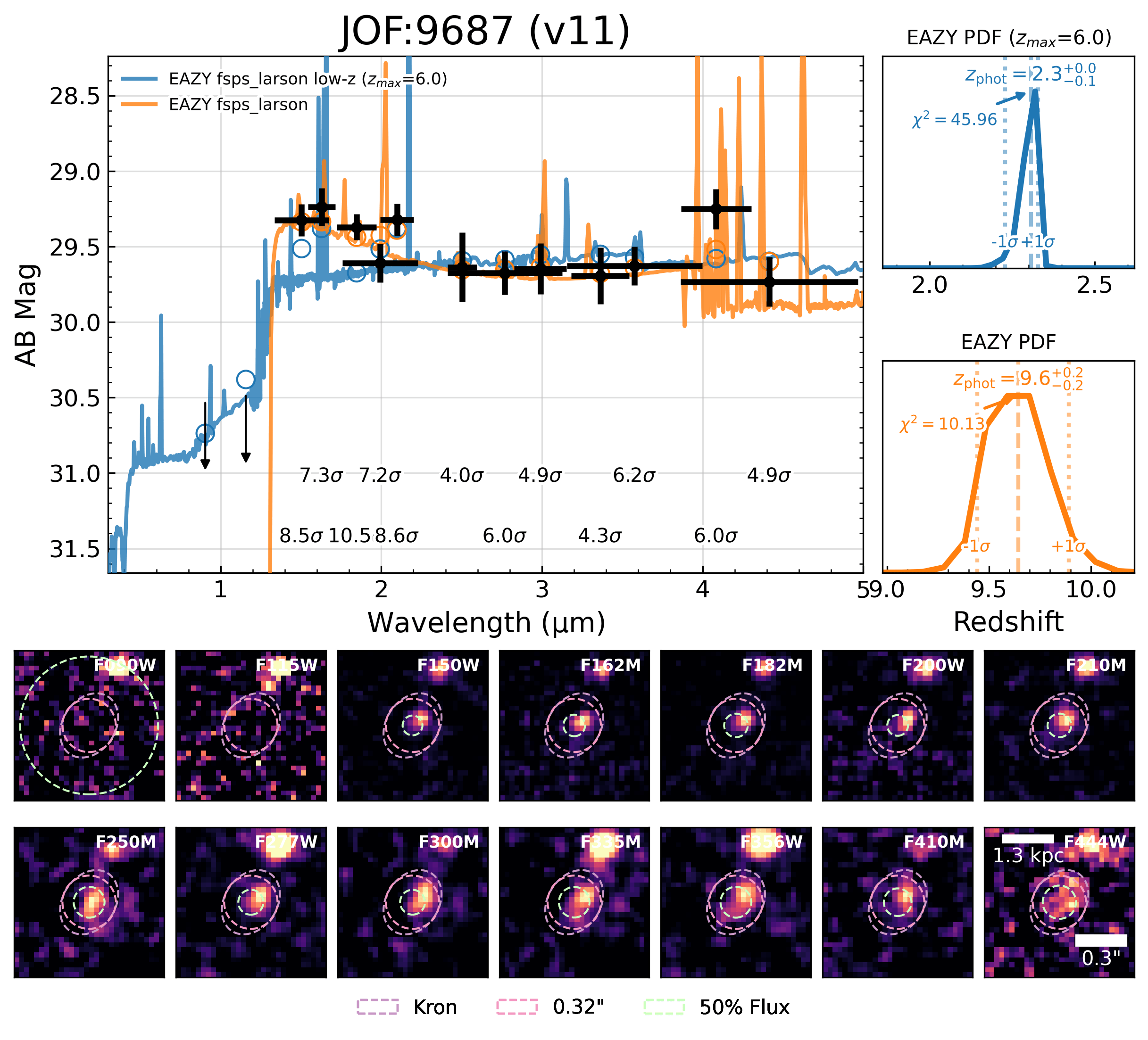} }}%
        \caption{A random selection of example SED fits for objects selected between $7.5<z<13.5$. Orange lines show the best high-z fit while the blue lines show the next best low-z fit at $z<6$, these are typically Balmer breaks or 1.6 micron bumps. ID's correspond to table rows from Table \ref{tab:fulllist}}%
\end{figure*}
\begin{figure*}\ContinuedFloat
    \subfloat[\centering Example $z=11$ source, ID 11919]{{\includegraphics[width=7.5cm]{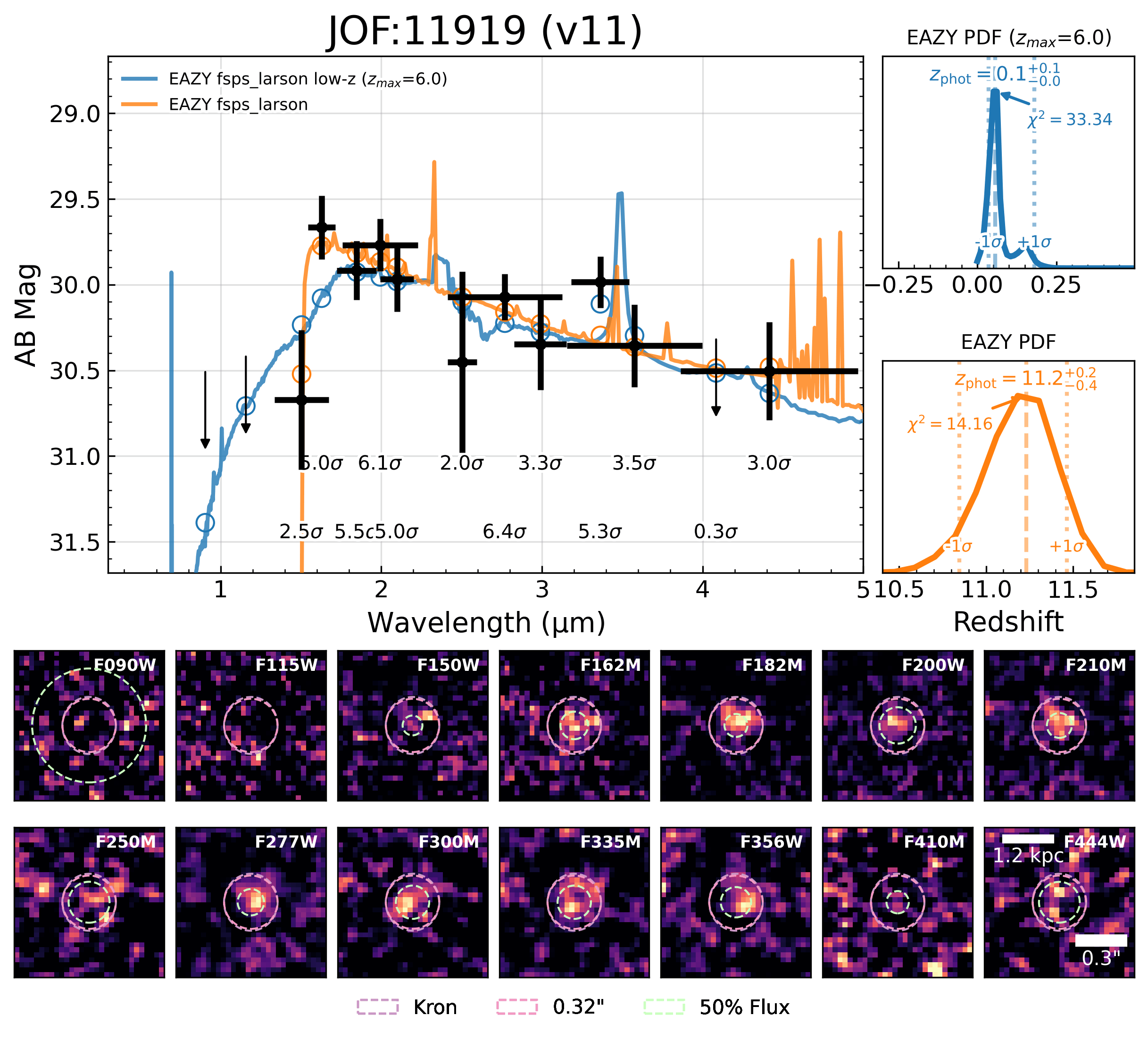} }}%
    \qquad
    \subfloat[\centering Example $z=11$ source, ID 1708]{{\includegraphics[width=7.5cm]{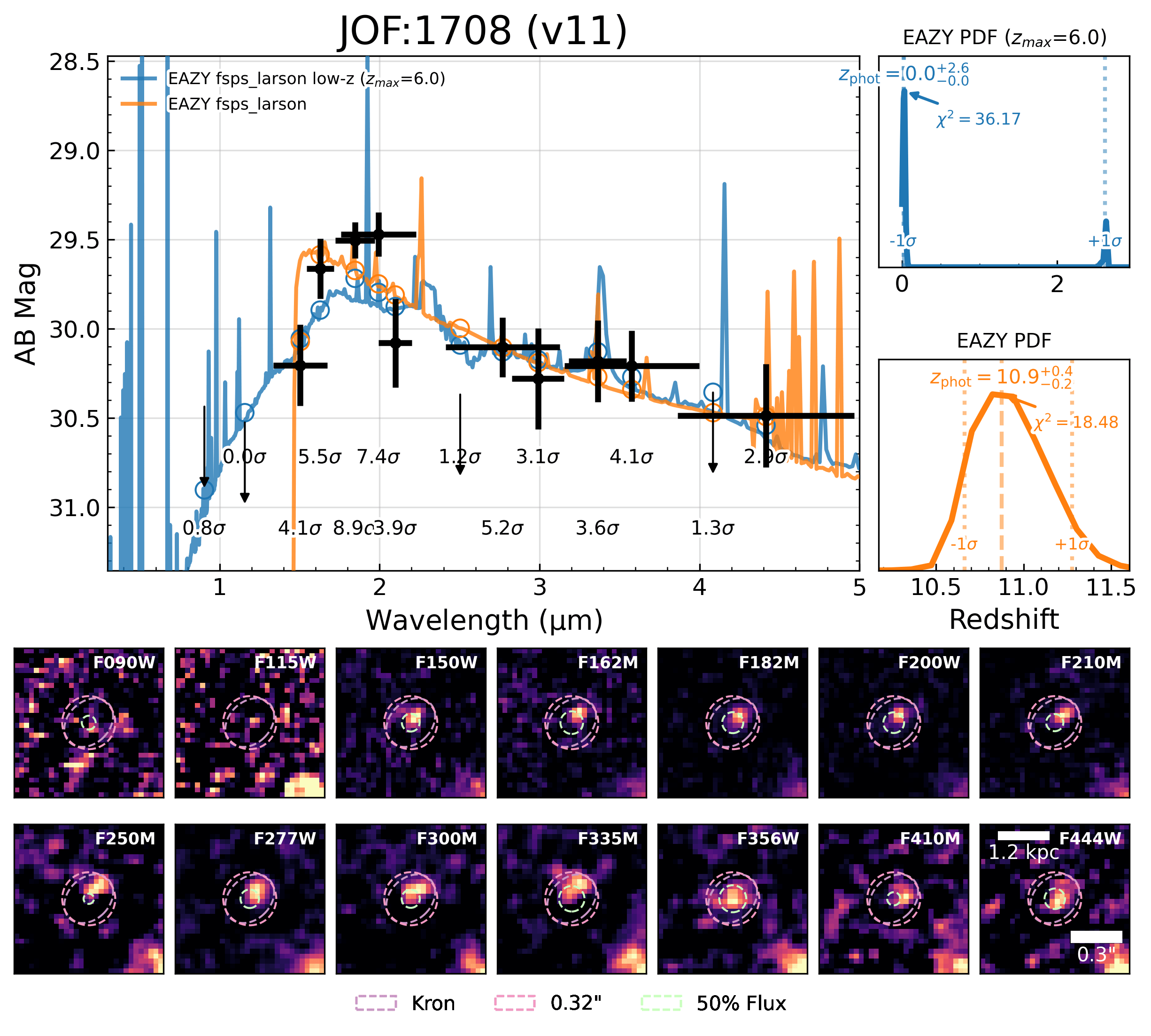} }}%
    \qquad
    \subfloat[\centering Example $z=12$ source, ID 8915]{{\includegraphics[width=7.5cm]{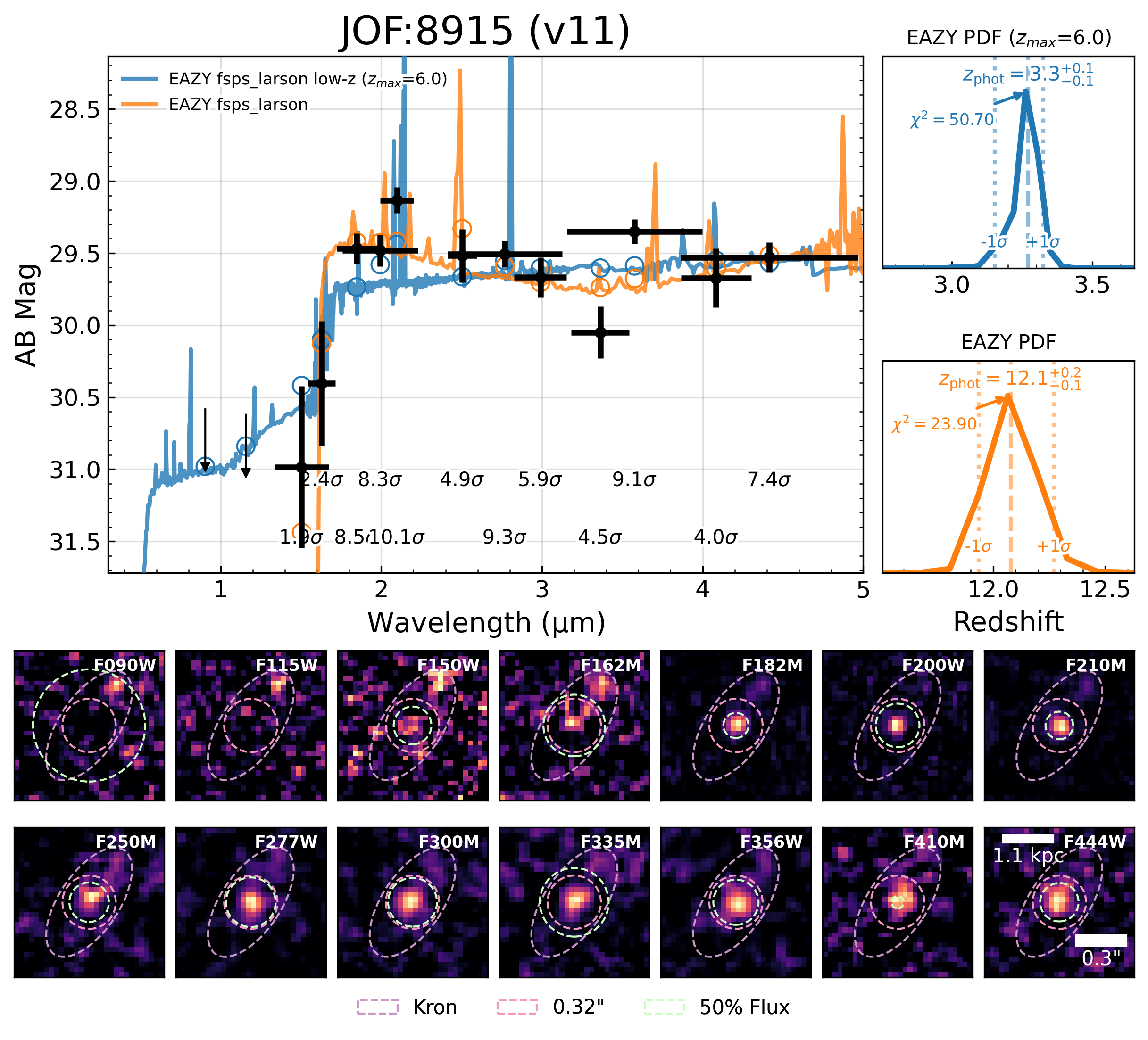} }}%
    \qquad
    \subfloat[\centering Example $z=13$ source, ID 16010]{{\includegraphics[width=7.5cm]{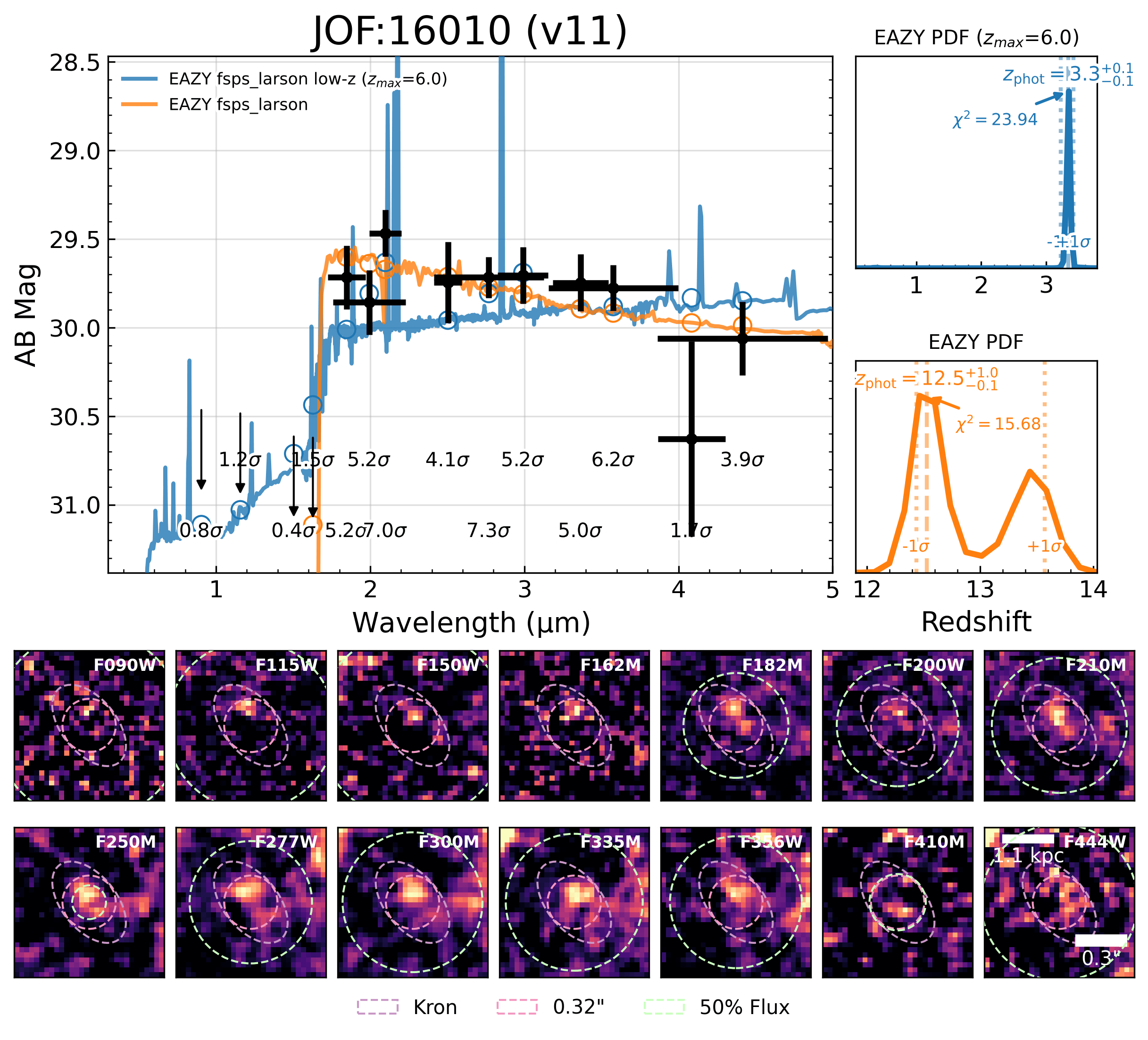} }}%
    \caption{A random selection of example SED fits for objects selected between $7.5<z<13.5$. Orange lines show the best high-z fit while the blue lines show the next best low-z fit at $z<6$, these are typically Balmer breaks or 1.6 micron bumps. ID's correspond to table rows from Table \ref{tab:fulllist}}%
    \label{fig:example}%
\end{figure*}


\bsp	
\label{lastpage}
\end{document}